\documentclass[11pt,preprint]{aastex}

\usepackage{ulem}

\accepted{17 Dec 2004}

\slugcomment{accepted for publication in {\it The Astrophysical Journal}}
\shorttitle{HVCs Toward PG\,1116+215}
\shortauthors{Ganguly et al.}

\begin{document}

\newcommand{\kms}{km\,s$^{-1}$}
\newcommand{\vlsr}{v_{\mathrm{LSR}}}
\newcommand{\acd}{N_{\mathrm{a}}}
\newcommand{\zabs}{z_{\mathrm{abs}}}

\title{Highly Ionized Gas in the Galactic Halo and the High Velocity Clouds
Toward PG\,1116+215\footnote{Based on observations made with the
NASA/ESA {\it Hubble Space Telescope}, which is operated by the
Association of Universities for Research in Astronomy, Inc., under
NASA contract NAS 5--26555. Also based on observations made with the
NASA-CNES-CSA {\it Far Ultraviolet Spectroscopic Explorer}, which is
operated for NASA by Johns Hopkins University under NASA contract NAS
5--32985.}}

\author{Rajib Ganguly\altaffilmark{1},
        Kenneth R.~Sembach\altaffilmark{1},
        Todd M.~Tripp\altaffilmark{2},
        Blair D.~Savage\altaffilmark{3}}
\altaffiltext{1}{Space Telescope Science Institute, 3700 San Martin Drive,
    Baltimore, MD  21218}
\altaffiltext{2}{Department of Astronomy, University of Massachusetts,
  Amherst, MA 01003}
\altaffiltext{3}{Department of Astronomy, University of Wisconsin-Madison,
    475 N. Charter Street, Madison, WI  53706}

\begin{abstract}
We have obtained high resolution FUSE and HST/STIS echelle
observations of the quasar {PG\,$1116+215$} ($z_{\mathrm{em}}=0.1765$,
$l=223\fdg36$, $b=+68\fdg21$). The semi-continuous coverage of the
ultraviolet spectrum over the wavelength range {916--2800\,\AA}
provides detections of Galactic and high velocity cloud (HVC)
absorption over a wide range of ionization species: {\ion{H}{1}},
{\ion{C}{2-IV}}, {\ion{N}{1-II}}, {\ion{O}{1}}, {\ion{O}{6}},
{\ion{Mg}{2}}, {\ion{Si}{2-IV}}, {\ion{P}{2}}, {\ion{S}{2}}, and
{\ion{Fe}{2}} over the velocity range {$-100<\vlsr<+200$\,\kms}. The
high dispersion of these spectra (6.5-20\,\kms) reveals that low
ionization species consist of five discrete components: three at low-
and intermediate-velocities ($\vlsr\approx-44,-7,+56$\,\kms), and two
at high velocities ($\vlsr\approx+100,+184$\,\kms).  Over the same
velocity range, the higher ionization species ({\ion{C}{3-IV}},
{\ion{O}{6}}, {\ion{Si}{4}}) - those with ionization potentials larger
than 40\,eV - show continuous absorption with column density peaks at
{$\vlsr\approx10$\,\kms}, the expected velocity of halo gas
co-rotating with the Galactic disk, and {$\vlsr\approx+184$\,\kms},
the velocity of the higher velocity HVC. The velocity coincidence of
both low and high ionization species in the {$\vlsr\approx+184$\,\kms}
HVC gas suggests that they arise in a common structure, though not
necessarily in the same gaseous phase. The absorption structure in the
high ionization gas, which extends to very low velocities, suggests a
scenario in which a moderately dense cloud of gas is streaming away
from the Galaxy through a hot external medium (either the Galactic
halo or corona) that is stripping gas from this cloud. The cloud core
produces the observed neutral atoms and low-ionization species. The
stripped material is the likely source of the high-ionization
species. Among the host of collisionally-ionized non-equilibrium
models, we find that shock-ionization and conductive interfaces can
account for the column density ratios of high ionization species. The
nominal metallicity of the neutral gas, using the {\ion{O}{1}} and
{\ion{H}{1}} column densities is {$[$O/H$]\sim-0.66$}, with a
substantial uncertainty due to the saturation of the {\ion{H}{1}}
Lyman series in the FUSE band. The ionization of the cloud core is
likely dominated by photons, and assuming the source of ionizing
photons is the extragalactic UV background, we estimate the cloud has
a density of {$10^{-2.7}$\,cm$^{-3}$} with a thermal pressure
{$p/k\approx24$\,cm$^{-3}$\,K}. If photons escaping the Galactic disk
are also included (i.e., if the cloud lies closer than the outer
halo), the density and thermal pressure could be higher by as much as
2\,dex. In either case, the relative abundances of O, Si, and Fe in
the cloud core are readily explained without departures from the solar
pattern. We compare the column density ratios of the HVCs toward the
{PG\,$1116+215$} to other isolated HVCs as well as Complex
C. Magellanic Stream gas (either a diffuse extension of the leading
arm or gas stripped from a prior passage) is a possible origin for
this gas and is consistent with the location of the high velocity gas
on the sky, as well as its high positive velocity, the ionization, and
metallicity.
\end{abstract}

\keywords{Galaxy: halo -- Galaxy: evolution -- ISM: abundances-- ISM: clouds}

\section{Introduction}
\label{sec:intro} Recent observations with the {\it Far
Ultraviolet Spectroscopic Explorer} (FUSE) have revealed a complex
network of highly-ionized high-velocity gas in the vicinity of the
Galaxy.  This new information demonstrates that the high-velocity
material is far more complex than originally thought and is
providing new insight into the formation and evolution of the
Milky Way.  The primary diagnostic of this gas is the
{\ion{O}{6}$\lambda1031.926$} line, which is seen in absorption at
velocities exceeding {$\sim100$\,\kms} in the Local Standard of
Rest in at least 60\% of the AGN/QSO sight lines observed in the
first few years of FUSE operations {\citep{wakker03}}.
{\citet{sembach03}} have attributed the high velocity \ion{O}{6}
to collisionally ionized gas at the boundaries between warm
circumgalactic clouds and a highly extended ($R \gtrsim 70$\,kpc),
hot ($T > 10^6$\,K), low-density ($n \lesssim 10^{-4}$\,cm$^{-3}$)
Galactic corona or Local Group medium. This result is supported by
detailed investigations of the relationship between the highly
ionized gas and lower ionization species (e.g., \ion{C}{2},
\ion{C}{4}, \ion{Si}{2-IV}) in high velocity cloud  Complex~C
{\citep{fox04}} and the high velocity clouds (HVCs) along the
sight line toward PKS\,2155-304 {\citep*{sembach99,csg04}}. An
alternate explanation for the origin of the high velocity
\ion{O}{6} -- photoionization in a low density plasma -- has also
been considered {\citep{nic02}} but is difficult to reconcile with
data available for the other {\ion{O}{6}} HVCs
{\citep{sembach04a,csg04}}.

Establishing the relationship of the high velocity \ion{O}{6} with
lower ionization gas and pinpointing its possible origins requires
high-resolution observations of other ionization stages with the
{\it Hubble Space Telescope} (HST).  We have obtained Space
Telescope Imaging Spectrograph (STIS) observations of the bright
quasar PG\,1116+215 ({$l = 223\fdg36$}, {$b = +68\fdg21$},
{$m_{\rm _V} \approx 15.2$}, {$z_{\mathrm{em}}=0.1765$}) to study
the high velocity \ion{O}{6} in a direction that is well away from
large concentrations of high velocity {\ion{H}{1}} observable in
21\,cm emission (e.g., the Magellanic Stream, Complexes~A, C, M).
High velocity gas in this sight line was first noted by
{\citet*{tls98}}, using HST observations with the Goddard High
Resolution Spectrograph. The sight line contains high velocity
\ion{O}{6} at {$\vlsr\approx+184$\,\kms} {\citep{sembach03}}. High
velocity gas in this general region of the sky is often purported
to be extragalactic in nature based upon its kinematical
properties {\citep[][-- but see {\citet{wakker04b}} for a rebuttal
to these arguments]{blitz99, nic03}}. PG\,1116+215 is on the
opposite side of the sky from the PKS\,2155-304 and Mrk\,509 sight
lines, which are the only other isolated HVCs to have had their
ionization properties studied in detail.  It therefore presents an
excellent case to test whether the ionization and kinematical
properties of the \ion{O}{6} and other ionization stages are
consistent with an extragalactic location. The sight line passes
through the hot gas of the Galactic halo as well as several
intermediate-velocity clouds located within a few kiloparsecs of
the Galactic disk {\citep[e.g.,][and references
therein]{wakker04a}}. Thus, observations of this single sight line
also provide self-contained absorption fiducials against which to
judge the character of the high-velocity absorption.

A complete spectral catalog of the FUSE and STIS observations for
the PG\,1116+215 sight line is presented in a companion study by
{\citet{sembach04b}}.  Information about the intergalactic
absorption-line systems along the PG\,1116+215 sight line can be
found in that paper. In \S\ref{sec:observations}, we present the
spectroscopic data available for the Galactic and high velocity
absorption in the sight line. In \S\ref{sec:kinematics}, we
provide a general overview of the absorption profiles and make
preliminary assessments of the kinematics of the Galactic
absorption and the high velocity gas. In \S\ref{sec:measure}, we
describe our methodology for line measurements (e.g., equivalent
widths, column densities) and present these for selected regions
of the absorption profiles. In addition, we present composite
apparent column density profiles for a sample of important
species. In the following sections, we discuss and analyze the
Galactic and intermediate velocity absorption (\S\ref{sec:mwgas}),
the high velocity absorption at {$\vlsr\approx100$\,\kms}
(\S\ref{sec:hvc100}), and the high velocity absorption at
{$\vlsr\approx184$\,\kms} (\S\ref{sec:hvc184}). Finally, we
discuss the implications of this study and summarize our findings
in \S\ref{sec:discussion} and \S\ref{sec:summary}, respectively.

\section{Observations \& Data Processing}
\label{sec:observations} We observed {PG~$1116+215$} with FUSE on
two separate occasions in April 2000 and April 2001. For all
observations {PG~$1116+215$} was aligned in the center of the LiF1
channel LWRS ($30\arcsec\times30\arcsec$) aperture used for
guiding.  The remaining channels (SiC1, SiC2, and LiF2) were
co-aligned throughout the observations.  The total exposure time
was 77 ksec in the LiF channels and 64 ksec in the SiC channels
after screening the time-tagged photon event lists for valid data.
We processed the FUSE data with a customized version of the
standard FUSE pipeline software ({\tt CALFUSE} v2.2.2).  The data
have continuum signal-to-noise ratios $S/N \sim 18$ and 14 per
0.07\,\AA\ (20--22 \kms) spectral resolution element in the LiF1
and LiF2 channels at 1050\,\AA, and $S/N \sim 8$ and 13 at
950\,\AA\ in the SiC1 and SiC2 channels. The zero-point velocity
uncertainty and cross-channel relative velocity uncertainties are
roughly {5\,\kms}. Further information about the acquisition and
processing of the FUSE data can be found in {\citet{sembach04b}}.
A description of FUSE and its on-orbit performance can be found in
articles by {\citet{moos00}} and {\citet{sahnow00}}.

We observed {PG~$1116+215$} with HST/STIS in May-June 2000 with
the E140M grating and {$0.2\arcsec\times0.06\arcsec$} slit for an
exposure time of 20 ksec.  We also obtained 5.6 ksec of E230M data
through the {$0.2\arcsec\times0.09\arcsec$} slit.  We followed the
standard data reduction and calibration procedures used in our
previous STIS investigations {\citep[see ][]{tripp01,sembach04b}}.
The STIS data have a spectral resolution of 6.5 \kms\ (FWHM) for
the E140M grating, and 10 \kms\ (FWHM) for the E230M grating, both
with a sampling of 2--3 pixels per resolution element. The
zero-point heliocentric velocity uncertainty is about 0.5 pixels,
or {$\sim1.5$\,\kms} for E140M, and {$\sim2.5$\,\kms} for E230M
{\citep{stis}}. The E140M spectra have $S/N\approx14-15$ per
resolution element at 1300\,\AA\ and 1500\,\AA. The E230M spectra
have $S/N\approx6.6-8$ per resolution element at 2400\,\AA\ and
2800\,\AA. For additional information about STIS, see
{\citet{stis98}}, {\citet{kimble98}}, and {\citet{stis}}. We plot
sample FUSE and STIS spectra in Figure~\ref{fig:data}. The three
panels are scaled to cover the same total velocity extent.
Interstellar absorption features are labelled, and high velocity
lines are indicated with offset tick marks.

We identified Galactic absorption features that were detected at
{$\geq3\sigma$} confidence using the method from {\citet{kpii}},
with atomic data from {\citet{morton03}}. In Figures~2a-f, we show
velocity-stacked flux profiles in the Local Standard of Rest
(LSR)\footnote{Henceforth, all velocities will be quoted in the
Local Standard of Rest frame.} for the metal-line transitions
detected in Galactic absorption. The dashed vertical lines at
100\,\kms\ and 184\,\kms\ in each panel mark the locations of the
high velocity absorption. The panels are ordered by atomic number
and ionization stage of the absorbing species. The detected
transitions range over a decade in ionization potential from
$\sim1-10$~Rydbergs.

\section{A General Tour Of the Absorption}
\label{sec:kinematics} The sight line toward {PG~$1116+215$} lies
at high Galactic latitude, {$l = 223\fdg36$} and {$b = +68\fdg21$},
where the transformation from heliocentric velocity to Local Standard
of Rest velocity is small. Using the {\citet{mb81}} definition for the
Local Standard of Rest, and canonical values for standard Solar
motion, we find $\vlsr = v_{\mathrm{helio}} + 1.0$\,\kms. (The
correction in transforming heliocentric velocities to Local Standard
of Rest velocity is $\vlsr = v_{\mathrm{helio}} + 0.6$\,\kms, if one
uses the conventions adopted by IAU Commission 33.) {\citet{wakker03}}
present an {\ion{H}{1}} 21 cm emission profile and report components
at {$\vlsr=-42$\,\kms}, which they identify with the S1 clump of the
Intermediate Velocity-Spur, and at {$\vlsr=-7$\,\kms}. For the
centroid velocity of the Intermediate-Velocity Spur,
{\citet{sembach04b}} report a velocity of {$\vlsr=-44$\,\kms} using
unsaturated lines from low-ionization species and {H$_2$}.  We adopt
the {\citet{sembach04b}} velocity in our analysis. As pointed out by
{\citet{kd96}}, {$\vlsr<0$} intermediate velocity gas in this
direction is inconsistent with pure Galactic rotation, and likely
originates from either infalling gas, turbulent clouds, or a Galactic
fountain. A simple model of uniform density, non-turbulent co-rotating
gas within 10\,kpc of the disk would produce absorption in the
velocity range {$0\lesssim \vlsr
\lesssim 15$\,\kms}.

The two velocity components detected in {\ion{H}{1}} 21\,cm
emission by {\citet{wakker03}} are readily visible at
{$\vlsr\lesssim10$\,\kms} in the low-ionization species like
{\ion{S}{2}}, which do not suffer from unresolved saturated
structure. In most neutral and low-ionization species, however,
these two components are strongly saturated and blended together.
At larger velocities ($10\lesssim\vlsr\lesssim 100$\,\kms), there
is an additional intermediate velocity component in the low
ionization species at {$\vlsr\approx56$\,\kms}, which is not
detected in the {\ion{H}{1}} 21\,cm emission [down to
$N$(\ion{H}{1})$\sim2\times10^{18}$\,cm$^{-2}$]. (Absorption at
this velocity is evident in the {\ion{H}{1}} Lyman series, which
we treat later.) This intermediate velocity component is readily
apparent in the strong lines of neutral (e.g., \ion{O}{1}) and
low-ionization (e.g., \ion{Mg}{2}, \ion{Si}{2}, \ion{Fe}{2})
species.

At velocities exceeding {$|\vlsr|=100$\,\kms}, there are two
absorption components, at {$\vlsr\approx+100$\,\kms} and
{$\vlsr\approx+184$\,\kms}. The {$\vlsr\approx+100$\,\kms}
absorption component is prominent in the {\ion{C}{2}} and
{\ion{Si}{3}} profiles, but noticeably absent in other
low-ionization (e.g., \ion{Mg}{2}, \ion{Fe}{2}) and neutral (e.g.,
{\ion{O}{1}}) species. The component is detected (at {$3\sigma$}
confidence) in the strongest {\ion{Si}{2}} line at 1260.422\,\AA,
but not in the weaker lines. Absorption at this velocity is also
detected in other moderate-ionization species (\ion{C}{3}) as well
as high ionization-species (\ion{C}{4}, \ion{Si}{4}, \ion{O}{6}).
The HVC at {$\vlsr\approx+184$\,\kms} is detected in a wide range
of neutral (\ion{O}{1}\footnote{We note that {\citet{sembach04b}}
identify a weak, intervening {Lyman $\alpha$} absorber at a
redshift {$z=0.0719$} which flanks the
{\ion{O}{1}$\lambda1302.168$} transition at
{$\vlsr\sim+210$\,\kms} on the {\ion{O}{1}} velocity scale.}),
low-ionization (\ion{C}{2}, \ion{N}{2}, \ion{Mg}{2}, \ion{Si}{2},
\ion{Fe}{2}), moderate-ionization (\ion{C}{3}, \ion{Si}{3}), and
high-ionization (\ion{C}{4}, \ion{Si}{4}, \ion{O}{6}) species.

In Figure~2 there are noticeable differences in the kinematics of
the high and low ionization species. The neutral and
low-ionization species typically break up into discrete components
(when the lines are not too strong) at the five velocities
discussed ($\vlsr\approx-44, -7, +56, +100, +184$\,\kms). By
contrast, the higher ionization species (\ion{C}{3}, \ion{C}{4},
\ion{O}{6}, \ion{Si}{4}) feature continuous absorption across the
entire low, intermediate, and high velocity range. The kinematic
distributions of these higher ions have an apparent bimodality,
with apparent optical depth peaks defining the low velocity
Galactic absorption and the high velocity absorption. The fluxes
do not return to the continuum level in between these absorbing
zones except for {\ion{Si}{4}} which may be due to (1) the lower
elemental abundance of silicon compared to carbon and oxygen, and
(2) the lower ionization of {\ion{Si}{4}} relative to {\ion{C}{4}}
and {\ion{O}{6}}. It is interesting to note, however, that at low
velocity the flux minimum of the {\ion{O}{6}} profile does not
coincide with the negative velocity component seen in
{\ion{Si}{4}} and {\ion{C}{4}} (the intermediate velocity S1
component at {$\vlsr\approx-44$\,\kms}), but rather the expected
location for Galactic halo absorption, $\vlsr\sim10$\,\kms. At the
velocity of the higher velocity HVC ($\vlsr\approx+184$\,\kms),
the high-ionization species coincide with the lower ionization
species. This velocity coincidence suggests that high and low
ionization absorption arise co-spatially in that component.
However, there are noticeable differences, as already mentioned,
in the general shapes of the high and low ionization species,
indicating absorption from a common structure with different
phases.

\section{Measurements}
\label{sec:measure}

We use apparent optical depth (AOD) techniques described by
{\citet{ss92}} to measure equivalent widths, integrated apparent
column densities, velocity centroids, and $b$-values for the five
components discussed above. Using the low and moderate-ionization
species (primarily \ion{S}{2}, \ion{Fe}{2}, and \ion{Si}{3}) as
guides, we chose the following integration ranges for the five
components\footnote{For our initial measurements, we combine the
low- and intermediate-velocity components into a single
measurement, and cope with unresolved saturated structure later.}:
$-100\leq\vlsr\leq-25$\,\kms\ for $\vlsr\approx-44$\,\kms;
$-25\leq\vlsr\leq+37$\,\kms\ for $\vlsr\approx-7$\,\kms;
$+37\leq\vlsr\leq+85$\,\kms\ for $\vlsr\approx+56$\,\kms;
$+85\leq\vlsr\leq+140$\,\kms\ for $\vlsr\approx+100$\,\kms; and
$+140\leq\vlsr\leq+230$\,\kms\ for $\vlsr\approx+184$\,\kms.

We first extracted a {$\pm500$\,\kms} window about the rest
wavelength of each line into separate files. We then determined a
local continuum for each line following the Legendre
polynomial-fitting method described by {\citet{ss92}}. We computed
the equivalent width of each transition of rest wavelength
{$\lambda_{\circ}$} for the two high velocity clouds and for the
Galactic/IVC absorption by performing the sum:
\begin{equation}
W_{\lambda} = {\lambda_{\circ} \over c}
\sum_{i=i_{\mathrm{min}}}^{i_{\mathrm{max}}} w_i [1 - I(v_i)]
\delta v_i,
\end{equation}
where the limits of the sum ($i_{\mathrm{min}}$\ and
$i_{\mathrm{max}}$) were chosen to encompass the specified
velocity range for the component, {$\delta v_i = v_{i+{1 \over 2}}
- v_{i-{1 \over 2}}$} is the full width of a bin, {$I(v_i)$} is
the normalized flux profile as a function of velocity, and {$w_i$}
is a weight that accounts for the fractional bins at the edges of
the summation.

The integrated apparent column density, AOD-weighted centroid
velocity, and $b$-value are all derived as moments of the apparent
optical depth distribution. To compute these quantities, we first
transform the normalized flux profiles to apparent column density
(ACD) profiles via:
\begin{equation}
\acd(v_i) = {{m_{\mathrm{e}} c} \over {\pi e^2}} {1 \over
{f\lambda_{\circ}}} \ln \left [ {1 \over {I(v_i)}} \right ],
\label{eq:naov}
\end{equation}
where $f$ is the oscillator strength of the transition. From the
ACD profiles, the desired quantities are computed via:
\begin{eqnarray}
\acd^{(j)} = \sum_{i=i_{\mathrm{min}}}^{i_{\mathrm{max}}} w_i v^j \acd(v_i) \delta v_i\\
\acd = \acd^{(0)} \nonumber \\
\langle v \rangle = \acd^{(1)}/\acd^{(0)} \nonumber \\
b^2 = 2 \times \acd^{(2)}/\acd^{(0)}. \nonumber
\end{eqnarray}
In the integrations involving moments of the apparent column density
for strongly saturated lines, we treat pixels with negative flux
(due to statistics) as having the flux equal to the RMS uncertainty
derived from the fit to the continuum regions adjoining the line. We
note that this causes the derived $\acd(v)$\ to be smaller than the
true $\acd(v)$\ (see eq.~\ref{eq:naov}) and the derived values of
$\acd$, $\langle v \rangle$, and $b$\ will also be affected (see
eq.~3). All other pixels are treated in the normal way, including
those with fluxes above the continuum level where the implied
apparent column density is negative. Since the moments of the
optical depth for weak/narrow features are very sensitive to the
choice of integration range, we only report mean velocities and
$b$-values for lines whose equivalent width exceeds five times its
{$1\sigma$} error.

In Tables~\ref{tab:hvc100measure}--\ref{tab:hvc184measure}, we
report measurements of the two high velocity components for the
observed transitions of various species. In each table, we list
the ion (column 1), transition rest wavelength (column 2),
transition strength ($\log f\lambda$, column 3), centroid velocity
(column 4), {$b$-value} (column 5), integrated apparent column
density (column 6), equivalent width (column 7), and integration
range (column 8) used to compute the aforementioned quantities.
The errors quoted for each quantity are {$1\sigma$} confidence
errors resulting from both Poisson noise and continuum placement
uncertainties. We also quote an additional error for the
integrated column density and equivalent width which characterizes
the systematic uncertainties based on our choice of integration
range. This latter error is derived by splitting the difference
between the addition and subtraction of 10\,\kms\ to the
integration range (5\,\kms\ on either side of the range). In
Table~\ref{tab:haloivcmeasure}, we provide the same measurements
for the the Galactic halo and intermediate velocity gas
(integrating over the velocity range -100 to +85\,\kms).

In Figure~3, we show the apparent column density profiles
constructed using equation~\ref{eq:naov}. For ions with more than
one transition detected, we have over-plotted the apparent column
density profiles. Apparent column density profiles from multiple
transitions of a given species provide a means of testing for
unresolved saturated structure, unidentified blends, and
unocculted flux (e.g., scattered light within the instrument, or
elevated background levels). In Table~\ref{tab:coladopt}, we
report our adopted column densities for the five absorption
components. These column densities were derived by computing the
variance-weighted mean of the integrated column densities of
transitions least affected by unresolved saturated structure or
other peculiarities. In the appendix, we provide a detailed
description of the transitions covered for each ion, and discuss
which transitions were used in the computations of the composite
apparent column density profiles and the adopted column densities.

In cases where multiple transitions from a given species exist, we
compute composite apparent column density profiles in order to
improve the data quality and maximize use of the available
information in a non-parametric way. Comparisons of these profiles
allows inspection of the kinematical similarities/differences that
exist between different species. Using the composite profile
information also avoids potential biases that can arise from simple
comparisons of integrated line widths calculated from equation~3.
Our general approach toward creating the composite profiles is to
first resample the individual ACD profiles (and their respective
error vectors) to common velocity bins using a cubic spline
interpolation {\citep{nrpress}} and then perform a variance-weighted
average for each velocity bin of the apparent column densities from
optically thin transitions (i.e., where the apparent optical depth,
$\tau_a(v_{\mathrm{i}})=\ln [1/I(v_{\mathrm{i}})]<1$). In cases
where all transitions are optically thick, we use the {$1\sigma$}
lower limit from the weakest transition. FUSE data are not used in
the composite profiles since the resolution of spectra are much
poorer than the STIS spectra. In cases where E230M data are
combined, we resample the ACD profiles to {5\,\kms} bins. When E140M
data are combined, we resample the ACD profiles to {3.5\,\kms} bins.
Our primary purpose in creating composites is for line shape
comparisons in the {$\vlsr\approx+184$\,\kms} HVC, not for column
density measurements. As such, we do not require a detailed
evaluation of the propagation of errors through the resampling
procedure. We have verified through visual inspection that the
composites provide a good facsimile of the underlying apparent
column density profiles.

In the appendix, we discuss the details of the transitions chosen
for each ion for the composite ACD computation and for the adopted
variance-weighted mean integrated apparent column density reported
in Table~\ref{tab:coladopt}. The final composite {$\acd(v)$}
profiles for ions detected in Galactic absorption are shown in
Figure~4. {\ion{H}{1}} was not included in this computation
because the higher order Lyman series lines detected in the FUSE
band are strong and suffer from blends with other lines. We return
to measurements of the {\ion{H}{1}} profiles below.

\section{Low and Intermediate Velocity Gas}
\label{sec:mwgas}

High ionization gas traced by \ion{Si}{4}, \ion{C}{4}, \ion{N}{5},
and \ion{O}{6} is common along high latitude sight lines that
extend several kiloparsecs or more through the Galactic thick disk
and halo. PG\,1116+215 lies in a region of the sky well above the
influence of spiral arm structure and outside nearby radio loops
that may contribute to some of the high ionization gas
{\citep*{sst97}}.  Most of the high ionization gas at low
velocities toward PG\,1116+215 occurs in the thick disk/halo
because the high latitude of the sight line ($b = 68\fdg21$)
ensures that only tenuous gas in the solar neighborhood is
intercepted within 100 pc of the Galactic disk. The amount of
\ion{O}{6} expected within $\sim100$\,pc of the Sun is
$\lesssim10^{13}$ cm$^{-2}$ {\citep{oeg04}}, or $\lesssim8$\% of
the total observed in the --44, --7, and +56 \kms\ components. The
sight line therefore allows an examination of the high ion column
density ratios in gas associated predominantly with the thick
disk/halo of the Galaxy.

The amount of \ion{O}{6} observed in the Milky Way thick-disk/halo
toward PG\,1116+215 over the velocity range from {-100} to
{+85\,\kms}, log~N(\ion{O}{6}) = 14.16 (see
Table~\ref{tab:haloivcmeasure}), is typical of that expected for a
plane-parallel gas layer with an exponential density distribution,
an \ion{O}{6} scale height of 2.3 kpc, a mid-plane density
$n_0$(\ion{O}{6}) $\sim 1.7\times10^{-8}$ cm$^{-3}$, and a
$\sim0.25$ dex enhancement as seen along high latitude sight lines
in the northern Galactic hemisphere {\citep{savage03}}. The other
high ions also have total column densities in the low and
intermediate velocity components that are typical for extragalactic
sight lines; we find $\log N$(\ion{C}{4})$=14.17$, $\log
N$(\ion{N}{5})$<13.14$, and {$\log N$(\ion{Si}{4})$=13.69$} in the
{$-100 \leq\vlsr\leq+85$\,\kms} velocity range.

The thick disk/halo absorption toward PG\,1116+215 consists of
three separate low-ionization components with continuous
absorption in the high ionization species over the velocity range
$-100\lesssim\vlsr\lesssim+85$\,\kms. At these velocities, the
\ion{C}{4} and \ion{Si}{4} profiles are quite similar, with
average column density weighted centroids of $-13$\ and
$\sim-15$\,\kms, respectively (see Table 3).  The widths of the
absorption features are also similar, with the \ion{C}{4} lines
being slightly broader: $b$(\ion{C}{4})$\sim 60$\,\kms\ and
$b$(\ion{Si}{4})$\sim 45$\,\kms. The centroids of the \ion{C}{4}
and \ion{Si}{4} are significantly different from the \ion{O}{6}
centroid, $\langle v \rangle \sim +18$\,\kms, indicating that
there are important differences in the distributions of these two
ions and the distribution of the low velocity \ion{O}{6}.  Most of
this difference arises from the lack of substantial \ion{O}{6} in
the $-44$ \kms\ component [$\log N$(\ion{O}{6})$<13.24$; see
Figure~4 and Table 4].  It is noteworthy that this is the only
component along the sight line that contains detectable H$_2$
absorption. Most IVCs are within one kiloparsec of the Galactic
disk {\citep{wakker01}}, so this component is probably closer than
some of the other low velocity \ion{Si}{4} and \ion{C}{4}
absorption features.

Strong low-ionization features, such as \ion{C}{2}
$\lambda1334.532$, have a negative velocity absorption cutoff at
essentially the same velocity as \ion{C}{4} or \ion{Si}{4}
($\vlsr\sim-90$\,\kms).  This indicates that the high and low
ionization species in the $-44$ \kms\ IVC are closely coupled
kinematically and, by inference, spatially. The
intermediate-ionization \ion{C}{3} $\lambda977.020$ line closely
approximates this behavior as well; its great strength suggests
that there may be a small amount of gas at slightly more negative
velocities. The column density ratio of
{$N$(\ion{C}{4})$/N$(\ion{Si}{4})$\approx3$} in this component is
typical of that for clouds in the general interstellar medium
{\citep[see ][]{sst97}}.  {\citet*{ssc94}} have suggested that the
constancy of the \ion{C}{4}-to-\ion{Si}{4}\ ratio along many
different directions through the Galactic disk and low halo can be
attributed to regulation of the ionization by conductive
interfaces, a result born out by their high-resolution GHRS data
that shows a close kinematical relationship between the high ions
and lower ionization velocity components along the HD\,167756
sight line. The same ratio of {$\sim3$} is found in the other low
and intermediate velocity features toward PG\,1116+215 as well.
The upper limits on \ion{N}{5} and \ion{O}{6} in the $-44$\,\kms\
component place limits on the age of the conduction front.   The
predicted strength of \ion{O}{6} is generally less than that of
\ion{C}{4} for conduction front ages {$\lesssim10^5$} years
{\citep*{bbf90}}.  Thus, if conduction is important in regulating
the \ion{Si}{4} and \ion{C}{4} column densities in this component,
the front must be in an early stage of evolution.

The ionization of the Galactic thick disk and halo gas toward
PG\,1116+215 is likely a hybrid of different collisional ionization
processes as no single model seems to be able to explain the high
ion column density ratios in all of the observed components. We list
the observed ratios of \ion{Si}{4}, \ion{C}{4}, and \ion{N}{5} to
\ion{O}{6} in Table~\ref{tab:cine} together with the corresponding
ratios predicted for different ionization mechanisms.  The sources
of the theoretical model predictions are listed in the footnotes of
Table~\ref{tab:cine} as well as the ranges of model parameters
considered in computing the column density ratios. All values listed
are appropriate for solar abundance gas, which should be a
reasonable approximation for the thick disk/halo and IVC components
considered here. Previous studies have shown that multiple
ionization mechanisms are required to explain the {\it total} high
ion column density ratios along sight lines through the halo
{\citep[e.g.,][]{ss92,ssl97,savage03,is04}}, and PG\,1116+215 is no
exception. This is not surprising given the range of ion ratios
observed in the three components. For example, both the
\ion{Si}{4}/\ion{O}{6} and \ion{C}{4}/\ion{O}{6} column density
ratios differ dramatically between the two components at --44 \kms\
and --7 \kms, while the \ion{C}{4}/\ion{Si}{4}\ ratio is {$\sim3$}
in both cases. Apparently, the high ionization gas is sufficiently
complex that no single process dominates the observed ionization
signature of the Galactic disk and halo gas along the sight line.

\section{High Velocity Gas at $\vlsr\approx+100$\,\kms}
\label{sec:hvc100}

As described in \S\ref{sec:kinematics}, there are clear signs of a
high velocity cloud at {$\vlsr\approx+100$\,\kms}. A discrete
feature at this velocity can be seen in the profiles of
{\ion{C}{2}}, {\ion{Si}{2}} (in the 1260.422\,\AA\ line), and
{\ion{Si}{3}}. The gas is not detected in any neutral ions covered
by the spectra; there are no indications of significant column
densities in {\ion{C}{1}}, {\ion{N}{1}}, or {\ion{O}{1}}.
Likewise, the cloud is not detected in any singly-ionized species
other than {\ion{C}{2}} and {\ion{Si}{2}}; there is no significant
column density in {\ion{N}{2}}, {\ion{Mg}{2}}, {\ion{P}{2}},
{\ion{S}{2}}, or {\ion{Fe}{2}}. Absorption in the {\ion{C}{3}}
977.020\,\AA\ line is also present, but it is affected by
unresolved saturation. There is significant column density at
{$\vlsr\approx+100$\,\kms} in the higher ionization species, but
it is difficult to associate the high-ionization gas with this
component for two reasons: (1) the kinematics of the
high-ionization species is peaked at the {$\vlsr\approx+184$} HVC
with a trailing absorption wing extending down to this velocity
(see \S\ref{sec:hvc184}); and (2) the (non-zero) minimum in the
apparent column density profiles occurs near this velocity. For
this reason, the column densities of high-ionization species
listed in Table~\ref{tab:hvc100measure} for this component should
be treated with care in the interpretation of the ionization of
this gas.

It is interesting to note that the {+100\,\kms} HVC has properties
that are quite similar to those of an intergalactic gas cloud at
$\zabs$ = 0.00530 observed toward 3C\,273
{\citep{tripp02,sem01}} that is located in the outskirts of the Virgo
cluster.  Like the PG\,1116+215 {+100\,\kms} HVC, the 3C\,273 absorber
is only detected in \ion{C}{2}, \ion{Si}{2}, and \ion{Si}{3}, and the
relative strengths of the lines are similar.  The \ion{H}{1} column
densities are also similar, although $N$(\ion{H}{1}) for the
{+100\,\kms} is rather uncertain due to blending with lower velocity
gas. Moreover, there are several galaxies within a few hundred
kiloparsecs of the 3C\,273 sight line at $z \approx$ 0.00530
{\citep[][and references therein]{tripp02,stocke04}}.  We also note
that the sight line to RX\,J1230.8+0115, which is $\sim 350
h_{70}^{-1}$ away from 3C\,273 in projection, also shows Virgo
absorption at $\zabs\approx$ 0.005, but with much stronger
high-ion absorption {\citep{ros03}}. The RX\,J1230.8+0115 Virgo
absorber is more analogous to the {+184\,\kms} HVC.  It appears that
the 3C\,273 and RX\,J1230.8+0115 Virgo absorbers are reminiscent of
Milky Way HVCs and could have similar origins.

\section{High Velocity Gas at $\vlsr\approx+184$\,\kms}
\label{sec:hvc184}

\subsection{Kinematics}

We now consider the detailed kinematics of the isolated HVC at
$\vlsr\approx+184$\,\kms. In Figure~\ref{fig:compacdhvc}, we show an
expanded version of the composite apparent column density profiles
for the HVC, with the detected species ordered by ionization
potential. There is a very clear progression in the kinematic
structure of the high velocity gas with ionization potential. There
are two distinct components that are aligned in velocity. A narrow
component comprises the {\ion{O}{1}}, and part of the {\ion{Mg}{2}},
{\ion{Si}{2}}, {\ion{Fe}{2}}, {\ion{C}{2}}, {\ion{N}{2}},
{\ion{Si}{3}}, and {\ion{Si}{4}} profiles. A broader component,
which becomes more prevalent with increasing ionization potential,
appears to produce all the {\ion{O}{6}}, {\ion{C}{4}}, and some
fraction of the other non-neutral ionization stages. In addition to
this, there is a tail of high ionization gas extending toward
smaller velocities.

The kinematics of the profiles are suggestive of a diffuse cloud, like
those producing the neutral and low-ionization species in the lower
velocity components, embedded within a hot, low-density external
medium, such as the Galactic halo or highly extended
corona. {\citet{savage03}} characterize the high-ionization Galactic
thick-disk/halo as asymmetrical plane-parallel patchy absorption with
mid-plane density {$n_{\circ}$(\ion{O}{6})$\sim10^{-8}$\,cm$^{-3}$}
(or about $n\gtrsim10^{-4}$, after ionization and abundance
corrections) and scale height of {$\sim2.3$\,kpc} (with a
$\sim0.25$\,dex column density excess near the north Galactic polar
region).  {\citet{sembach03}} define the Galactic corona as a more
diffuse ($n\lesssim10^{-4}$\,cm$^{-3}$), hot ($T\sim10^6$\,K), and
highly extended ($\gtrsim50$\,kpc) envelope of gas surrounding the
Galaxy.  For the {PG\,$1116+215$} HVC, one possible description of the
gas would be that the embedded clouds produce the observed
low-ionization absorption, while the interaction between the cloud(s)
and the corona or halo produces the higher ionization absorption. With
this isolated high velocity cloud, we can examine the abundances and
ionization structure of the gas.

\subsection{Abundances}
\label{sec:hvcabund}

An important diagnostic in constraining the origin of high velocity
gas is its metallicity. Since high metallicities are not expected for
primordial/unprocessed material, a high metallicity argues against an
IGM origin and favors a Galactic origin. In the high velocity cloud
toward {PG~$1116+215$}, we start with the assumption that low
ionization species co-exist in a neutral phase of gas. In such a
phase, {N(\ion{O}{1})/N(\ion{H}{1})} can be used as a metallicity
indicator, O/H, since {\ion{O}{1}} and {\ion{H}{1}} have nearly
identical ionization potentials and are strongly coupled through
charge exchange reactions {\citep{fs71}}. The ionization corrections
in transforming {N(\ion{O}{1})} to a total oxygen column density and
{N(\ion{H}{1})} into a total hydrogen column density cancel out over a
wide range of ionization conditions with the minor assumption that
both ions span similar thicknesses within the absorbing medium
{\citep{tripp03}}.

\subsubsection{The O\,{\sc i} Column Density}
\label{sec:184oicol}

A direct integration of the apparent column density profile of the
{\ion{O}{1}$\lambda1302.169$} line over the velocity range
{$140<\vlsr<192$\,\kms} yields {$\log
[N$(\ion{O}{1})cm$^{-2}]=13.82\pm0.03$}.  The absorption in this
velocity range is flanked by a weak, intervening {Lyman $\alpha$}
absorber at a redshift {$z=0.0719$} {\citep[{$\vlsr\sim210$\,\kms}
on the {\ion{O}{1}} velocity scale, see][]{sembach04b}}.  This is
the only high velocity {\ion{O}{1}} line present in the STIS
spectrum. The column density estimate is consistent with the
non-detections of weaker {\ion{O}{1}} lines in the noisier and
lower resolution FUSE spectra (e.g.,
\ion{O}{1}$\lambda988.773,\lambda1039.230$).

A slightly higher value of {$N$(\ion{O}{1})} is obtained if the
weak absorption attributed to the redshifted Lyman {$\alpha$}
absorber is included in the velocity range 140--230\,\kms. We
believe this added absorption is unlikely to be {\ion{O}{1}} in
the HVC because the implied {\ion{O}{1}} profile has a shape
inconsistent with the shapes of other neutral and low ionization
species present in the HVC. Comparison of the flanking absorption
with the primary absorption in the HVC reveals that the implied
{\ion{O}{1}} ratio in the two components (flanking/primary) would
be many times higher than inferred from the weak lines of species
that are the dominant ionization stages in neutral media (e.g.,
\ion{Mg}{2}, \ion{Si}{2}, \ion{Fe}{2}). It would also be
moderately inconsistent with the shape of the
{\ion{C}{2}$\lambda1036.337$} profile.

The width of the {\ion{O}{1}$\lambda1302.169$} line is comparable
to the weak absorption lines produced by other low ionization
species (e.g., {\ion{Si}{2}}, \ion{Fe}{2} - see Figures~2c and
{\ref{fig:compacdhvc}}). The {\ion{O}{1}} line has a maximum
apparent optical depth, {$\tau_{\mathrm{a}}^{\mathrm{max}}\sim1$}.
This optical depth is very similar to the maximum optical depth in
the {\ion{Si}{2}$\lambda1304.370$} line, for which unresolved
saturated structure is not present as evidenced by the good
agreement in the {$N_{\mathrm{a}}(v)$} profiles for the
{\ion{Si}{2}$\lambda1304.370$} and {\ion{Si}{2}$\lambda1526.707$}
lines (Figure~3). Therefore, if the velocity structure of the
{\ion{O}{1}} is similar to that in the core of the
{\ion{Si}{2}$\lambda1304.370$} line, the value of
{$N_{\mathrm{a}}$({\ion{O}{1}})} should show little diminution due
to unresolved saturated structure.

A rigorous upper limit on the {\ion{O}{1}} column density can be
placed by estimating the effects that a narrow component indicative
of a temperature near 1000 K (i.e., $b\sim1$\,\kms) might have on
the observed profile shapes.  Using the core of the
{\ion{Si}{2}$\lambda1304.370$} and {$\lambda1526.707$} lines as a
guide, we find that such a narrow component could be present without
violating the good agreement in the {\ion{Si}{2}}
{$N_{\mathrm{a}}(v)$} profiles if the intrinsic central optical
depth of the component is less than {$\sim3.5$}.  Translating this
optical depth to the {\ion{O}{1}$\lambda1302.169$} line implies that
the {\ion{O}{1}} column density in the hypothetical narrow component
could be as high as about {$6\times10^{14}$\,cm$^{-2}$} and would
account for {$\sim13$\,m\AA} (roughly half) of the equivalent width
of the observed line.  Thus, the presence of a very narrow component
could in principle increase the nominal {\ion{O}{1}} column density
estimate by a factor of 1.5 without violating other observational
constraints. While we think it unlikely that such a narrow component
exists within the {\ion{O}{1}} profile since there is no evidence of
{\ion{C}{2}*} absorption or neutral species (\ion{Mg}{1},
\ion{C}{1}) that would favor its presence, we nevertheless adopt a
conservative logarithmic {\ion{O}{1}} column density range of
13.79-13.91, or {$\log N$({\ion{O}{1}})$=13.82_{-0.03}^{+0.09}$} in
our discussions of the {\ion{O}{1}} column density below. This value
is reported in Table~\ref{tab:coladopt}.

\subsubsection{The H\,{\sc i} Column Density}

From the direct integration of the {\ion{H}{1}} Lyman {$\lambda$} line
in the FUSE spectrum, and the non-detection of the {\ion{H}{1}} 21 cm
emission from {\citet{wakker03}}, the HVC {\ion{H}{1}} column density
must lie in the range {$(5.2-200)\times10^{16}$\,cm$^{-2}$}. To
further constrain this range, we take two approaches to measuring the
{\ion{H}{1}} column density -- profile-fitting the higher order Lyman
series, and a curve of growth fit to the {\ion{H}{1}} equivalent
widths reported in Table~\ref{tab:hvc184measure}.

In our first approach, we have attempted to fit the higher order
Lyman series detected in the FUSE spectrum. Since the higher order
Lyman series lines of the {$\vlsr\approx+184$\,\kms} HVC are
blended with both the {\ion{H}{1}} lines from intermediate
velocity and Galactic absorption, and with several {\ion{O}{1}}
transitions, we first considered the available constraints on the
kinematics, column densities, and intrinsic line widths from these
lines prior to the fitting the {\ion{H}{1}} profile of the HVC. In
our consideration, it is only important to construct a model which
accurately reproduces the shapes of the absorption profiles.
Consequently, in constructing the model, we started with the
information already provided by the {\ion{H}{1}} 21\,cm emission
profiles from {\citet{wakker03}}, the high velocity {\ion{O}{1}}
column density derived above, and the kinematic information for
the low-ionization lines from \S\ref{sec:kinematics}. Thus, our
model contains five kinematic components at velocities
{$\vlsr\approx-44, -7, +56, +100$}, and +184\,\kms. From a fit to
the {\ion{H}{1}} 21\,cm emission, {\citet{wakker03}} report
{\ion{H}{1}} column densities of {$10^{19.83}$\,cm$^{-2}$} and
{$10^{19.70}$\,cm$^{-2}$}, and $b$-values of {15.5,\kms} and
{14.7\,\kms} for the {$\vlsr\approx-44$\,\kms} and
{$\vlsr\approx-7$\,\kms} components, respectively. To further
reduce the number of free parameters, we fixed the {\ion{O}{1}}
column densities of these two components to that implied by a
solar metallicity: {$\log N$(\ion{O}{1})$=16.49$} for
{$\vlsr\approx-44$\,\kms}, and {$\log N$(\ion{O}{1})$=16.36$} for
{$\vlsr\approx-7$\,\kms}.

Before proceeding with a fit to the higher order Lyman series in
the FUSE spectrum, we first used the above information on the
kinematics, column densities, and line widths to fit the
{\ion{O}{1}$\lambda1302.168$} profile in the STIS E140M spectrum.
The purpose of this fit was to constrain the {\ion{O}{1}} line
widths for all components, and the {\ion{O}{1}} column densities
for the {$\vlsr\approx+56$\,\kms} and {100\,\kms} components, in
addition to determining the optimal velocities of all components.
We fixed the {\ion{O}{1}} column density of the
{$\vlsr\approx+184$\,\kms} component to the value derived in the
previous section. We also added a weak Lyman {$\alpha$} feature at
{$z=0.0719$} in order to deblend the high velocity {\ion{O}{1}}
absorption {\citep[see][]{sembach04b}}. In this preliminary fit,
we synthesized the profile assuming the components were subject to
Voigt broadening, and convolved the resulting profile with a
normal distribution having a full-width at half maximum intensity
of {6.5\,\kms} to mimic the STIS instrumental resolution. We
varied all remaining parameters (column densities, $b$-values, and
component velocities) to produce a least-squares fit to the data.
We found that the component at {$\vlsr\approx100$\,\kms} did not
contribute significantly to the {\ion{O}{1}} profile, so this
component was removed from the {\ion{O}{1}} fit (as expected from
the non-detection, see \S\ref{sec:hvc100}). Conversely, we found
that a component was needed at {$\vlsr\approx+26$\,\kms} to
provide a good match to the observed profile.

With this fit of the {\ion{O}{1}} profile in hand, we proceeded to
model the higher order Lyman series lines (in particular, the
Lyman {$\eta-\mu$} lines detected in the FUSE SiC2a spectrum). In
this {\ion{H}{1}} model, we fixed all the parameters from the fit
to the {\ion{O}{1}} profile [component velocity, $b$(\ion{O}{1}),
$N$(\ion{O}{1})]. We also fixed the {\ion{H}{1}} column densities
of the {$\vlsr\approx-44, -7, +27$, and +56\,\kms} components to
the values implied by a solar metallicity: $\log
N$(\ion{H}{1})$=19.83, 19.70, 17.29$, and 17.58, respectively. We
added the {$\vlsr\approx100$\,\kms} component back in, fixing the
velocity, but allowing the {\ion{H}{1}} column density and
$b$-value [as well as the $b$-values of the components not
detected by {\citet{wakker03}}] to vary. Instead of allowing the
{\ion{H}{1}} column density of the {$\vlsr\approx+184$\,\kms} HVC
to freely vary, we considered two extreme cases, the limits
implied by the direct integration of the Lyman {$\lambda$} profile
[$\log N$(\ion{H}{1})$\geq16.73$], and the non-detection in the
{\citet{wakker03}} 21\,cm spectrum [$\log
N$(\ion{H}{1})$\leq18.3$]. We fixed the {\ion{H}{1}} column
density of the HVC at these two values and allowed the $b$-value
to vary. These new models were synthesized assuming
Voigt-broadened components and then convolved with a normal
distribution having a full-width at half-maximum intensity of
{20\,\kms} to mimic the FUSE instrumental resolution. The
least-squares optimized fit parameters are listed in
Table~\ref{tab:hifit} (along with the corresponding {\ion{O}{1}}
parameters from the preliminary fit). In Figure~\ref{fig:lyman}
(panels a and c), we show the best-fit models for the two extreme
cases, along with the same model without the contribution of the
HVC, for the higher order Lyman series. In addition, we show the
contribution of the {\ion{O}{1}} absorption. The differences in
the two extreme cases are minor, with the most significant change
arising in the Lyman {$\mu$} transition. Neither model can be
clearly ruled out at the {$2\sigma$} level. We therefore consider
this full range a 95\% confidence interval for the {\ion{H}{1}}
column density.

The model used in panel b was derived using our second approach
toward constraining the {\ion{H}{1} column density - a curve of
growth fit to the {\ion{H}{1}} equivalent widths reported in
Table~\ref{tab:hvc184measure}. Assuming that the profiles are
formed via a single Voigt-broadened component with a column
density {$N$} and Doppler width {$b$}, the equivalent widths
{$W_\lambda$} can be predicted and fit to the observations. In
Figure~\ref{fig:cog}, we plot {$W_\lambda/\lambda$} versus
{$f\lambda$} for the Lyman series lines detected in the FUSE SiC2
and LiF1 channels using the equivalent widths reported in
Table~\ref{tab:hvc184measure}. In the figure, we plot two sets of
error bars, one for the {$1\sigma$} statistical$+$continuum
placement error (solid), and one reflecting an additional
{$10$\,\kms} uncertainty in the choice of integration range
(dashed). The column density and $b$-value that best fit the
plotted equivalent widths are {$\log
N$(\ion{H}{1})$=17.82_{-0.14}^{+0.12}$}, {$b=14.7\pm0.4$\,\kms},
and the curve of growth is overplotted in the figure. The central
panel of Figure~\ref{fig:lyman} shows the full model using these
parameters for the {\ion{H}{1}} HVC absorption. In addition, two
other curves of growth using the model parameters of the two
extreme cases considered above are also overplotted. From
Figure~\ref{fig:cog}, it appears that the two extreme cases listed
in Table~\ref{tab:hifit} are unable to reproduce the observed
equivalent widths of the stronger Lyman series lines (e.g., Lyman
$\beta$, $\gamma$, $\delta$). Visual inspection of the synthetic
profiles for lower order {\ion{H}{1}} Lyman series lines revealed
that the high velocity side of the absorption profiles are
reproduced well. However, it is conceivable that some of the high
velocity absorption is missed due to the lower velocity cutoff in
the integration (140\,\kms) and therefore not incorporated into
the measurements. We adopt the curve-of-growth results and
{$1\sigma$} errors in our following estimation of (O/H).

\subsubsection{The O/H Abundance}

From the preceding analysis, we have constrained the {\ion{O}{1}}
and {\ion{H}{1}} column densities to: $\log
N$(\ion{O}{1})$=13.82_{-0.03}^{+0.09}$, and $\log
N$(\ion{H}{1})$=17.82_{-0.14}^{+0.12}$. Using these adopted values,
we estimate that the metallicity of the neutral phase of the high
velocity gas is [O/H]$=-0.66_{-0.16}^{+0.39}$, or about
{$0.14-0.42Z_\odot$} ({$1\sigma$} confidence\footnote{We have
propagated the uncertainties in both the column density
measurements, as well as the 0.05~dex uncertainty in the solar
abundance of oxygen reported by \cite{asplund04}. The {$2\sigma$}
confidence interval for the metallicity of the neutral phase is
$0.1-1.3Z_\odot$.}). This estimate assumes that the neutral gas
arises in a single phase of gas, which is a reasonable approximation
considering the shape of the {\ion{O}{1}} {$1302.169$\,\AA} line
shape. If the $b$-values obtained in the fitting process arise from
a contribution of thermal broadening from a Maxwellian distribution
and turbulent broadening (which can be characterized by a Gaussian
of width $b_{\mathrm{turb}}$), we can constrain the separate
contributions of these broadening mechanisms: $b^2 = {{2 k T} \over
{m}} + b_{\mathrm{turb}}^2$, where {$k$} is the Boltzmann constant,
and {$m$} is the mass of the atom. From the inferred $b$-values of
{\ion{O}{1}} and {\ion{H}{1}}, we place the following constraints on
the two values: $T=11400\pm800$\,K,
$b_{\mathrm{turb}}=1.9\pm0.7$\,\kms. These quantities should be
considered rough gauges of the temperature and turbulence in the
neutral gas since the absorption may contain substructure that is
blended in velocity.

In addition to the metallicity, we can consider the abundances of
other elements relative to oxygen in the neutral phase of the high
velocity gas. Of the ions with transitions observed, the following
are expected to dominate in a neutral phase: {\ion{C}{2}},
\ion{N}{1}, \ion{O}{1}, \ion{Mg}{2}, \ion{Si}{2}, {\ion{P}{2}},
\ion{S}{2}, \ion{Fe}{2}. (Most of these are singly ionized stages,
for which the ionization potential for creation is below
13.6\,eV.) If the relative abundances in this neutral phase
followed a solar pattern (Table~\ref{tab:solabund}) with no
ionization corrections the predicted elemental column densities
implied by the {\ion{O}{1}} column density
[$N$(X)=$N$(\ion{O}{1})$\cdot$(X/O)$_\odot$] would be $\log
N$(C)$=13.75$, $\log N$(N)$=13.09$, $\log N$(Mg)$=\log
N$(Si)$=12.70$, $\log N$(P)$=10.73$, $\log N$(S)$=12.43$, and
$\log N$(Fe)$=12.61$. Comparison of these scaled column densities
with the column densities reported in Table~\ref{tab:coladopt}
reveals that the column density for every detected singly-ionized
species listed above {\it exceeds} the expected column density for
neutral gas. (The column density upper limits on neutral and other
singly-ionized species covered by our dataset are not sufficiently
restrictive to place useful constraints on relative abundances.)
There are only two possible scenarios that can explain this: (1)
the relative elemental abundances deviate from the solar pattern,
or (2) ionized gas contributes significantly to the observed
column densities. While the former scenario might explain the
relative abundance of iron to oxygen with the invocation of
different enrichment scenarios, it is unlikely to explain the
relative abundances of oxygen, magnesium, and silicon, which are
all {$\alpha$}-process elements. We conclude that there are
significant ionization corrections for this low-ionization
material which must be considered in order to constrain the
relative elemental abundances of the HVC. To first order, we can
roughly estimate the magnitude of this ionization correction by
considering the {\ion{Si}{2}} column density, since the apparent
column density profile of {\ion{Si}{2}} appears to trace that of
{\ion{O}{1}} (see Figure~\ref{fig:compacdhvc}). Using the
predicted and observed column densities for {\ion{Si}{2}}, we
estimate that ionized gas contributes {$\gtrsim90$\%} ({$3\sigma$}
confidence) of the observed column density. This implies a
substantial ionization correction is needed to transform
{\ion{Si}{2}} to a total silicon abundance in the cloud. The
correction could, in fact, be larger if the HVC contains dust,
since the ionization correction only accounts for the gas phase
atoms. In the next section, we examine various scenarios to
explain the ionization of this gas.

\subsection{Ionization}
\label{sec:ionize}

As we have pointed out in the previous sections, in order to infer
relative abundances of metals and address, for example, the
possible presence of dust, it is imperative to consider the
effects of ionization on the observed column density ratios.
Understanding the ionization is important as such an analysis
yields information on the structure of the gas, the mass contained
in the neutral and ionized gas, and constraints on the possible
locations of the gas.

In the previous section, we estimated the approximate contribution
of the neutral phase (where {\ion{O}{1}} is produced) to the
observed column densities of low-ionization species, motivating
the need for an additional low-ionization phase. We can apply a
similar first order comparison of the amount of high-ionization
gas producing the {\ion{O}{6}} to the neutral gas producing the
{\ion{O}{1}}. The relative amounts of gas in each phase can be
estimated though the column densities of {\ion{O}{6}} and
{\ion{O}{1}} corrected for their relative ionization fractions
($f$), and abundances [(O/H)]:
\begin{equation}
{{N(\mathrm{H})_{\mathrm{HIG}}} \over
{N(\mathrm{H})_{\mathrm{NG}}}} = {{N(\mathrm{O~VI})
f_{\mathrm{O~VI}}^{-1} \mathrm{(O/H)}_{\mathrm{HIG}}^{-1}} \over
{N(\mathrm{O~I}) f_{\mathrm{O~I}}^{-1}
\mathrm{(O/H)}_{\mathrm{NG}}^{-1}}},
\end{equation}
where the subscripts HIG and NG denote high-ionization gas and
neutral gas, respectively. We assume that the oxygen abundances of
the two phases are similar
[(O/H)$_{\mathrm{HIG}}$ $\approx$ (O/H)$_{\mathrm{NG}}$] and that all
the oxygen in the neutral gas is in the form of {\ion{O}{1}}
($f_{\mathrm{O~I}}\approx1$). The ionization fraction of
{\ion{O}{6}} rarely exceeds 20\% {\citep{sembach03,ts00}}, so we
can place a lower limit on amount of gas in the high-ionization
phase relative to the neutral phase:
${{N(\mathrm{H})_{\mathrm{HIG}}} \over
{N(\mathrm{H})_{\mathrm{NG}}}} \gtrsim 0.2^{-1} \times
10^{14.00}/10^{13.82} = 7.6$.

For a more detailed understanding of the ionization of the high
velocity gas, we consider the kinematics of the HVC to motivate
the scenario under which the gas is ionized. The apparent column
density profiles shown in Figure~\ref{fig:compacdhvc}, which are
ordered in ionization potential of the depicted species, provide a
starting point. The complex kinematics of the profiles easily rule
out absorption originating in a medium having a single density and
temperature. It is physically impossible to explain, for example,
the drastically different kinematics of {\ion{O}{1}} and
{\ion{O}{6}} in such a phase. However, the velocity alignments of
the profiles suggest a common origin. This rules out scenarios
such as pure photoionization by the extragalactic background in a
low density plasma {\citep[e.g.,][]{nic02}}. Furthermore, variants
of this type of scenario also have difficulty explaining both the
ionization of the HVC and the kinematical properties of the high
velocity absorption lines.

We propose that the kinematics and ionization of the high velocity
gas are qualitatively consistent with a diffuse cloud that is moving
away from the Galactic disk and is being stripped by a hot,
low-density external medium such as the Galactic halo or corona. A
schematic depiction of the cloud structure is shown in
Figure~\ref{fig:cartoon}.  The neutral gas resides in the core of
the cloud, with some low ionization gas surrounding it. A leading
``shock'' may exist depending on the density contrast between the
cloud and external medium. [It is unlikely to be a classical shock
as the velocity of cloud is of the same order as the sound speed of
a hot ($>10^6$\,K), low density ($\lesssim10^{-4}$\,cm$^{-3}$)
external medium.] As the stripped matter equilibrates with the
external medium, it will heat up and slow down. For a fully
self-consistent and proper comparison of the observed profiles with
such a scenario, one needs a detailed hydrodynamical simulation
{\citep[e.g.,][]{qm01,ml04} which is beyond the scope of this paper.
Instead, we focus on three key observational features of the
absorption that support this scenario - the core
($162\lesssim\vlsr\lesssim192$\,\kms), the leading edge
($192\lesssim\vlsr\lesssim210$\,\kms) of the low ionization gas
(which may produce some moderate ionization gas), and finally, the
trailing wing ($125\lesssim\vlsr\lesssim184$\,\kms) of the
highly-ionized gas. The regions depicted in Figure~\ref{fig:cartoon}
are further illustrated in Figure~\ref{fig:acdcomp1}, where we
overlay a subset of the apparent column density profiles shown in
Figure~\ref{fig:compacdhvc} and shade the regions to be discussed in
the following sections.

\subsubsection{The Core}

The apparent column density profile of the {\ion{O}{1}} indicates
that there is a low-ionization/neutral core of absorbing gas. From
the profile shapes of the other species, we can associate the
following apparent column densities (and {$3\sigma$} limits) in
the velocity range {$162\leq\vlsr\leq192$\,\kms} with this phase
of gas: {$\log N$(\ion{O}{1})$=13.79\pm0.03$}, {$\log
N$(\ion{C}{2})$>14.20$}, {$\log N$(\ion{Mg}{2})$>13.0$}, {$\log
N$(\ion{Si}{2})$=13.76\pm0.04$}, and {$\log
N$(\ion{Fe}{2})$=13.32\pm0.07$}. As we indicated in
\S\ref{sec:hvcabund}, the Doppler widths of the {\ion{O}{1}} and
{\ion{H}{1}} allow us to measure the the temperature of this phase
if line widths are due to a combination of thermal and turbulent
broadening. The temperature of {$T \sim 10^4$\,K} implies that the
dominant ionization mechanism in this phase is photoionization,
since purely collisional processes would yield high ionization
fractions of neutral species whose ionization potentials fall
below 1~Rydberg. For example, at a temperature of {11,400\,K}, the
ionization fraction of {\ion{C}{1}} assuming collisional
ionization balance is 85\% {\citep{ds96}}, which would imply a
column density of {$\log N$(\ion{C}{1})$=13.63$} if the C/O
relative abundance is solar.

In photoionization models, there are two parameters that determine
the ionization structure of a gaseous slab - the ionizing
spectrum, and the gas density. (To a lesser degree, the
metallicity of the gas can also play a role since cooling
processes are related to metal content.) We assume here that the
dominant source of ionizing photons is the extragalactic
background and that the shape of this background is accurately
given by the computation of {\citet{hm96}} at {$z=0$}. We
normalize this background to
{$J_\nu(\mathrm{1~Ryd})=10^{-23}\,\mathrm{erg~
cm^{-2}~s^{-1}~Hz^{-1}~ster^{-1}}$} \citep[e.g., ][]{scott02} and
compute a grid of models with varying gas densities using the
Cloudy photoionization code {\citep{hazy}}. At each grid point, we
assume a plane-parallel geometry, and tune the thickness of the
ionized slab to reproduce the {\ion{H}{1}} column density
(assuming that the bulk of the {\ion{H}{1}} arises in this cloud
core). In addition, we tune the metallicity of the gas to
reproduce the observed {\ion{O}{1}} column density. (For slabs
with $\log n>-3$, the metallicity is determined directly from the
observed {\ion{O}{1}}/{\ion{H}{1}} ratio. For lower density slabs,
ionization corrections become important.) We summarize the results
of these calculations in Figure~\ref{fig:pie}. The thickness of
the slab is shown in the bottom panel as a dashed curve with the
scale shown on the right axis. The metallicity is shown in the
bottom panel as a solid curve with the scale shown on the left
axis. The bottom axis of each panel shows the ionization parameter
at the surface of slab, defined as the number density of ionizing
photons per hydrogen atom: $U = n_{\gamma}/n_{\mathrm{H}}$. On the
top axis, we show the gas density.

Assuming the relative metal abundances occur in their solar
proportions (Table~\ref{tab:solabund}), we show in the top panel of
Figure~\ref{fig:pie} the predicted column density curves {\ion{C}{2}},
{\ion{N}{1}}, {\ion{O}{1}}, {\ion{Mg}{2}}, {\ion{Si}{2}}, and
{\ion{Fe}{2}}. We overplot the integrated column densities of those
species at the location where they intersect the predicted curve with
larger symbols. The solid portions of the model curves indicate the
range over which the model column densities satisfy the observed
column density constraints ({$1\sigma$} range for measured values and
lower limits, {$3\sigma$} range for upper limits). There are three
competing effects that determine the shapes of these curves.  First,
the ionization fraction of {\ion{H}{1}} decreases with decreasing
density, so the required thickness of the slab (and the total amount
of gas) increases. Second, for densities lower than
{$10^{-3}$\,cm$^{-3}$}, the relative ionization correction between
{\ion{O}{1}} and {\ion{H}{1}} becomes important, resulting in larger
metallicities. Third, the ionization fractions of the species shown
(that is, neutral and singly-ionized species) decrease with decreasing
density. The first two effects serve to increase the predicted column
density of metal species as the gas density decreases, while the third
effect tends to decrease the column density of neutral and
low-ionization species.

Since the kinematics of {\ion{Si}{2}} are similar to {\ion{O}{1}},
and since both oxygen and silicon are $\alpha$-process elements,
it is reasonable to use the comparison of the column densities of
those species to constrain the density (and thereby the level of
ionization) of the core. If the models are an accurate description
of this phase, we constrain the density of the gas to {$\log
n_{\mathrm{H}}\sim-2.7$} ($U\sim-3.8$), corresponding to a thermal
pressure of {$p/k\sim24$\,K cm$^{-3}$} and an absorber thickness
of {$\sim1$\,kpc}. Under the approximation of a spherical
geometry, the gas mass of this phase is
{$\sim10^{4.4}$\,M$_{\odot}$}. (This estimate should treated
lightly since the geometry of the cloud is uncertain, and a
plane-parallel geometry was assumed in the model.) At this
density, and within the range of densities allowed by the
{\ion{Si}{2}}, all of the other column density constraints shown
are satisfied.

If the HVC is close to the Galactic disk, hot O and B stars can
provide an additional source of ionizing radiation
{\citep[e.g.,][]{csg04,sembach03,bm99,bm01}}. In this case, both
the normalization and shape of the ionizing spectrum will change,
favoring softer photons which will ionize neutral species whose
ionization potentials lie below 1~Rydberg. In turn, this would
require a larger density to yield a similar ionization parameter.
Consequently, if the HVC lies closer to the Galactic disk, then
the internal thermal pressure of the core is larger than derived
above, and the cloud core is also smaller. The addition of a
35,000\,K Kurucz model atmosphere with a normalization of
{$J_\nu(\mathrm{1~Ryd})=10^{-21}\,\mathrm{erg~
cm^{-2}~s^{-1}~Hz^{-1}~ster^{-1}}$} {\citep*{bp01,wvw02}} to the
extragalactic background would require a gas density of
{$\sim0.1$\,cm$^{-3}$} ($U\sim-4$) in order to provide the proper
shielding to explain the observed column densities considered
here. (Solar relative abundances are still sufficient.) In this
case, the thermal pressure in the cloud core is {$p/k\sim10^3$\,K
cm$^{-3}$} and the cloud size is {$\sim30$\,pc}.

\subsubsection{The Leading Edge}

The apparent column density profiles of the low-, moderate- and
high-ionization species show significant amounts of gas at
velocities beyond the cloud core out to about
{$\vlsr\approx+230$\,\kms}. The profiles in the velocity range
show a very striking trend between the maximum velocity of
detectable column density and the ionization potential of the
species. In Figure~\ref{fig:acdcomp1}, where we overlay the
apparent column density profiles of {\ion{O}{1}}, {\ion{Si}{2}},
{\ion{Fe}{2}}, {\ion{Si}{4}}, {\ion{C}{4}}, and {\ion{O}{6}}, this
trend is readily apparent within the region labelled ``Leading
Edge.'' While the high ionization {\ion{O}{6}} apparent column
density profile extends to {$\vlsr=230$\,\kms}, lower ionization
species clearly cut off at lower velocities, implying a strong
ionization velocity-gradient.

In the context of the proposed model of gas stripped from a cloud,
it is difficult to place where this higher-velocity edge arises
without a detailed understanding of the velocity field. It is
attractive to place the gas in front of the cloud, where there is
a direct interaction of the cloud with the hot external medium,
and perhaps a weak bow shock {\citep[e.g., ][]{qm01,ml04}}
depending on the relative velocity and density differential of the
interacting medium. The existence, relative importance, and
detailed velocities of such a shock are dependent on the speed of
the cloud through the medium, the density contrast, and the dark
matter content of the cloud. If the hot external medium has a
density of {$\sim10^{-4}$\,cm$^{-3}$} and temperature
{$\sim10^6$\,K} {\citep[e.g., the Galactic corona,][]{sembach03}},
then the cloud is moving through the external medium at
approximately the sound speed (i.e., with a Mach number of order
unity) and has a density contrast of about
{$n_{\mathrm{core}}/n_{\mathrm{external}}\sim30$}, using the
line-of-sight velocity and the core density inferred from
photoionization by the extragalactic background. If the cloud is
travelling through the halo, where the density of the external
medium is {$\sim10^{-3}$\,cm$^{-3}$}, then the density contrast is
{$\sim100$} (using the inferred density from photoionization by
the combination of the extragalactic background and Galactic
starlight). This is within the ranges of density contrasts
considered in recent models of the density evolution of clouds
moving through hot, low density media {\citet{ml04}}.

\subsubsection{The Trailing Wing}

The trailing tail of gas at velocities smaller than the core
absorption, $\vlsr<184$\,\kms, features statistically significant
column densities in high ionization species, {\ion{O}{6}},
{\ion{C}{4}}, {\ion{C}{3}}, and {\ion{Si}{4}}
(Figure~\ref{fig:acdcomp1}) down to $\vlsr=125$\,\kms. As with the
leading edge, the trailing tail also appears to show a trend of
increasing ionization with velocity separation (relative to the
core). To further examine this trend, we overlay in
Figure~\ref{fig:acdcomp2} the apparent column density profiles of
the {\ion{O}{6}$\lambda1031$}, {\ion{C}{4}$\lambda1548$} (top
panel), and {\ion{Si}{4}$\lambda1393$} (bottom panel). The profiles
have been scaled to roughly match the {\ion{O}{6}} apparent column
density at {$\vlsr=184$\,\kms} for ease of comparison, and they
shown with a sampling of two bins per resolution element. While the
FUSE data of the {\ion{O}{6}} profile have poorer resolution than
the STIS data of the {\ion{C}{4}} and {\ion{Si}{4}} profiles, the
observed trend of increasing ionization with velocity separation is
not a resolution effect, as the wing spans several FUSE resolution
elements. In addition to overlaying the apparent column density
profiles, we compute the
{$N_{\mathrm{a}}$(\ion{C}{4})$/N_{\mathrm{a}}$(\ion{O}{6})} and
{$N_{\mathrm{a}}$(\ion{Si}{4})$/N_{\mathrm{a}}$(\ion{O}{6})} ratios
as a function of velocity and overplot those as points in the same
panels. The points were computed by integrating over the velocity
range encompassed by the each bin of the {\ion{O}{6}} profile (each
bin corresponds to half of a FUSE resolution element, or about
10\,\kms). (We did not use the composite profiles for this
exercise.) The points are shown with {$1\sigma$} error bars when the
integrated apparent column densities are larger than three times
their errors, and the absorbed flux is larger than the $1\sigma$\
RMS in the continuum. Arrows are drawn on points which represent
$3\sigma$\ confidence limits. The ionization trend is clear, with
both ratios having a peak value at the core of the profile, and
decreasing (as the {\ion{O}{6}} column density becomes larger
relative to the other two ion column densities) with decreasing
velocity. In the context of the stripped-gas model we have proposed,
it is easier to place this lower-velocity gas, since the gas that is
stripped should take on the velocity and physical conditions of the
hot external medium.

\section{Discussion}
\label{sec:discussion}

A high velocity cloud at {$\vlsr\approx+184$\,\kms} in the high
Galactic latitude sight-line toward the quasar {PG\,$1116+215$} is
detected in absorption over a wide range of ionization species from
neutral {\ion{O}{1}} to high-ionization {\ion{O}{6}}. The kinematic
alignment of all detected species implies that these arise from a
common structure. Such a structure must contain multiple ionization
phases, since it is physically impossible for {\ion{O}{1}} and
{\ion{O}{6}} to arise from a single phase. From the high-quality
FUSE and HST/STIS data, we are able to infer a number of properties
for this isolated cloud. The detection of {\ion{O}{1}} is
fortuitous, since it provides a robust constraint (in tandem with
{\ion{H}{1}} information) on the metallicity of neutral gas within
the cloud. We estimate a metallicity of
{[O/H]$=-0.66_{-0.16}^{+0.39}$}. The systematic uncertainty is quite
large, unfortunately, since the Lyman series lines lie on the flat
part of the curve of growth. Nevertheless, comparison of the of the
{\ion{O}{1}} and {\ion{Si}{2}} column densities implies that the
neutral phase contributes no more than 10\% of the observed
low-ionization gas. Furthermore, comparison of the {\ion{O}{6}} to
{\ion{O}{1}} implies a highly ionized to neutral gas ratio of {\it
at least} $\sim$8. In summary, the relative proportions of gas in
the neutral, low-ionization, and high-ionization phases are roughly
1:10:8. This implies that substantial ionization corrections are
necessary to convert ion column densities to total elemental column
densities. We have proposed a possible scenario whereby a dense
cloud of gas is streaming through the Galactic corona, and have
pointed out various kinematic features - a core of neutral gas, a
leading edge of ionized gas, and trailing tail of highly-ionized gas
- to lend credence to this scenario. Detailed hydrodynamic modelling
to properly and self-consistently account for photoionization and
non-equilibrium collisional processes is required to test whether
this scenario can account for specific kinematic shapes and column
densities of the ions presented.

Ultimately, we would like to constrain the location of the
high-ionization high-velocity clouds, the origin of the gas, the
ionization mechanism, and mass contained therein. Many models have
been proposed, ranging from expanding superbubbles which originate
from the Galactic disk, to Local Group gas raining down on the
Milky Way, to a local filament of warm-hot intergalactic medium
encompassing the Local Group. With the addition of this work to
that of {\citet*{csg04}} and {\citet{fox05}}, we have a detailed
examination of the high-ionization high velocity clouds toward
five sight-lines ({PG\,1116+215}, Markarian~509, {PKS\,2155--304},
{HE\,0226--4110}, and {PG\,0953+414}), which are not readily
associated with one of the structures detected in {\ion{H}{1}}
21\,cm emission (e.g., the Magellanic Stream, Complex C, etc.).
{\citet{sembach03}} and {\citet{fox05}} note that the sight line
toward {HE\,0226--4110} passes {$10\fdg8$} from the
{$N$(\ion{H}{1} 21\,cm)$=2.0\times 10^{18}$\,cm$^{-2}$} contour
presented by {\citet{morras00}}. {\cite{fox05}} report {$\log
N$(\ion{H}{1})$\approx16.3$} for the HVCs detected in the sight
line. Thus, it is possible that those HVCs are associated with an
extension of the  Magellanic Stream. While nine HVCs is probably
insufficient to draw gross conclusions about the general
population of high-ionization HVCs, it is interesting to compare
the conclusions of {\citet{csg04}}, {\citet{fox05}}, and this
work.

We first note that the HVCs observed toward {PG\,1116+215} (as
well as those observed toward HE\,0226--4110 and PG\,0953+414)
have large positive velocities with respect to the Local Standard
of Rest, in contrast to the HVCs toward {PKS\,2155--304} and
{Mrk\,509}, and Complex C which have large negative velocities.
{\citet{csg04b}} have examined a simple model of gas falling
toward the Galactic center to consider the effects of Galactic
rotation on the observed sky distribution of velocities in the
Local Standard of Rest. They find that gas falling with a velocity
of {50\,\kms} will appear at high positive velocities
($\vlsr\geq100$\,\kms) for sightlines in the range {$225^\circ
\lesssim l \lesssim 320^\circ$}, {$|b|\lesssim 50^\circ$}. The
sight lines toward {PG\,1116+215}, {HE\,0226-4110}, and
{PG\,0953+414} lie at more extreme latitudes, so the observed
velocities imply that the clouds are travelling away from the
Galactic center. As further noted by {\citet{csg04b}}, these
velocities could be reproduced through the inclusion of a Galactic
fountain, although the low (albeit uncertain) metallicity of the
gas makes such an association difficult.

However, we note that all of the properties of the {+184\,\kms} HVC
follow naturally if this gas cloud is related to the Magellanic
Stream (MS). There is strong evidence that the MS has a
tidally-stripped, leading arm {\citep[e.g., ][]{putman98}}, and
{PG\,1116+215} is in the general direction of the leading arm.  In
this case, the positive velocity of the gas, the metallicity, and
the indications that the HVC is interacting with the ambient halo
all fit together: (1) the leading arm is expected to have positive
velocities in this direction, (2) the HVC metallicity is bracketed
by the metallicities observed in the ISM of the Magellanic Clouds,
[Zn/H]$_{\mathrm{SMC}}\sim$-0.6 \citep{welty97},
[Zn/H]$_{\mathrm{LMC}}\sim$-0.3 \citep{welty99}, and is similar to
the metallicity of the MS leading arm {\citep{lu98,sembach01}}, and
(3) it is known from other sight lines that the MS is interacting
with the halo/corona of the Milky Way {\citep{sembach03}}.  The
Magellanic Stream shows complex 21 cm emission at multiple
velocities {\citep[][and references therein]{wakker01}}, so some of
the other absorption components observed toward PG\,1116+215 could
also be related to the MS (e.g., the {+100\,\kms} HVC). We note that
the PG\,1116+215 sight line is outside of the region of the leading
arm that is easily recognized in 21 cm emission maps, but this could
be due to ionization and/or dissipation of the leading arm, which
causes the \ion{H}{1} column density to drop below the 21 cm
detection threshold. It is also possible that the {+184\,\kms} HVC
is not strictly ``leading arm'' material, but rather is gas that was
lost by the Magallanic Clouds during a previous orbit. In this case,
the passage of time could have reduced $N$(\ion{H}{1}), for example,
by ionization.

A possible counter-argument to an association with the leading arm
of the Magellanic Stream is the relative abundance pattern observed
in the low-ionization species. Our photoionization model of the
neutral and singly-ionized species in the cloud core shows that the
observed column densites are readily explained with solar relative
abundances. While the overall metallicity of the cloud is consistent
with that of the MS leading arm, there are marked differences in the
observed dust content, and depletion patterns.  {\citet{sembach01}}
report the detection of {H$_2$} in the MS leading arm in Lyman and
Werner absorption against the spectrum of the bright Seyfert 1
galaxy NGC\,3783 ($l=287\fdg5$, $b=23^\circ$). No molecular hydrogen
is detected in the PG\,1116+215 {+184\,\kms} HVC. In the MS leading
arm, Si and Fe show signs of appreciable depletion by 0.2 and 0.9
dex, respectively. Such a change to the relative abundances of Si
and Fe cannot explain the observed column densities in the
PG\,1116+215 {+184\,\kms} cloud core, if the photoionization model
is accurate. It is important to note, however, that the
low-latitude sight line to NGC\,3783 passes through a dense region
of the leading arm where the contribution of ionized gas to the
observed column densities of singly-ionized species is small
($\lesssim10-15$\%). The gas in the high-latitude PG\,1116+215
{+184\,\kms} cloud core, on the other hand, is predominantly ionized
and more diffuse. \citet{sembach01} estimate a density of
$n_{\mathrm{H}}=3.3$\,cm$^{-3}$\ for the H$_2$\ in the low-latitude
gas, which is at least 10 times larger than the density in the
{PG\,1116+215} {+184\,\kms} cloud core inferred from our
photoionization model. In the PG\,1116+215 {+184\,\kms} HVC, we
observe $N$(\ion{Fe}{2})$/N$(\ion{Si}{2})$=0.41$, whereas in the MS
leading arm toward NGC\,3783, {\citet{sembach01}} observed
$N$(\ion{Fe}{2})$/N$(\ion{Si}{2})$=0.18$. This difference is most
readily explained by a combination of ionization and differing
degrees of heavy metal incorporation into dust grains.  With the
available data, it is not possible to unambiguously distinguish
between these two effects, both of which modify N(Fe II)/N(Si II) in
lower density environments where the ionization level is typically
higher and the grain destruction is usually more complete than in
denser, mainly neutral regions. Thus, it is still possible that the
observed differences may be explained by the destruction of dust
grains, dissociation of molecules and ionization as the MS leading
arm gas continues to interact with the hotter external medium,
particularly if the high-velocity gas toward PG\,1116+215 (and
perhaps PG\,0953+414) is remnant stream material from an earlier
(older) stream passage.

We can compare the gross characteristics of the highly-ionized HVCs
by considering various column density ratios. In
Table~\ref{tab:cdr}, we reproduce the column density ratios reported
for the high velocity clouds toward {PKS\,2155--304}, and {Mrk\,509}
{\citep[][see their Table 5 for notes]{csg04}}, and for Complex C as
observed along the {PG\,1259+593} sight line
{\citep*{csg03,fox04,sembach04b}}. We add three additional high
velocity clouds from {\citet{fox05}}, and two high velocity clouds
from this work as well as the three Galactic ISM/IVC components
observed along the {PG\,1116+215} sight line. For comparison, we
plot these ion ratios in Figure~\ref{fig:coldensrat} (see
Table~\ref{tab:cloudkey} for the ordering of clouds in the figure).
The column density ratios for the HVCs toward {PG\,1116+215},
{HE\,0226--4110}, {PG\,0953+414}, {PKS\,2155--304} and {Mrk\,509}
tend to favor higher ionization gas, as evidenced by the
{\ion{O}{1}/\ion{O}{6}} constraints, in stark contrast to the
low/intermediate-velocity clouds toward {PG\,1116+215} and Complex
C. There is a clear distinction between Complex C (and the
low/intermediate velocity gas) and the high-ionization HVCs, with
the former being dominated by {\ion{H}{1}} and the latter being
dominated by {\ion{H}{2}}. This separation is also apparent in the
{\ion{Si}{2}/\ion{Si}{4}} ratio.

A possible argument against a Galactic origin for the highly-ionized
HVCs would be a comparison of the $N$(\ion{C}{4})$/N$(\ion{Si}{4})
column density ratio. As reported by {\citet{sst97}} in a study of
Radio Loops I and IV, this ratio remains fairly constant close to
and within the Galactic disk: $\log
N$(\ion{C}{4})$/N$(\ion{Si}{4})$=0.58_{-0.30}^{+0.18}$. Of the nine
high-ionization HVCs, only three are actually inconsistent with this
range, the {$\vlsr\approx+145$\,\kms} HVC toward {HE\,0226--4110},
the {$\vlsr\approx-270$\,\kms} HVC toward {PKS\,2155--304}, and the
{$\vlsr\approx-240$\,\kms} HVC toward {Mrk\,509}. {\citet{csg04}}
note that the {$N$(\ion{C}{4})$/N$(\ion{Si}{4})} ratio for the HVCs
toward {PKS\,2155--304} and {Mrk\,509} are significantly higher than
the ratio for gas within the Galactic disk or low halo, and argue
against a location close to or within the disk. The same ratio for
the {$\vlsr\approx+184$} high velocity cloud is not only consistent
with the ratio reported by {\citet{sst97}}, but is also consistent
with the ratio reported by {\citet*{csg03}} for Complex C. However,
as summarized by {\citet{fox04}}, there are several collisional
processes that can potentially explain the observed range of
{\ion{C}{4}/\ion{Si}{4}}. Thus, for the majority of the
highly-ionized HVCs that have been observed to date, we cannot rule
out an origin near (or within) the Galactic disk based on the
$N$(\ion{C}{4})$/N$(\ion{Si}{4}) column density ratio.

Further comparison of the observed {\ion{O}{6}/\ion{C}{4}} column
density ratios to the predicted ranges for the collisional processes
considered by {\citet{fox04}} reveals that conductive interfaces and
shock ionization can account for observed ratios for both the
highly-ionized HVCs (with one exception - {$\vlsr\approx-300$\,\kms}
toward {Mrk\,509}) and Complex C (see Table~\ref{tab:cine} for a
summary of the model parameters and ranges considered). Thus, while
the origins of HVCs may be varied (as indicated by the differences in
abundance patterns of the neutral/low-ionization species), there may
be a common mechanism (or mechanisms) for the production of the
high-ionization species. In particular, the interaction of the HVCs
with a hot medium such as the Galactic halo or corona can account for
the ionization patterns observed.  The ratios of high ionization
species for the highly-ionized HVCs are inconsistent with radiative
cooling processes (see Table~\ref{tab:cine}), which is presumably the
dominant process in the collapse of a large scale filament. This lends
additional credence to the idea that at least some of the
high-ionization HVCs reside either in the outer Galactic halo or the
more extended, lower-density corona (as opposed to the Local Group or
the WHIM). It is unlikely that the gas lies within a few kiloparsecs
of the Galactic disk, since the ion column density ratios favor higher
ionization species as compared to the low/intermediate-velocity gas in
this same sight line. For this reason, it is also more likely that the
highly ionized HVCs result from infalling material or tidally stripped
material than from gas ejected from the Galactic disk by supernova
explosions.

\section{Summary}
\label{sec:summary}
The primary results of this study are as follows:
\begin{itemize}
{\item We have obtained high resolution FUSE and HST/STIS echelle
observations of the quasar {PG\,$1116+215$}. The semi-continuous
coverage of the ultraviolet spectrum over the wavelength range
{916--2800\,\AA} provides detections of Galactic and high velocity
absorption over a wide range of ionization species: {\ion{H}{1}},
\ion{C}{2-IV}, {\ion{N}{1-II}}, {\ion{O}{1}}, {\ion{O}{6}},
{\ion{Mg}{2}}, {\ion{Si}{2-IV}}, {\ion{P}{2}}, {\ion{S}{2}}, and
{\ion{Fe}{2}}.}
{\item The high spectral resolution of these data (6.5-20\,\kms)
yields kinematic information for the Galactic, intermediate, and
high velocity absorption over the velocity range
{$-100<\vlsr<+300$\,\kms}, which provides an important {\it a
priori} starting point in the analysis of the absorption profiles.
We have constructed composite apparent column density profiles for
species where multiple unblended transitions are available within
a given instrumental setup. These are valid,
instrumentally-smeared profiles with a large dynamic range in
column density. In particular, we are able to fully recover the
apparent column density profiles of {\ion{Si}{2}} and
{\ion{Fe}{2}} at {6.5\,\kms} and {10\,\kms} velocity resolution,
respectively.}
{\item In the low and intermediate ionization species, we detect
two high velocity clouds at {$\vlsr\approx100$\,\kms} and
{$\vlsr\approx184$\,\kms}. The {$\vlsr\approx100$\,\kms} component
is detected as an absorption feature in the
{\ion{C}{2}$\lambda1334.532$}, {\ion{Si}{2}$\lambda1260.422$} and
{\ion{Si}{3}$\lambda1206.500$} transitions. Blended and/or
saturated absorption at this velocity exists in the other higher
ionization species (i.e., \ion{C}{3}, \ion{C}{4}, \ion{O}{6},
\ion{Si}{4}), while no detectable absorption exists in other low
ionization or neutral species. The high velocity cloud at
{$\vlsr\approx184$\,\kms} is detected over a large range of
ionization species, from neutral {\ion{O}{1}} to highly ionized
{\ion{O}{6}}, with striking kinematics differences.}
{\item The apparent column density profile of {\ion{O}{1}} in the
{$+184$\,\kms} HVC has a narrow discrete component that is
apparent in other low ionization species. In the singly-ionized
species, there also appears to be a shelf of column at slightly
higher (i.e., more positive) velocities than the neutral core
traced by {\ion{O}{1}}. In the high ionization species (e.g.,
\ion{C}{4}, \ion{O}{6}), the profiles are broad, and asymmetric,
with a tail of absorption extending toward lower velocities.}
{\item Since there is essentially no relative ionization correction
between {\ion{O}{1}} and {\ion{H}{1}}, we have attempted to measure
the metallicity of the neutral HVC gas using column density ratios
of these two species. The higher order {\ion{H}{1}} Lyman series
lines suffer from saturation, and therefore do not tightly constrain
the {\ion{H}{1}} column density. The best fit curve of growth and
profile fit analysis yields an optimal metallicity of
[O/H]$=-0.66_{-0.16}^{+0.39}$~(1$\sigma$).}
{\item The {\ion{Si}{2}}/{\ion{O}{1}} ratio provides a measure of
the contribution of the neutral gas to the observed column density
of low-ionization species. For solar relative abundances of Si and
O, we estimate that at most 10\% of the {\ion{Si}{2}} column
density can be attributed to the neutral gas that produces the
{\ion{O}{1}}.}
{\item A simple model of gas with density
{$10^{-2.7}$\,cm$^{-3}$}, a thickness of $\sim1$\,kpc, and solar
relative metal abundances for O, Si, and Fe photoionized by the
Haardt-Madau extragalactic spectrum is able to explain the
observed column densities (or limits) of the neutral and
low-ionization species in the velocity range
{$\vlsr=$162--192\,\kms} (i.e., in the cloud core). The addition
of ionizing photons from the Galactic disk increases the density
and thermal pressure estimates by about a factor of 50, and
decreases the absorber thickness by about a factor of 30.}
{\item We have examined the relative contributions of neutral and
highly-ionized gas in the high velocity cloud by considering the
ratio {\ion{O}{6}} with respect to {\ion{O}{1}}. In the total
velocity range 140--230\,\kms, the integrated column densities
indicate that there is at least $\sim$8 times more highly ionized
gas than neutral gas, assuming that highly-ionized gas has a similar
metallicity to the neutral gas.}
{\item The qualitative features of the apparent column densities as
function of both velocity and ionization suggest an absorbing
structure in which a moderately dense cloud of gas is heading away
from the Galactic center (or in orbit) and passing through a hot
tenuous medium (e.g., the Galactic halo or corona), which is
stripping gas off the cloud. The denser core of the cloud gives rise
to the neutral species and some low-ionization species. The front of
the cloud faces away from the observer at higher velocities and is
the prime site where gas is stripped. This front edge produces the
more moderate ionization species (e.g., \ion{Si}{3}, \ion{C}{3}), as
well as some high ionization species (\ion{Si}{4}, \ion{C}{4},
\ion{O}{6}). The stripped gas interacts with the hot external
medium, takes on the velocity and ionization of that medium, and
gives rise to a high-ionization (e.g., \ion{O}{6}) ``wake'' trailing
the cloud at lower velocities.}
{\item In considering this model, it is important to distinguish
whether the cloud is travelling through the Galactic halo or the
more extended and diffuse Galactic corona, since the ability of the
ambient medium to strip gas from the cloud depends on the relative
density. If the cloud is travelling through the Galactic corona, its
velocity is roughly the sound speed, and the density contrast
between the neutral core and the external medium is roughly 30:1. If
the cloud is travelling through the Galactic halo, the contrast is
of order 100:1.}
{\item To further examine the feasibility of the non-equilibrium
scenarios in explaining the absorbing gas, we suggest that future
numerical simulations of the evolution of clouds moving through a
tenuous medium be accompanied by an exploration of the direct
observables (e.g., column densities, line widths, line
asymmetries) as a function of ionization stage. This can lead to
the formulation of key diagnostics (e.g., column density ratios
and line profile shape differences) that can be used to understand
the relative importance of different hydrodynamical processes in
producing the high velocity gas.}
{\item The ionization structure, kinematical complexity, and
column density ratios of the ionic species observed in this high
velocity cloud rule out the photoionization models suggested by
{\citet{nic02}} as an explanation for {\ion{O}{6}} HVCs in the
Local Group {\citep[e.g.,][]{nic02}}. Rather, this HVC is part of
a dynamical system that likely involves complex interactions with
the outer regions of the Galaxy, as has been suggested by other
authors {\citep[e.g., ][]{sembach03,tripp03,fox04}}.}
{\item We suggest that the leading arm of the Magellanic Stream may be
the origin of this gas based on the location of the high velocity gas
on the sky, and its high positive velocity. The ionization level and
the metallicity of the high velocity gas core, and the kinematics of
the higher ionization gas, are also consistent with such an origin.
Alternatively, the gas could arise from Stream gas from a prior orbit
or from a tidally disrupted dwarf galaxy unrelated to the Magellanic
Clouds.  Further studies of the metallicity and ionization of the
positive high velocity gas along other sight lines in this region of
the sky could help to clarify the origin(s) of the gas.}
\end{itemize}

\acknowledgments This work is based on data obtained for the
Guaranteed Time Team by the NASA-CNES-CSA FUSE mission operated by the
Johns Hopkins University. Partial financial support has been provided
by NASA contract NAS5-32985 and Long Term Space Astrophysics grant
NAG5-3485 (KRS). TMT appreciates support from NASA through grant
NNG04GG73G.

\appendix
\section{Description of Individual Ions}

\begin{itemize}
\item[\ion{C}{1}:] In the STIS/E140M spectrum, we cover four
strong transitions of the {\ion{C}{1}} ion at {1277.245\,\AA},
{1328.833\,\AA}, {1560.309\,\AA}, and {1656.928\,\AA}. The relative
values of {$f\lambda$} of these lines (with the exception of
{1560.309\,\AA}) have been investigated by {\citet{jt01}}, who find
systematically higher values of $f\lambda$ for lines with smaller
published $f\lambda$. We adopt their values over those listed in
{\citet{morton03}} for the {1277.245\,\AA}, {1328.833\,\AA}, and
{1656.928\,\AA} lines. [The line at {1560.309\,\AA} was not
considered in the {\citet{jt01}} analysis, and we adopt the
{\citet{morton03}} data.] An additional transition at {945.191\,\AA}
is also covered by the FUSE data. There is a weak unidentified
absorption feature {$\sim$160\,\kms} blueward of the {1277.245\,\AA}
line, but it does not affect any of our measurements. In addition,
there is a weak, broad Lyman {$\alpha$} line at {$z=0.0928$}, which
is blended with the {1328.833\,\AA} line, but it was easily removed
in the continuum-fitting process. In the top panel of Figure~3a, we
overlay the apparent column density profiles of these transitions.
Absorption by this ion is only visible in the
{$\vlsr\approx-44$\,\kms} intermediate velocity component, where the
core apparent optical depth in all transitions is less than unity.
The agreement between ACD profiles is reasonable, given the weakness
of the lines and the noisiness of the data. The integrated column
density of the {$\vlsr\approx-44$\,\kms} absorption is greatly
affected by the choice of integration range for this same reason
(weak lines and noisy data), so we restrict the integration range of
this component to {$-60\leq\vlsr\leq-35$\,\kms}. The {1656.928\,\AA}
and {1277.245\,\AA} lines yield identical integrated column
densities, while the {1560.309\,\AA} and {1328.833\,\AA} lines yield
slightly smaller and larger integrated column densities,
respectively. We use the three lines with self-consistent
line-strengths ({1277.245\,\AA}, {1328.833\,\AA}, and
{1656.928\,\AA}) to compute our adopted column densities (and
limits) for all integration ranges.
\item[\ion{C}{2}:] There are two transitions of the {\ion{C}{2}}
ion at {1036.337\,\AA} and {1334.532\,\AA} which are covered by
our data. Both are detected, the {1036.337\,\AA} line in the FUSE
spectra, and the {1334.532\,\AA} line in the STIS-E140M and both
suffer from unresolved saturated structure and blends with the
associated excited state lines. We use only the {1334.532\,\AA}
line for our measurements, and we can derive only lower limits on
{$N$(\ion{C}{2})} in most of the components.
\item[\ion{C}{3}:] The only transition of {\ion{C}{3}} covered is
the strong {977.020\,\AA} line in the FUSE spectrum. The line
saturates at an apparent column density per velocity interval of
{$5.1 \times 10^{11}$} cm$^{-2}$\ (\kms)$^{-1}$, and is
effectively black (i.e., with an absorbed flux less than the
$1\sigma$\ above the zero flux level) at an apparent column
density per velocity interval of {$1.2 \times 10^{12}$} cm$^{-2}$\
(\kms)$^{-1}$. In all of the integration ranges, the observed flux
is less that $1\sigma$\ above the zero flux level, and we adopt
lower limits on integrated column densities.
\item[\ion{C}{4}:] The resonant doublet at
$\lambda\lambda$1548.204, 1550.781 is detected in the STIS E140M
spectrum (see Figure~\ref{fig:data}). The {1548.204\,\AA} falls at
the edges of two adjacent echelle orders and the data are
interlaced to produce the profile shown in Figure~2a and for the
subsequent analysis. There is agreement in the integrated apparent
column densities between the two transitions over the velocity
range encompassing the low- and intermediate velocity gas,
{$-100\leq\vlsr\leq+85$\,\kms}. Integration over the velocity
ranges associated with the two high velocity clouds yield slightly
discrepant apparent column densities, although they are consistent
within the uncertainties. For the {$\vlsr\approx+100$\,\kms}
component, the integrated apparent column densities indicate
possible unresolved saturated structure, but this is unlikely to
be the case as the apparent optical depths are low. We use both
transitions in computing the composite, but use only the
{1548.204\,\AA} transition for our adopted column densities in all
velocity intervals.
\item[\ion{N}{1}:] There are several strong triplets of the
{\ion{N}{1}} ion, with the strongest and best separated at
{$\lambda\lambda$1199.550, 1200.223, 1200.710}. The two {\ion{H}{1}}
components detected in the {\citet{wakker03}} 21\,cm emission
profile are detected in this ion, although they are heavily
saturated and blended together. Our ``standard'' treatment for
recovering the composite apparent column density profile (described
above) using all three transitions is able to recover a clean
profile with unresolved saturated structure over those two
components and provide clear non-detections in the other three
components. For our adopted {\ion{N}{1}} column densities, we use
the {1200.710\,\AA} transition for the low-velocity clouds and the
{$\vlsr\approx+100$\,\kms} HVC since it is unaffected by blending
and is the weakest of the three transitions (thus providing the best
lower limits in the regions of unresolved saturation). For the
{$\vlsr\approx+184$\,\kms} HVC, we use the {1200.223\,\AA} line to
derive the best upper limit, since it is unaffected by blends over
the integration range and is stronger than the {1200.710\,\AA}
transition.
\item[\ion{N}{2}:] Only the {1083.994\,\AA} line is covered by our
dataset. It is detected in the SiC1 channel of the FUSE spectrum.
The line is clearly saturated over the integration ranges of the
{$\vlsr\approx-44$\,\kms} and {$\vlsr\approx-7$\,\kms} components
and may contain some unresolved saturated structure in the
{$\vlsr\approx+184$\,\kms} HVC component and we adopt lower limits
for the column density for these components. The absorption is not
saturated over the integration of the {$\vlsr\approx+56$\,\kms}
component and we quote a value over the integration range
{+37\,\kms} to {+85\,\kms}. It is not detected at
{$\vlsr\approx+100$\,\kms} and we adopt an upper limit on the
column density.
\item[\ion{N}{5}:] The strong doublet of the high-ionization
{\ion{N}{5}} ion at {$\lambda\lambda$1238.821,1242.804} is covered
in the STIS E140M data and is not detected in any component. The
nominal location of the {1238.821\,\AA} line falls on two orders
of the echelle, like the {\ion{C}{4}$\lambda1548.204$} line. We
use the {1238.821\,\AA} transition to compute the best upper
limits on the {\ion{N}{5}} column density for all integration
ranges.
\item[\ion{O}{1}:] There are several {\ion{O}{1}} lines in the
FUSE band which suffer from blends with transitions from other ions
(e.g., the {\ion{H}{1}} Lyman series). In the STIS E140M band, we
detect the  {1302.169\,\AA} line. The low- and intermediate velocity
gas in this line suffers from unresolved saturation. The line also
falls on two orders of the echelle. We use the {1302.169\,\AA}
profile for all integrated column densities, reporting lower limits
for the low-velocity components, an upper limit for the {+100\,\kms}
HVC and a measurement for the {+184\,\kms} HVC. See
\S\ref{sec:184oicol} for further examination of possible saturation
effects in the \ion{O}{1} column density for this HVC.
\item[\ion{O}{6}:] The strong doublet at {1031.926\,\AA}, and
{1037.617\,\AA} is clearly detected in the FUSE band, and has been
presented by {\citet{wakker03}}. The {1031.926\,\AA} line is free
of blends, while the {1037.617\,\AA} line is blended with {H$_2$}
from the {$-44$\,\kms} intermediate-velocity gas. Thus, we only
use the {1031.926\,\AA} line for our measurements, and do not
attempt to combine the two lines to form a composite ACD profile.
Since \ion{O}{6} is not detected in the {-44\,\kms} component, we
report an upper limit on the column density.
\item[\ion{Mg}{1}:] The strongest line from {\ion{Mg}{1}} at
{2852.963\,\AA} ($\log f\lambda=3.718$) is not covered by our dataset;
our STIS E230M spectrum cuts off at 2819.13\,\AA.  However, the weaker
lines at {2026.477\,\AA} and {1707.061\,\AA} are covered. These are
not detected, and we use the stronger {2026.477\,\AA} line to
determine the limiting column densities for all components. The
{2026.477\,\AA} line falls on two adjacent orders of the STIS E230M
echelle spectrum.
\item[\ion{Mg}{2}:] The strong doublet at {2796.354\,\AA}, and
{2803.532\,\AA} is detected in the STIS E230M spectrum. The velocity
range {$-100\leq\vlsr\leq+40$} suffers from blended saturated
structure (the velocities covering the {$\vlsr\approx-44$\,\kms} and
{$\vlsr\approx-7$\,\kms} components), and the component at
{$\vlsr\approx+56$\,\kms} is resolved, but saturated. For these
components we quote lower limits on the integrated apparent column
density using the {2803.532\,\AA} line. The high velocity gas at
{$\vlsr\approx+100$\,\kms} is not detected, while the high velocity
gas at {$\vlsr\approx+184$\,\kms} is marginally saturated. We use
the {2796.354\,\AA} transition to compute the best upper limit on
the column density for the {+100\,\kms} component, and combine the
measurements from both transitions for the {$+184$\,\kms} component.
\item[\ion{Si}{2}:] There are five transitions covered by the STIS
E140M spectrum, which range in line strength {$2.051\leq\log
f\lambda\leq3.172$} at 1190.416\,\AA, 1193.416\,\AA,
1260.422\,\AA, 1304.370\,\AA, and 1526.707\,\AA. The transitions
suffer from blended saturated structure in the velocity ranges
which include the {$\vlsr\approx-44$\,\kms} and
{$\vlsr\approx-7$\,\kms} components. For these two components, we
quote limiting column densities from the integration of the
{1304.370\,\AA} line, which is detected in adjacent echelle
orders. Likewise, we also use this line to quote a measurement of
the {$\vlsr\approx+184$\,\kms} high velocity component, since it
is the least affected by unresolved saturated structure. The high
velocity cloud at {$\vlsr\approx+100$\,\kms} is only detected in
the {1260.422\,\AA} transition and we adopt the integrated column
density from this line.
\item[\ion{Si}{3}:] The only transition of {\ion{Si}{3}} covered
by our dataset is at 1206.500\,\AA. The components at
$\vlsr\approx-44$\,\kms, $-7$\,\kms, and +56\,\kms are strongly
saturated and blend together. The component at
{$\vlsr\approx+100$\,\kms} is optically thick in this transition,
as is the component at {$\vlsr\approx+184$\,\kms}. We treat the
integrated column densities in all components as lower limits.
\item[\ion{Si}{4}:] The strong doublet at {1393.760\,\AA}, and
{1402.773\,\AA} is covered and detected in the STIS E140M spectrum.
The {1402.773\,\AA} line falls in two adjacent echelle orders in the
velocity region {$\vlsr\leq20$\,\kms}. The {1402.773\,\AA} profile
shows some minor inconsistencies with the {1393.760\,\AA} profile.
In particular, there is an excess of absorption at
{$\vlsr\approx10$\,\kms}. The feature extends over a few pixels and
appears to be real as it exists independently in the two echelle
orders. It is possible that an unidentified absorption line (e.g., a
weak, narrow {Lyman $\alpha$} line) contaminates the {1402.773\,\AA}
line. For this reason, we use only the {1393.760\,\AA} in our
computation of the column density of the {-7\,\kms} component. For
other components, we combine the measurements from both transitions.
\item[\ion{P}{2}:] The strongest transition of {\ion{P}{2}} which
is covered by our dataset is the {963.800\,\AA} line ($\log
f\lambda=3.148$). This line is clearly detected in the FUSE band, but
is blended with the \ion{N}{1}$\lambda\lambda$ 963.990, 963.626,
963.041 triplet and is not useful for our purposes. The strongest
transition that is free of blends is the {1152.818\,\AA} line which is
covered by both FUSE, and STIS data. The line is detected in the FUSE
spectrum, but not in the STIS spectrum due to poor signal-to-noise in
that region of our STIS spectrum. The next strongest lines available
in the higher resolution STIS spectra are at {1301.874\,\AA} and
{1532.533\,\AA} (E140M). The low- and intermediate-velocity components
at {$\vlsr\approx-44$\,\kms} and {$-7$\,\kms} are detected in the
{1152.818\,\AA} line. No other components are detected. No absorption
is detected in the {1532.533\,\AA} line, and the absorption in the
expected region of the {1301.874\,\AA} line is dominated by the
Galactic \ion{O}{1} 1302.619\,\AA\ absorption. (Given the strength of
the \ion{P}{2} {1152.818\,\AA} line, there is no reason expect
blending issues with the Galactic {\ion{O}{1}} measurements.) We use
the {1152.818\,\AA} line for the column density determinations in all
components (measurements for the {-44\,\kms} and {-7\,\kms}
components, upper limits for the other components).
\item[\ion{S}{1}:] The strongest lines from {\ion{S}{1}} which are
covered by our spectra are at 1295.653\,\AA, and 1425.030\,\AA.
Both of these are covered in E140M spectrum but are not detected.
We use the stronger {1425.030\,\AA} line to obtain upper limits on
the integrated column densities.
\item[\ion{S}{2}:] The only transitions of {\ion{S}{2}} at
wavelengths redward of the Lyman limit are covered and detected in
the E140M spectrum. The {\ion{S}{2}} $\lambda\lambda$1250.578,
1253.805, 1259.518 lines all show a resolved two component
structure at velocities {$\vlsr\approx-44$\,\kms} and
{$\vlsr\approx-7$\,\kms}. No other velocity components are
detected. The {1259.518\,\AA} line has an odd kinematic structure
over the velocities spanned by these two components which is not
consistent with the other two lines. The structure is apparent in
only one order, but we find nothing to suggest that it should not
be trusted (e.g., there is nothing odd in the error spectrum). The
two detected components contain unresolved saturated structure in
this line, so we do not use this transition in our adopted column
density measurements. In addition, the integration range for the
{$\vlsr\approx-44$\,\kms} component in the {1250.578\,\AA} line is
contaminated by {Lyman $\alpha$} at $z=0.02845$. Comparison of the
apparent column density profiles of the {1250.578\,\AA} and
{1253.805\,\AA} lines indicates possible unresolved saturated
structure in the {1253.805\,\AA} line. So we quote the integrated
column density of the {1253.805\,\AA} line as a lower limit. For
the {-7\,\kms} component, we combine the integrated apparent
column densities of the {1250.578\,\AA} and {1253.805\,\AA} lines.
The integration range for the undetected high velocity cloud at
{$\vlsr\approx+184$} in the {1253.805\,\AA} line is contaminated
by another {Lyman $\alpha$} line at $z=0.03223$. Only the
{1250.578\,\AA} line is used for the adopted column density upper
limit for this component and the {+100\,\kms} and {+56\,\kms}
components.
\item[\ion{Fe}{2}:] A suite of {\ion{Fe}{2}} transitions is
covered in our STIS E230M dataset. In particular, we detect all
transitions in the wavelength range 2249.8--2600.2\,\AA\ with line
strengths exceeding {$\log f\lambda\geq0.6$}. We also detect the
transition at 1608.451\,\AA\ in the E140M spectrum. The
transitions with line strengths exceeding {$\log
f\lambda\geq1.87$} suffer from unresolved saturated structure in
the velocity range {$-100\leq\vlsr\leq+20$\,\kms}. We combine the
integrations over the 2260.781\,\AA\ and 2249.877\,\AA\ lines to
determine composite column densities for the
{$\vlsr\approx-44$\,\kms} and {$\vlsr\approx-7$\,\kms} components.
The component at {$\vlsr\approx+56$\,\kms} is detected in
transitions with {$\log f\lambda\geq1.87$}. We combine all
transitions to compute a composite integrated column density for
this component. The high velocity component at
{$\vlsr\approx+100$\,\kms} is not detected in any transition (up
to $\log f\lambda=2.882$), and we use the {2382.765\,\AA} line to
compute an upper limit on the integrated column density. For the
high velocity cloud at {$\vlsr\approx+184$\,\kms}, there are
inconsistencies in the apparent column density profiles to note
which affect the computation of a composite column density. The
profiles from the 2382.765\,\AA\ and 2600.173\,\AA\ lines may be
affected by unresolved saturated structure, although they yield
identical integrated column densities. The 2344.214\,\AA\ and
2374.461\,\AA\ lines yield systematically (and significantly)
smaller and larger integrated column densities, respectively, than
the 2382.765\,\AA\ and 2600.173\,\AA\ lines. If the 2382.765\,\AA\
and 2600.173\,\AA\ lines are included in the computation of a
variance-weighted column density, then they dominate the
computation, $\log \acd$(\ion{Fe}{2})$=13.41\pm0.03$. However, if
only the 1608.451\,\AA\ and 2586.650\,\AA\ lines are used, then we
find a slightly larger, though more uncertain, value $\log
\acd$(\ion{Fe}{2})$=13.46\pm0.09$. We adopt this latter value for
the integrated column density.
\end{itemize}

\bibliographystyle{apj}

\clearpage

\begin{figure*}[ht!]
\figurenum{1}
\epsscale{0.8}
\plotone{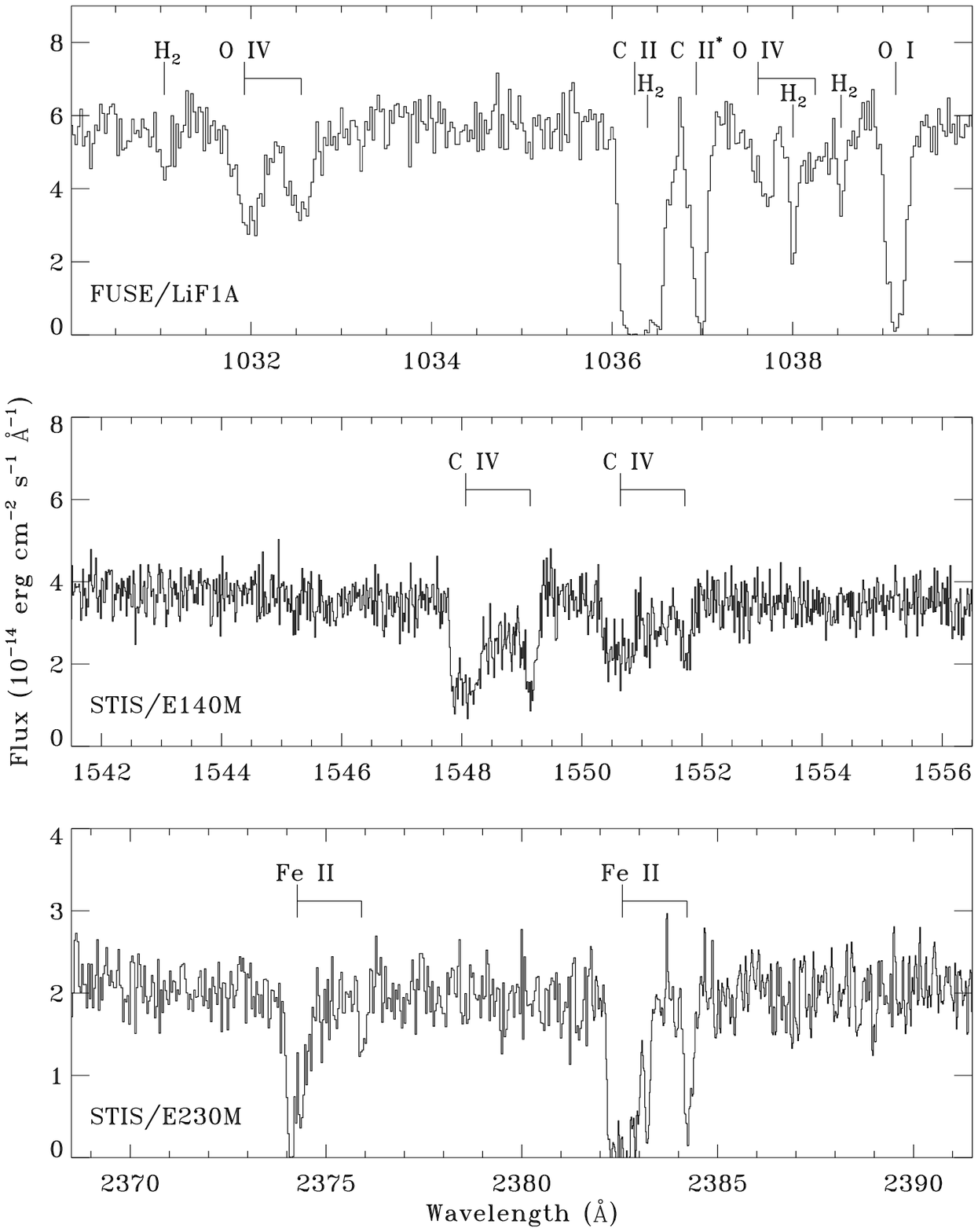}
\protect\caption[Sample Spectra]{Portions of the spectra from the
three observations used in this paper - FUSE (top panel),
HST/STIS-E140M (middle panel), HST/STIS-E230M (bottom panel). The
plotted regions are centered around key transitions from each
observation, and the wavelength scale shown spans about
{3300\,\kms} in all three panels. In each panel, the Galactic ISM
profiles are labelled with the absorbing species, and the
corresponding {+184\,\kms} HVC absorption are indicated by the
offset tick marks.} \label{fig:data}
\end{figure*}

\clearpage

\begin{figure*}[ht!]
\figurenum{2a}
\vglue -0.3in
\epsscale{0.60}
\plotone{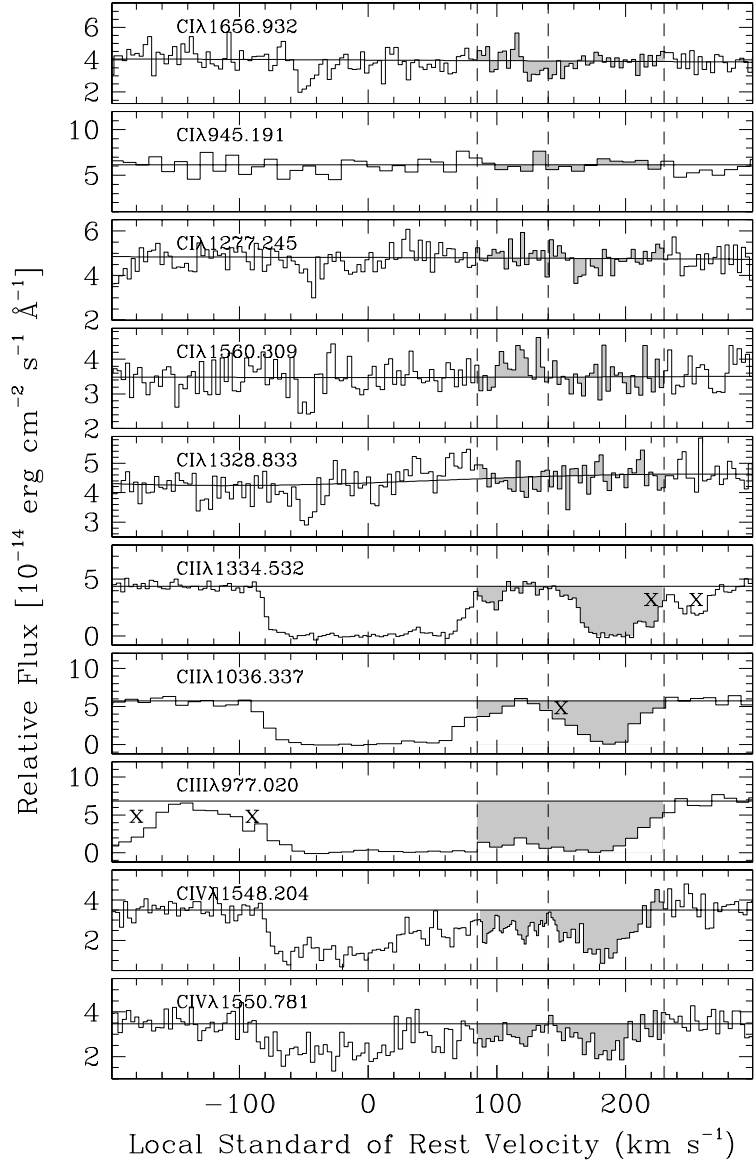}
\vglue -0.1in
\protect\caption[Flux Profiles]{Flux profiles of selected
transitions of carbon ions versus Local Standard of Rest velocity.
The profiles are ordered, top to bottom, by increasing ionization
stage and decreasing transition strength (within each stage). The
velocity range 100--230\,\kms, over which high velocity absorption
is expected (though not necessarily detected), is shaded. The dashed
vertical lines at {$\vlsr=85$}, 140, and 230\,\kms\ mark the
integration limits for the HVCs. Regions affected by blends with
other absorption features are marked with an 'x'. Note that the
scale for the flux axis changes from panel to panel. (See the notes
in the appendix and {\citet{sembach03}} for identifications of the
blends.) Transitions below {1184\,\AA} are detected in the FUSE
spectrum; transitions in the range {1184--1720\,\AA} are detected in
the STIS E140M spectrum; transitions above {2000\,\AA} are detected
in the STIS E230M spectrum.}
\end{figure*}

\clearpage

\begin{figure*}[ht!]
\figurenum{2b}
\vglue -0.3in
\epsscale{0.73}
\plotone{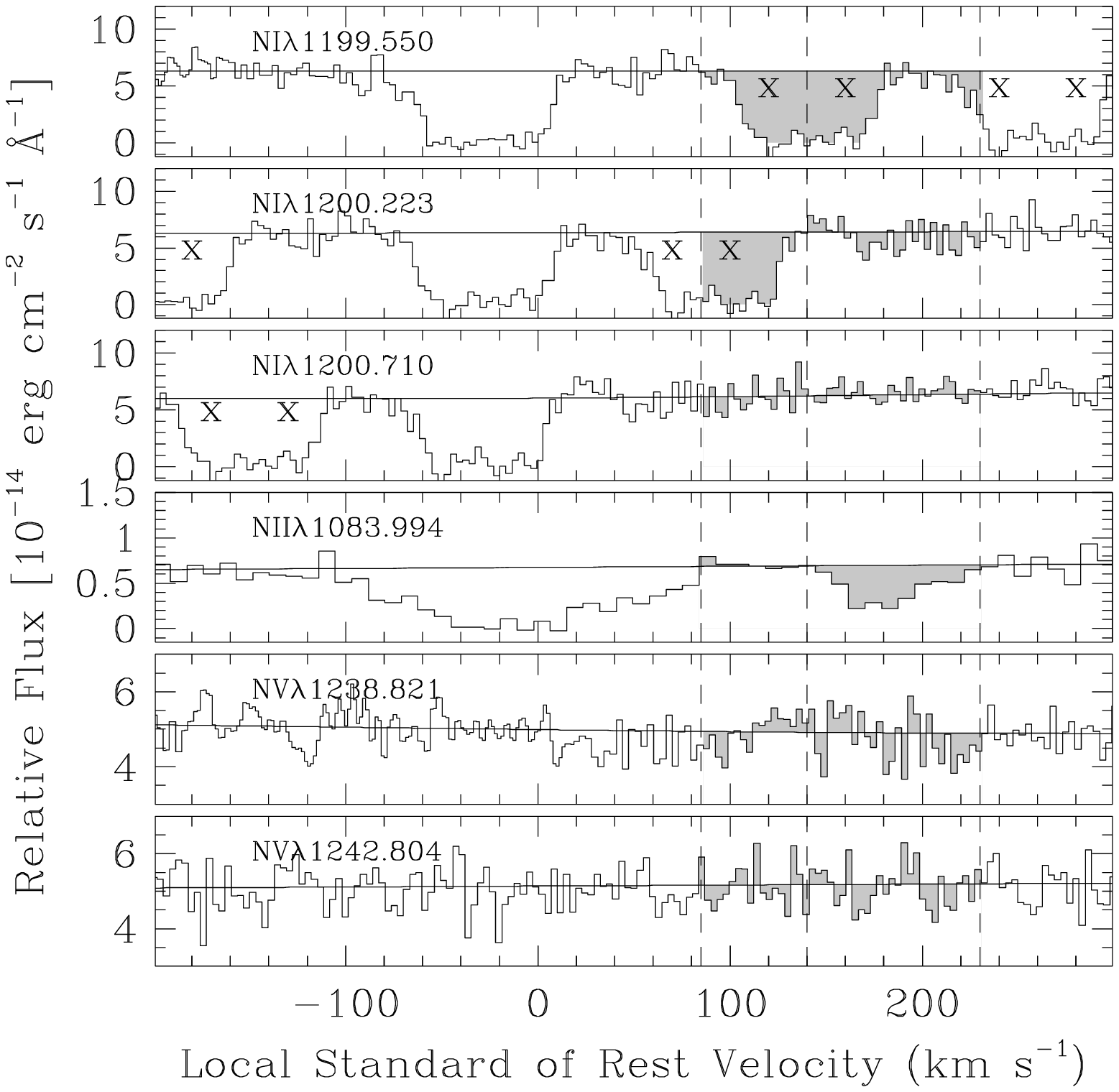}
\vglue -0.1in
\protect\caption[Flux Profiles]{Same as Figure~2a, but for the
ions of nitrogen.}
\end{figure*}

\clearpage

\begin{figure*}[ht!]
\figurenum{2c}
\vglue -0.3in
\epsscale{0.73}
\plotone{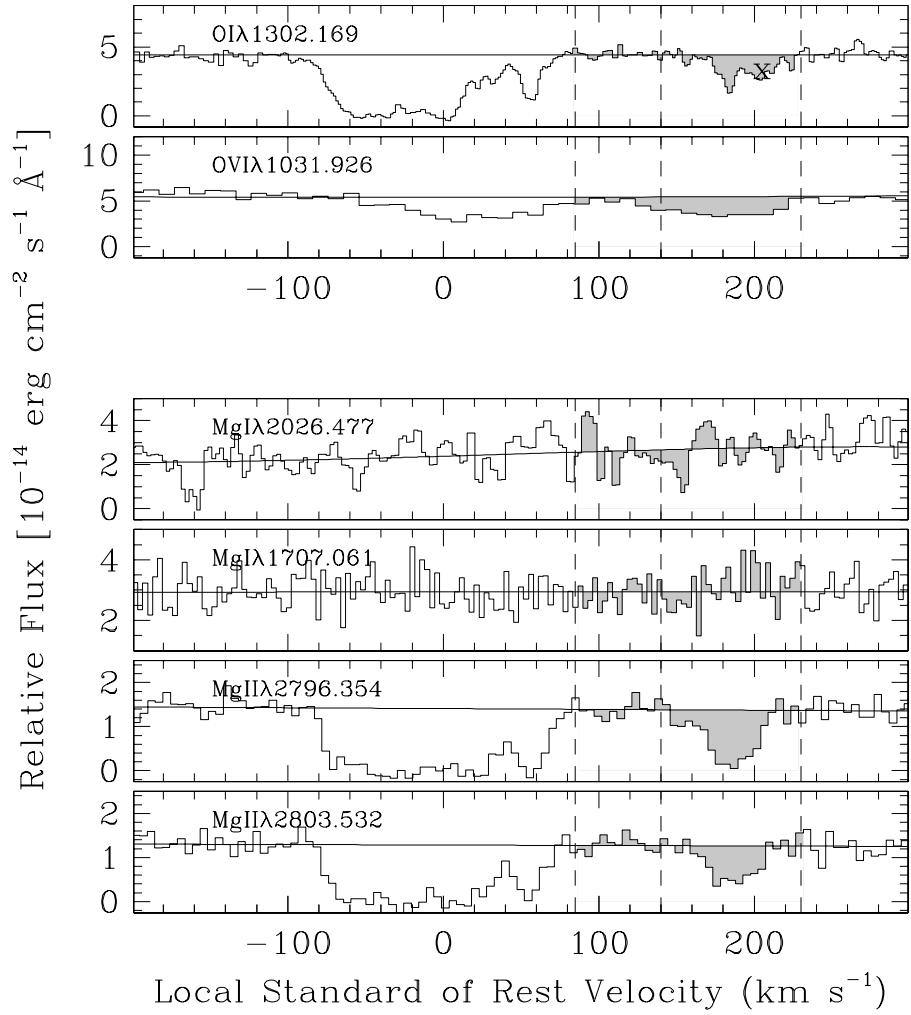}
\vglue -0.1in
\protect\caption[Flux Profiles]{Same as Figure~2a, but for the
ions of oxygen and magnesium.}
\end{figure*}

\clearpage

\begin{figure*}[ht!]
\figurenum{2d}
\vglue -0.3in
\epsscale{0.73}
\plotone{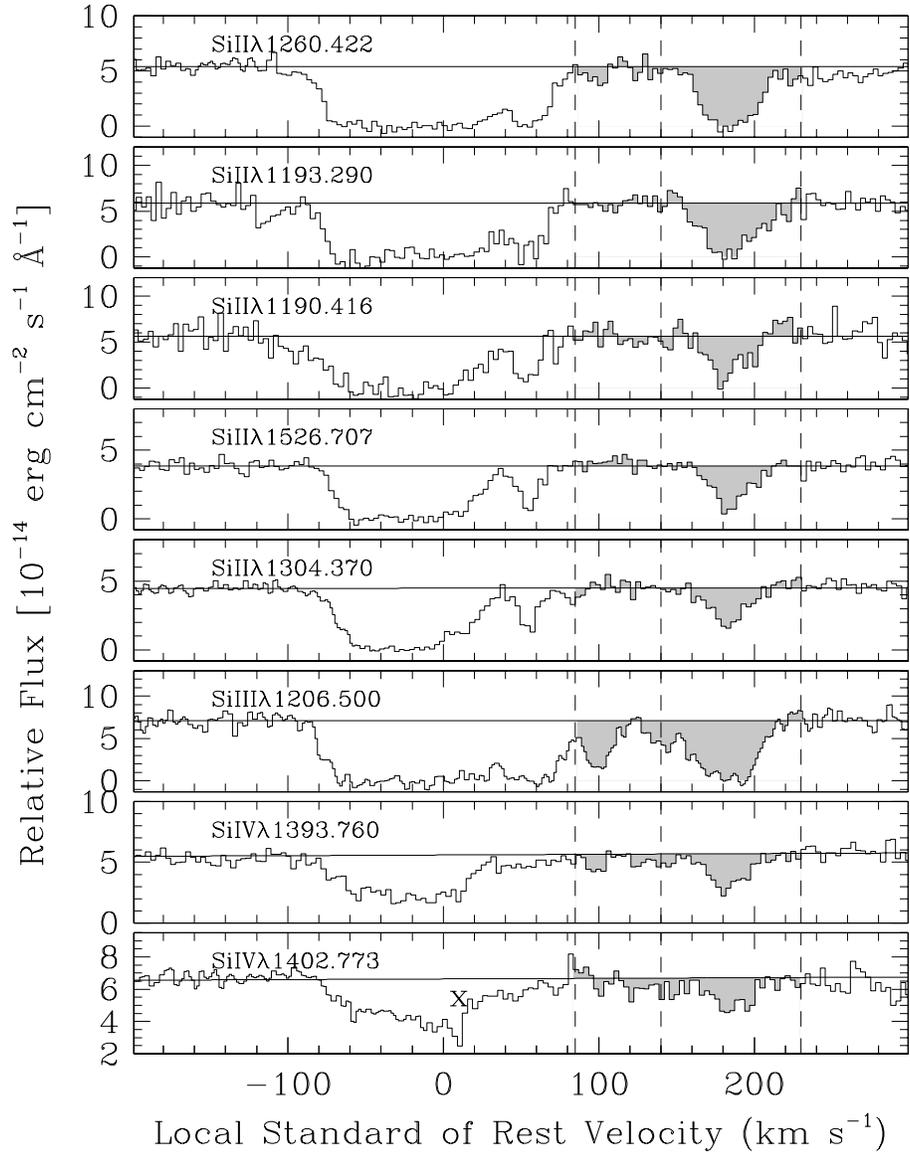}
\vglue -0.1in
\protect\caption[Flux Profiles]{Same as Figure~2a, but for the
ions of silicon.}
\end{figure*}

\clearpage

\begin{figure*}[ht!]
\figurenum{2e}
\vglue -0.3in
\epsscale{0.73}
\plotone{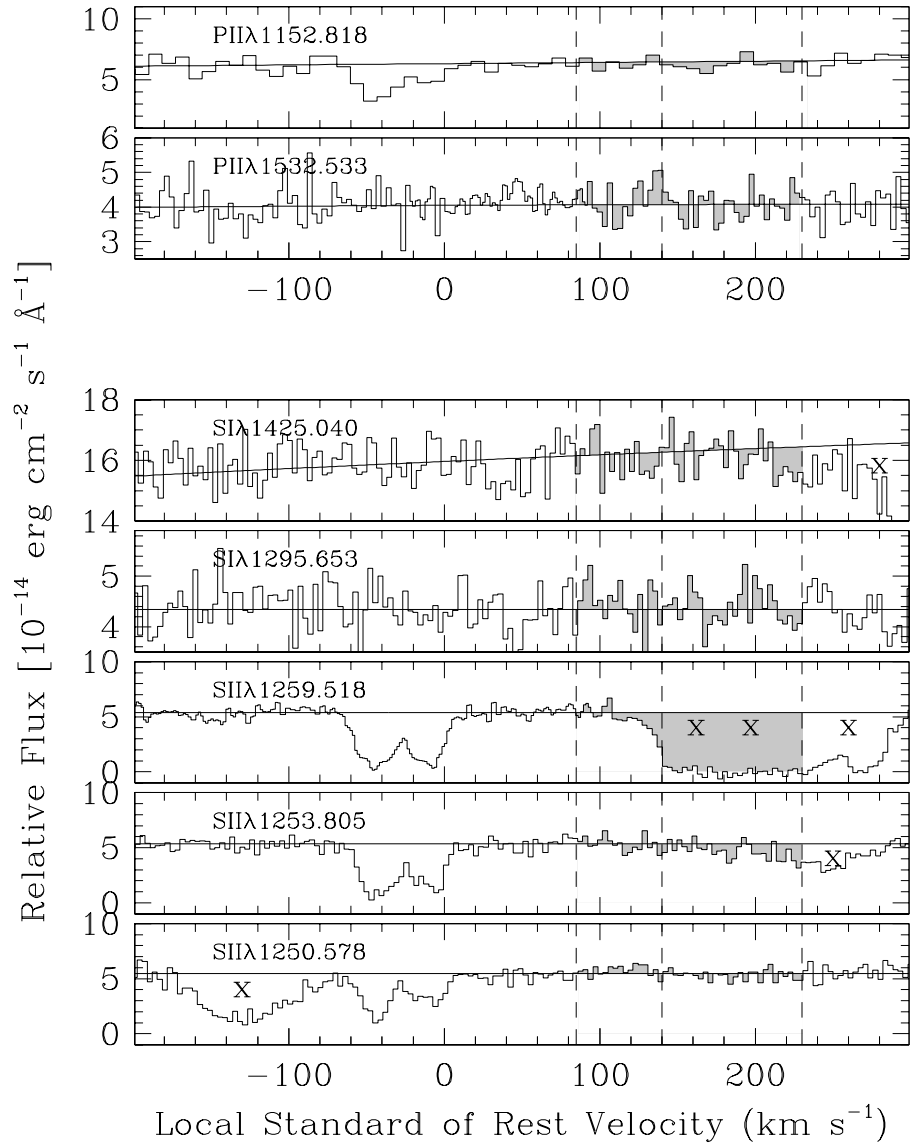}
\vglue -0.1in
\protect\caption[Flux Profiles]{Same as Figure~2a, but for the
ions of phosphorus and sulfur.}
\end{figure*}

\clearpage

\begin{figure*}[ht!]
\figurenum{2f}
\vglue -0.3in
\epsscale{0.73}
\plotone{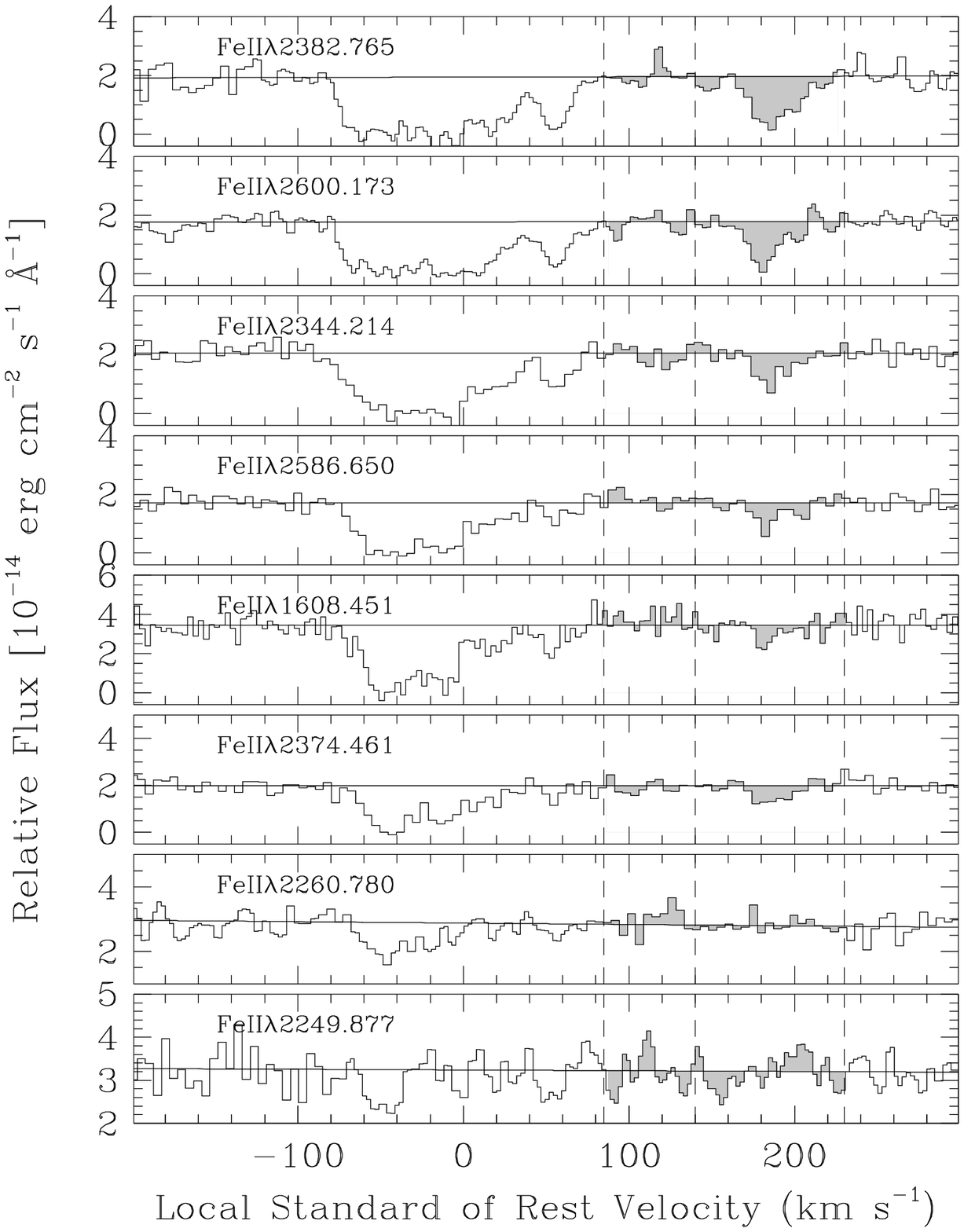}
\vglue -0.1in
\protect\caption[Flux Profiles]{Same as Figure~2a, but for the
ions of iron.}
\end{figure*}

\clearpage

\begin{figure*}[ht!]
\figurenum{3a}
\begin{center}
\epsscale{0.9}
\vglue -0.3in
\rotatebox{0}{\plotone{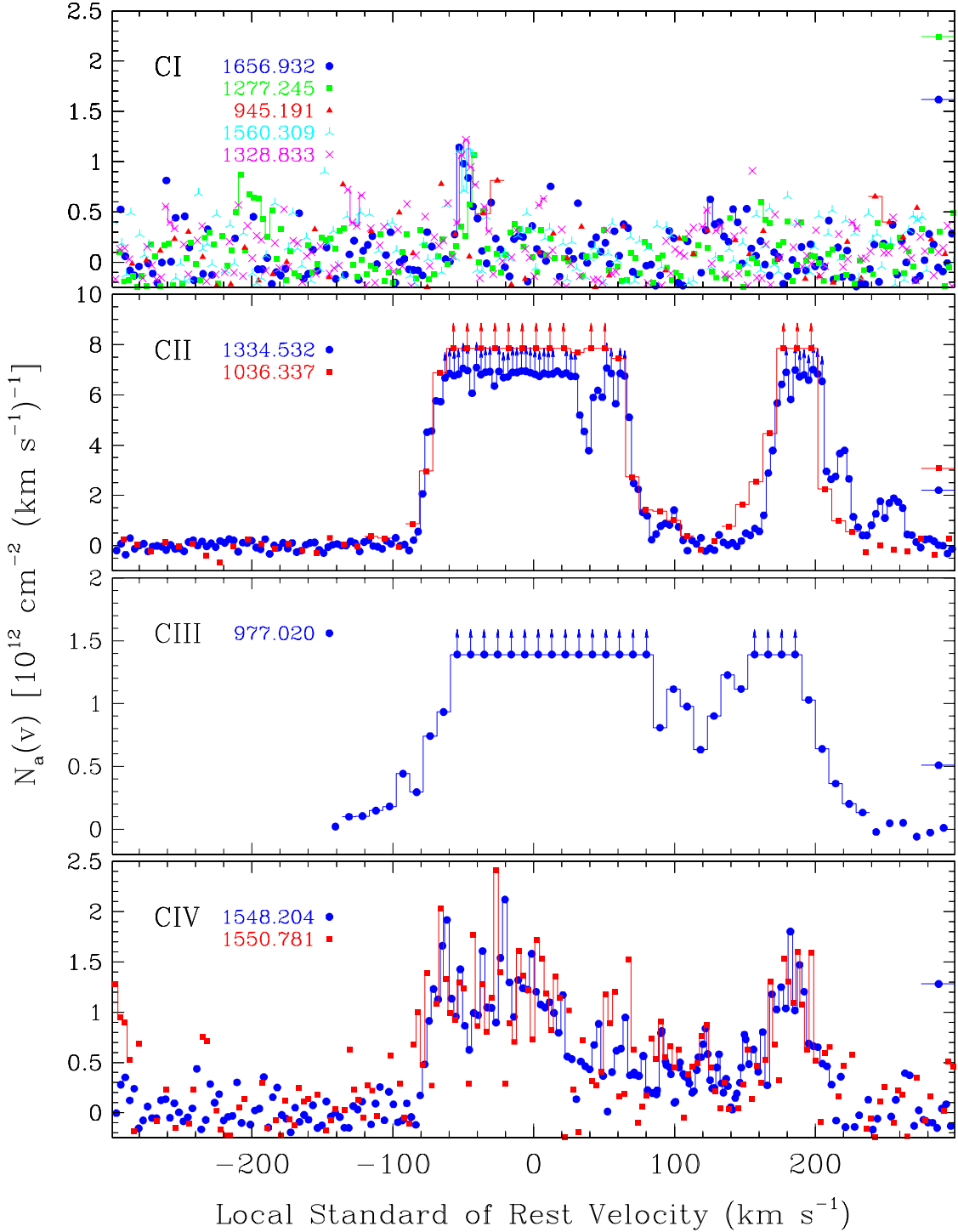}}
\end{center}
\vglue -0.3in \protect\caption[Apparent Column Density
Profiles]{In the above panels, we show apparent column density
profiles versus Local Standard of Rest velocity for a sample of
transitions detected in Galactic absorption. Each panel depicts
the apparent column density profiles for a given species, with the
profile transitions color-coded and over-plotted. A histogram line
connects bins where absorption is detected, and arrows indicate
bins where the absorbed flux is consistent with zero. A horizontal
bar is shown on the right side of the panels at the value of
{$\acd(v)$} for which the transition reaches an apparent optical
depth of unity. For clarity, pixels which are blended with other
transitions are not shown (e.g., each member of the {\ion{N}{1}}
triplet was cleaned to avoid blends with the other two members). }
\label{fig:acdplot1a}
\end{figure*}

\clearpage

\begin{figure*}[ht!]
\figurenum{3b}
\begin{center}
\epsscale{0.9}
\rotatebox{0}{\plotone{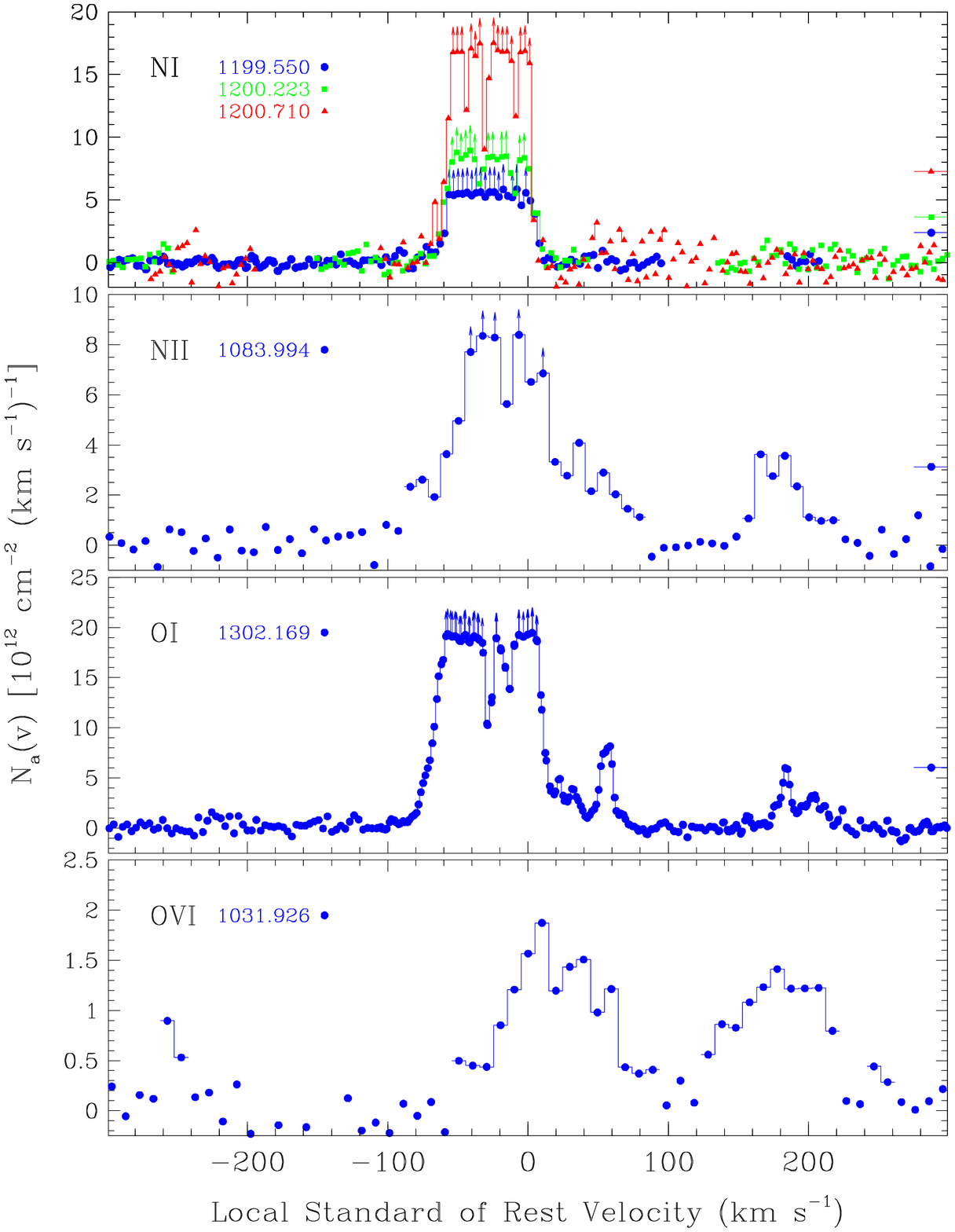}}
\end{center}
\protect\caption[Apparent Column Density Profiles]{Same as
Figure~\ref{fig:acdplot1a}}
\label{fig:acdplot1b}
\end{figure*}

\clearpage

\begin{figure*}[ht!]
\figurenum{3c}
\begin{center}
\epsscale{0.9}
\rotatebox{0}{\plotone{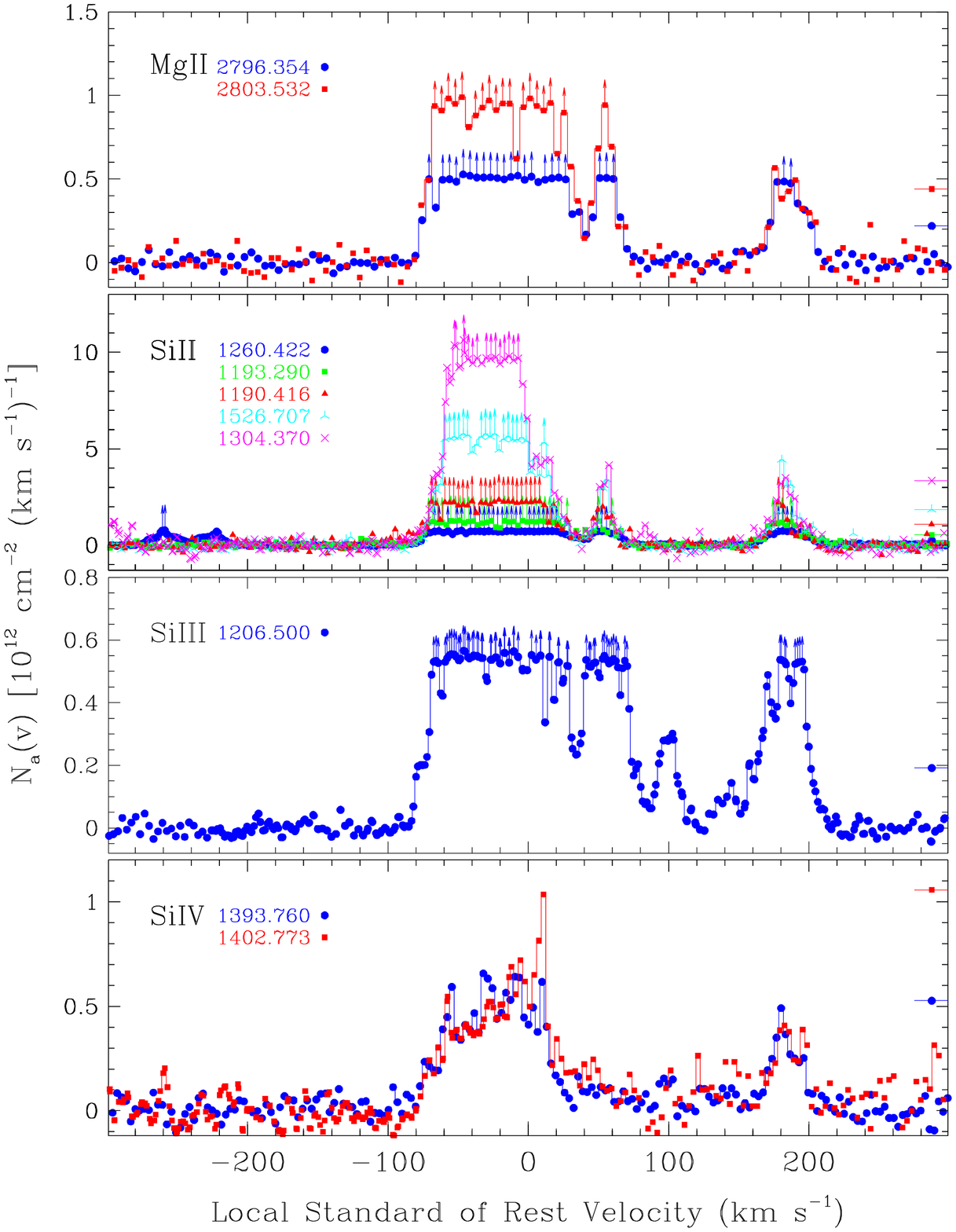}}
\end{center}
\protect\caption[Apparent Column Density Profiles]{Same as
Figure~\ref{fig:acdplot1a}}
\label{fig:acdplot1c}
\end{figure*}

\clearpage

\begin{figure*}[ht!]
\figurenum{3d}
\begin{center}
\epsscale{0.9}
\rotatebox{0}{\plotone{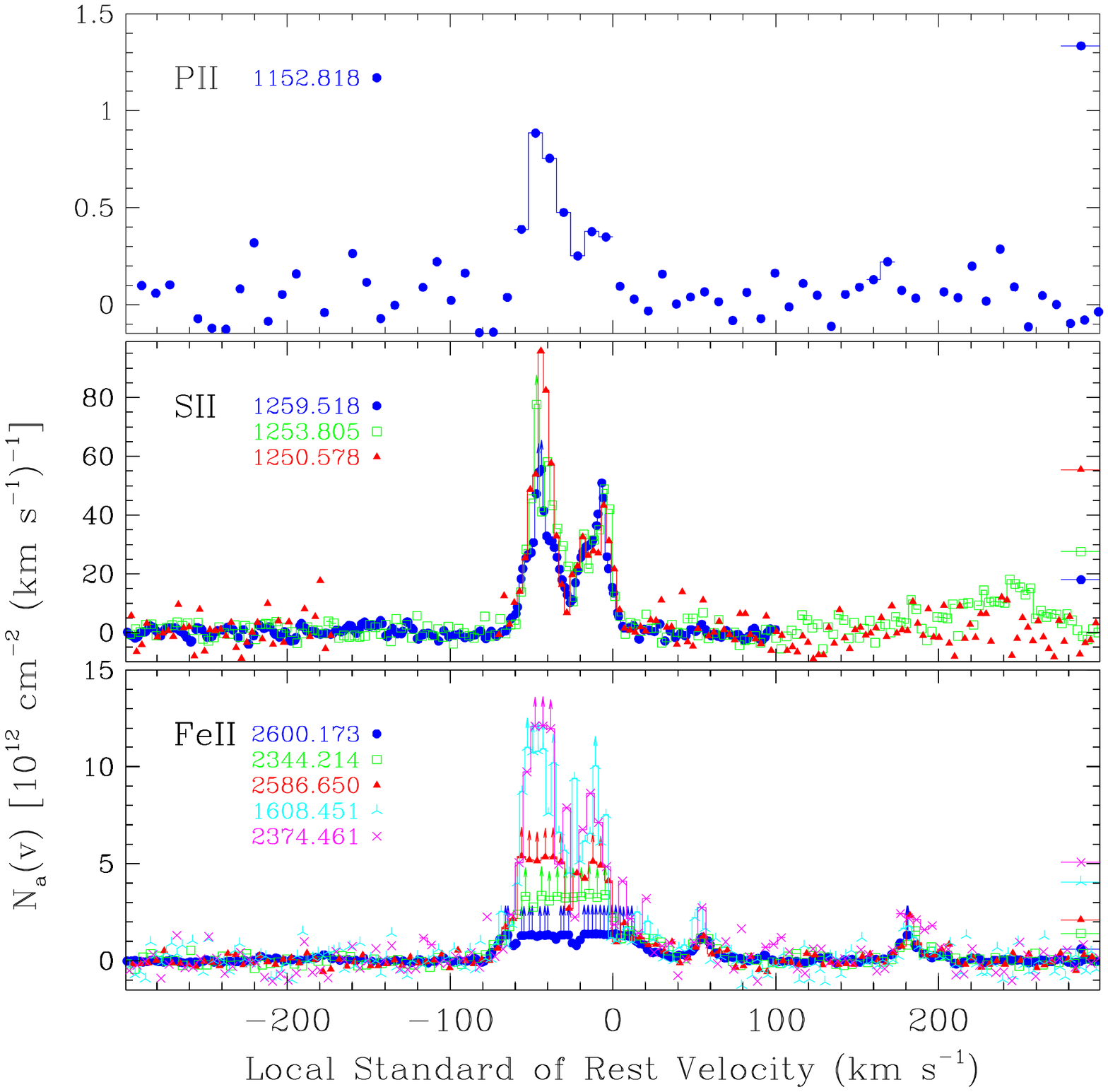}}
\end{center}
\protect\caption[Apparent Column Density Profiles]{Same as
Figure~\ref{fig:acdplot1a}}
\label{fig:acdplot1d}
\end{figure*}

\clearpage

\begin{figure*}[ht!]
\figurenum{4a}
\begin{center}
\epsscale{0.9}
\vglue -0.5in
\rotatebox{0}{\plotone{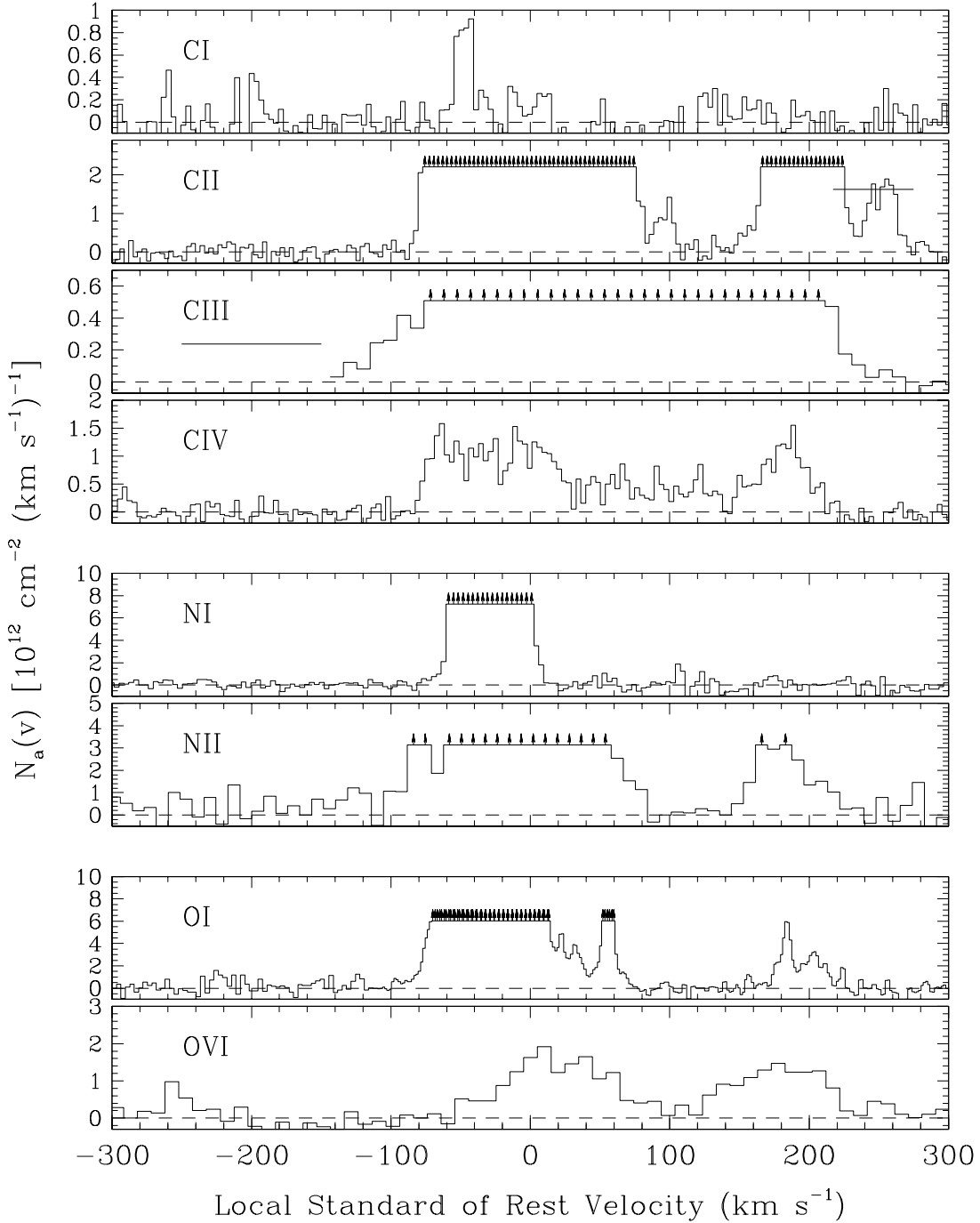}}
\end{center}
\vglue -0.5in
\protect\caption[Composite ACDs]{In the above panels, we show the
composite apparent column density profiles of all the species
detected in Galactic absorption, ordered by atomic number and
ionization stage. Velocity bins where the apparent optical depth
of even the weakest available transition is larger than unity are
indicated with arrows. The horizontal bar in the {\ion{C}{2}}
panel at {250\,\kms} indicates the velocity range of a spurious
feature caused by absorption by {\ion{C}{2}*} at $v<0$\kms.
Similarly, the horizontal bar in the {\ion{C}{3}} panel at
{-250\,\kms} indicates the location of the
{\ion{O}{1}$\lambda976.448$} absorption which has been removed for
clarity. For velocity bins where the apparent optical depth of the
weakest available transition is greater than unity, we quote a
lower limit on the apparent column density. See the text for a
listing of transitions used in computing the profiles.}
\label{fig:compacd1a}
\end{figure*}

\clearpage

\begin{figure*}[ht!]
\figurenum{4b}
\begin{center}
\epsscale{0.9}
\rotatebox{0}{\plotone{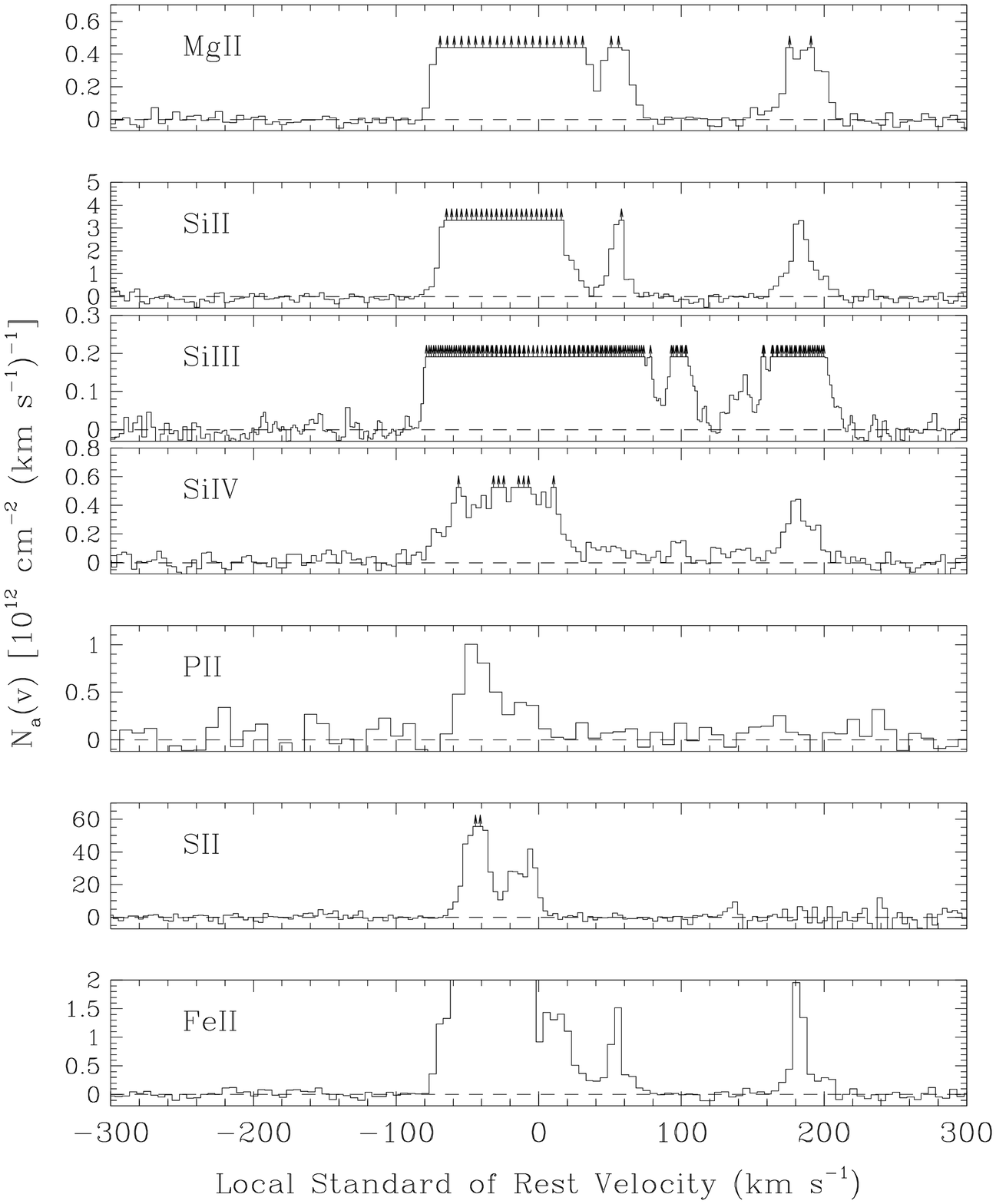}}
\end{center}
\protect\caption[Composite ACDs]{Same as Figure~\ref{fig:compacd1a}}
\label{fig:compacd1b}
\end{figure*}

\clearpage
\begin{figure*}[ht!]
\figurenum{5}
\begin{center}
\epsscale{0.75}
\rotatebox{270}{\plotone{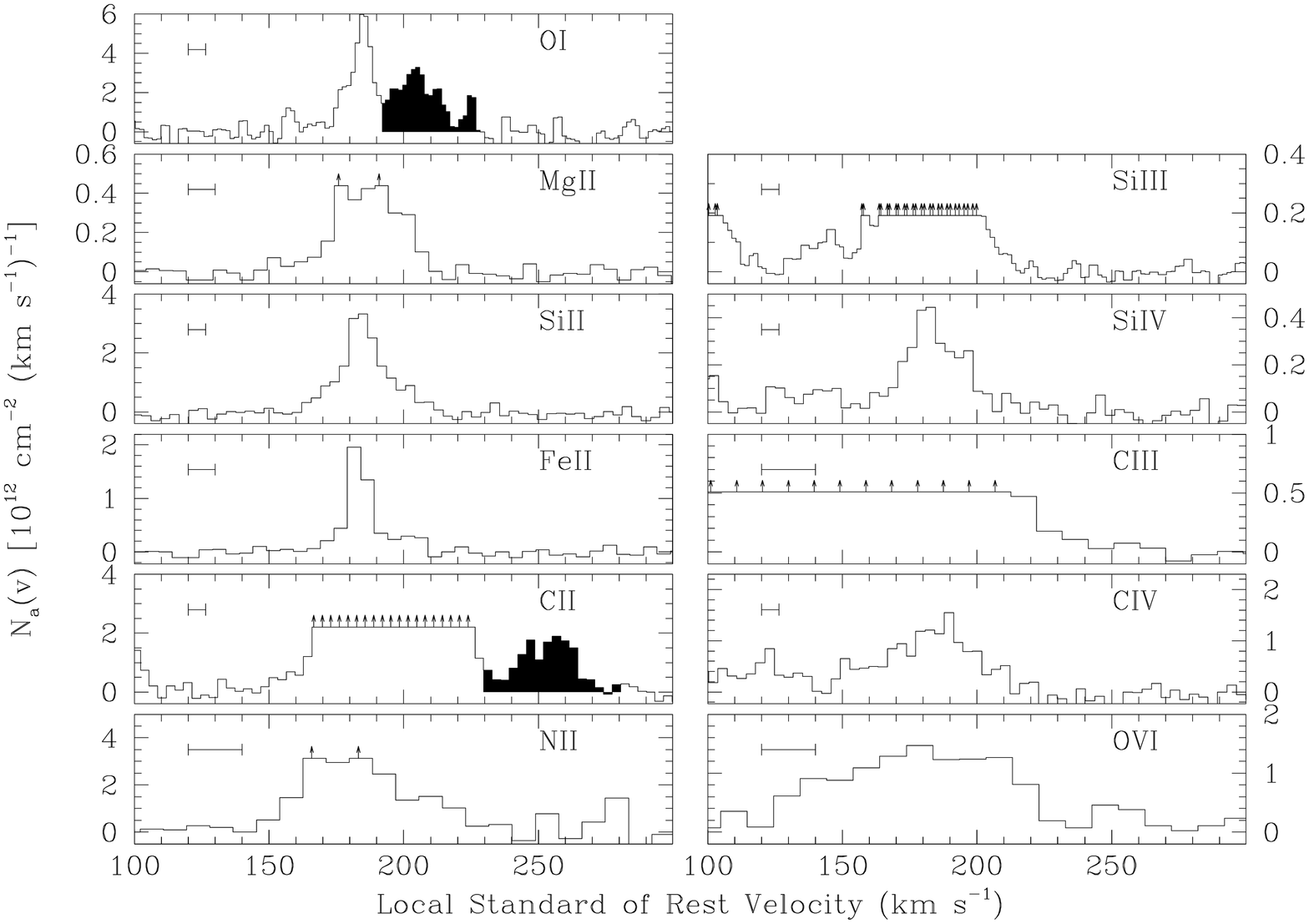}}
\end{center}
\protect\caption[High Velocity ACDs]{In the above panels, we show
the composite apparent column density profiles of the species
detected in the high velocity cloud at {$\vlsr\approx+184$\,\kms}
ordered by ionization potential. Velocity bins where even the
weakest available transition of a species are saturated are
indicated with arrows.  The horizontal bar in the top left corner
of each panel gives the full-width at half-maximum intensity of
the instrumental spread function. The shaded region in the
{\ion{O}{1}} panel marks the velocity where the absorption is
blended with a weak Lyman $\alpha$\ line at {$z=0.0719$}.
Similarly, the shaded region in the {\ion{C}{2}} profile marks the
velocity where the absorption is blended with {\ion{C}{2}*}
absorption in the {$\vlsr\approx-44$\,\kms} component.}
\label{fig:compacdhvc}
\end{figure*}

\clearpage

\begin{figure*}[ht!]
\figurenum{6}
\epsscale{0.80}
\begin{center}
\vglue -0.3in
\rotatebox{0}{\plotone{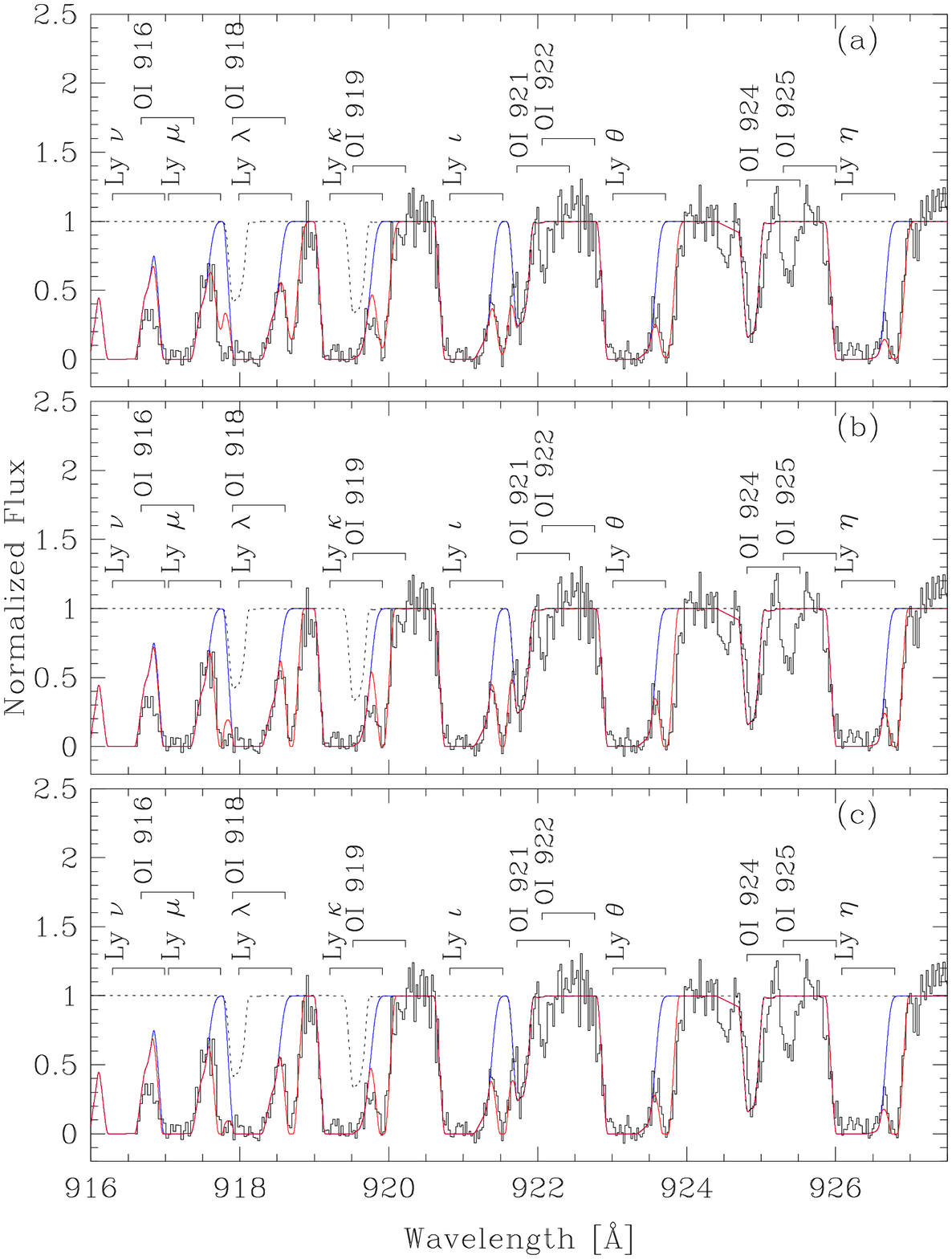}}
\end{center}
\vglue -0.3in
\protect\caption[H\,{\sc i} Fits]{Normalized flux
versus wavelength for the high order {\ion{H}{1}} Lyman series and
{\ion{O}{1}} transitions detected in the FUSE SiC2A channel in the
wavelength range 916--927.5\,\AA. The flux is plotted as a histogram
with {10\,\kms} bins. The dashed (blue) curve in each panel shows the
model of the intermediate velocity and Galactic absorption as
described in the text. The dotted curve shows the contribution to this
curve from {\ion{O}{1}} absorption as implied by the fit to the
{1302.168\,\AA} transition. The thick solid (red) curve in each panel
shows the same model with the inclusion of a high velocity component
at {$\vlsr\approx184$\,\kms} under three scenarios: (a) assuming the
{\ion{H}{1}} column density derived from direct integration of the
Lyman {$\lambda$} line, $\log N$(\ion{H}{1})$=16.73$; (b) assuming the
model derived from fitting the curve of growth, $\log
N$(\ion{H}{1})$=17.82$; and (c) assuming an {\ion{H}{1}} column
density at the detection limit of the {\citet{wakker03}} 21\,cm
emission spectrum, $\log N$(\ion{H}{1})$=18.3$. The parameters of
these models are listed in Table~\ref{tab:hifit}.}
\label{fig:lyman}
\end{figure*}

\clearpage
\begin{figure*}[ht!]
\figurenum{7}
\epsscale{0.75}
\begin{center}
\rotatebox{-90}{\plotone{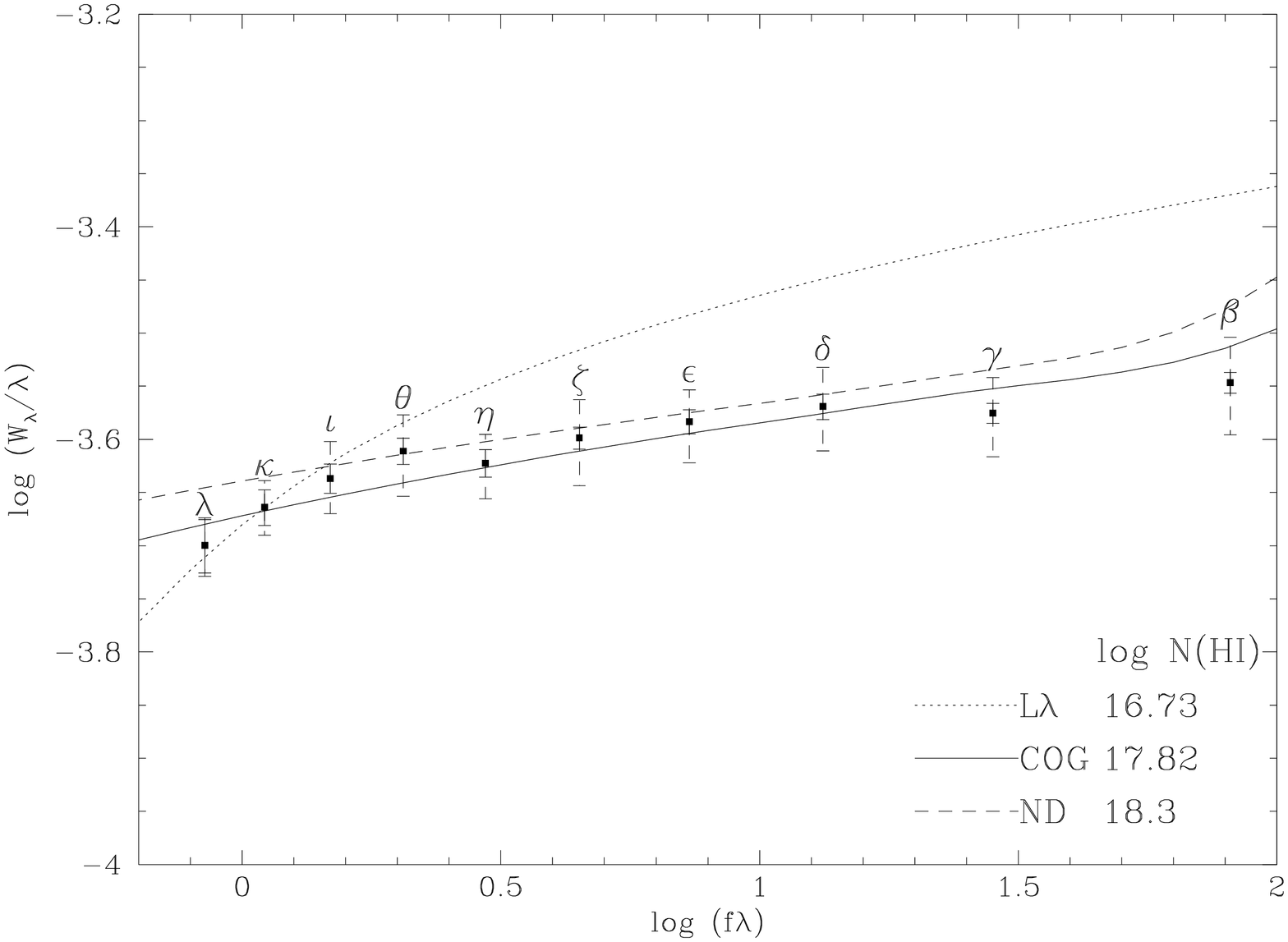}}
\end{center}
\vglue -0.5in
\protect\caption[H\,{\sc i} Curve of Growth]{This
plot shows the Lyman series equivalent width measurements of the high
velocity cloud ({$W_\lambda/\lambda$} vs. {$f\lambda$}) from
Table~\ref{tab:hvc184measure} for transitions detected in the FUSE
SiC2A and LiF1A detector segments. The solid error bars indicate the
{1$\sigma$} confidence uncertainties resulting from statistical and
continuum placement errors. The dashed error bars indicate the
additional systematic uncertainty from measuring the equivalent widths
across the velocity range 140--230\,\kms. Three curves of growth are
overplotted corresponding to the best-fit curve (COG) and the two
extreme cases considered in the profile fitting (L$\lambda$: lower
limit from direct integration of Lyman $\lambda$\ with a $b$-value of
26\,\kms; ND: upper limit from non-detection of 21\,cm emission and a
$b$-value of 14.2\,\kms).}
\label{fig:cog}
\end{figure*}

\clearpage
\begin{figure*}[ht!]
\figurenum{8}
\begin{center}
\epsscale{1.01}
\rotatebox{0}{\plotone{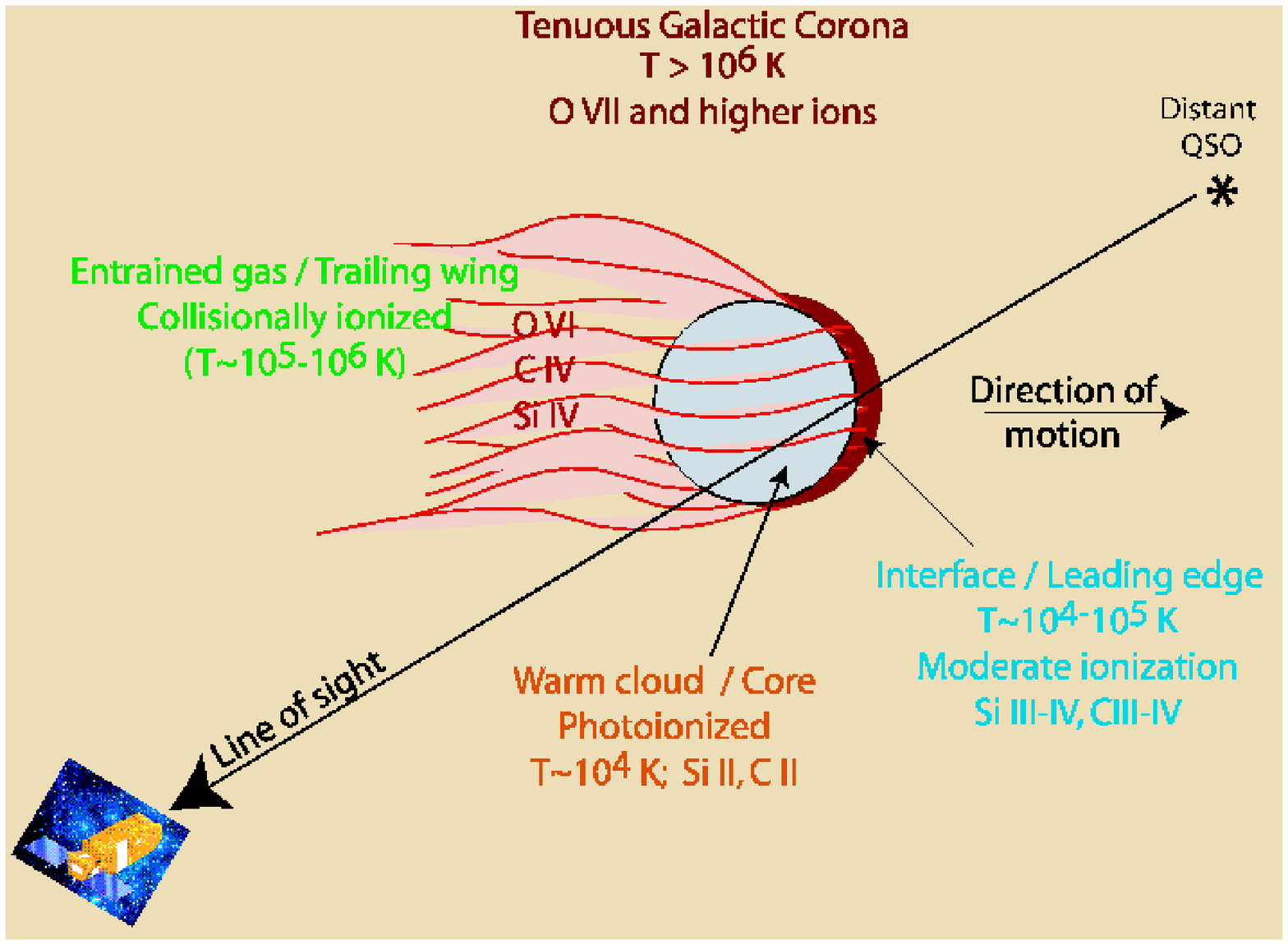}}
\end{center}
\vglue -0.3in
\protect\caption[Cartoon]{Schematic illustration of
the model proposed to explain the high-velocity absorption observed at
$\vlsr\approx+184$\,\kms. A cloud of gas with temperature
$\approx10^4$\,K is moving through a hotter ($>10^6$\,K), low-density
medium such as the Galactic corona. The interaction between the
external medium and the cloud strips material off the cloud. This
stripped material becomes entrained in the external medium, adopting
its temperature and velocity over time. We draw a possible sight line
and label locations for the possible production of low-, moderate-,
and high-ionization species corresponding to the velocity regions
discussed in {\S\ref{sec:ionize}}. See {\citet{qm01}} and
{\citet{ml04}} for the results of detailed hydrodynamic simulations of
this scenario.}
\label{fig:cartoon}
\end{figure*}

\clearpage
\begin{figure*}[ht!]
\figurenum{9}
\epsscale{0.8}
\begin{center}
\rotatebox{0}{\plotone{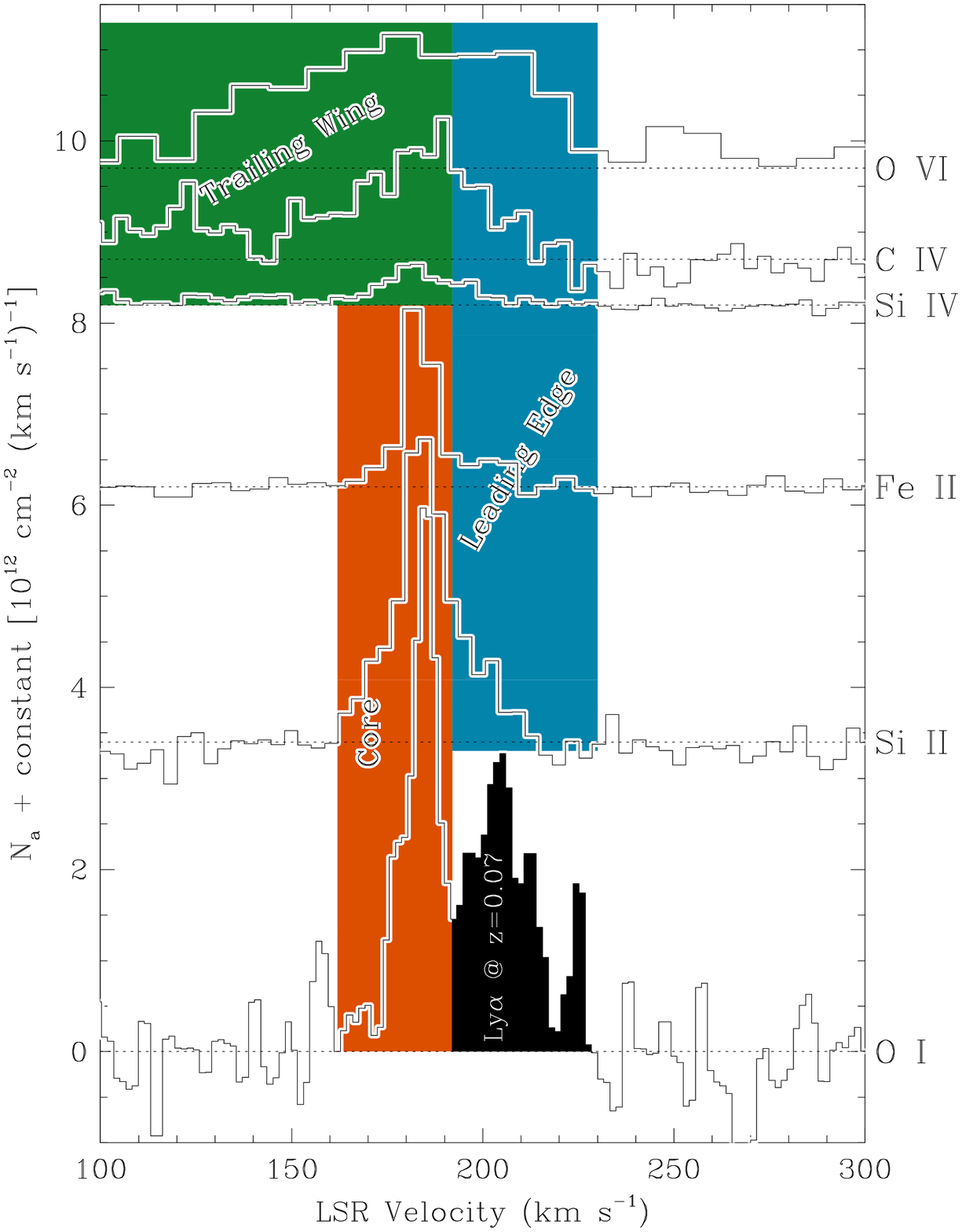}}
\end{center}
\vglue -0.5in \protect\caption[ACD Comparison 1]{In the above
panels, we overlay the apparent column density profiles
{\ion{O}{1}}, {\ion{Si}{2}}, {\ion{Fe}{2}}, {\ion{Si}{4}},
{\ion{C}{4}}, and {\ion{O}{6}}. The profiles are offset for clarity,
and are shown with a sampling of the two bins per resolution
element. We shade and label the three kinematic regions discussed in
the text - ``Core,'' ``Leading Edge,'' and ``Trailing Wing.'' The
velocities in {\ion{O}{1}} which are blended with a weak Lyman
{$\alpha$} cloud at {$z=0.0719$} is shaded in black.  The kinematic
distribution of the profiles as a function of ionization is readily
apparent.} \label{fig:acdcomp1}
\end{figure*}

\clearpage
\begin{figure*}[ht!]
\figurenum{10} \epsscale{0.8}
\begin{center}
\rotatebox{0}{\plotone{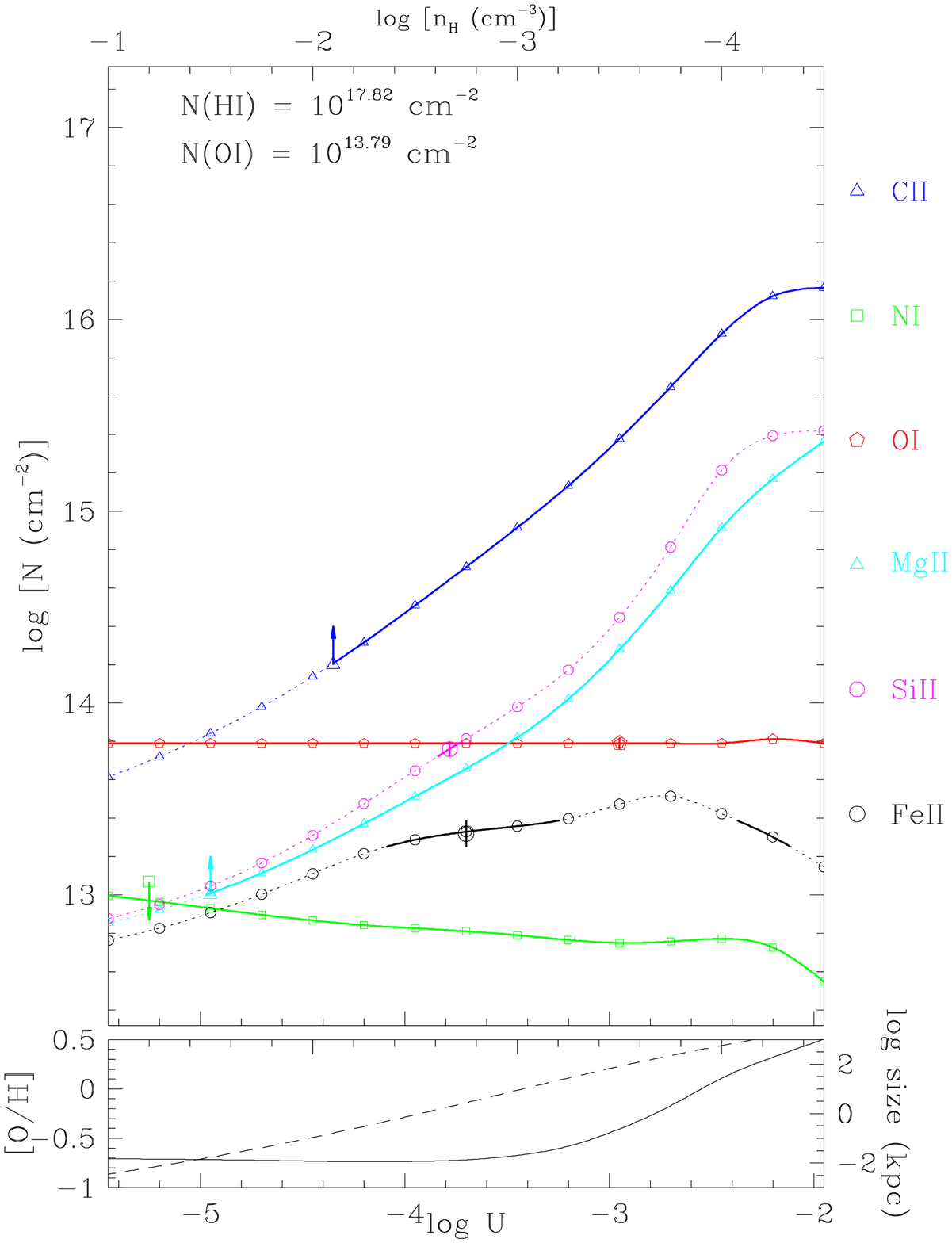}}
\end{center}
\vglue -0.5in
\protect\caption[Photoionization Model]{Bottom
panel: metallicity (solid line; left axis) and thickness (dashed line;
right axis) of a photoionized, plane-parallel slab of gas that
reproduces the observed {\ion{H}{1}} and {\ion{O}{1}} column densities
of the core region of the {+184\,\kms} HVC. Top panel: predicted
column densities of various species assuming a solar relative
abundance pattern. The solid portions of the curves correspond to the
allowed {$1\sigma$} range in column density. The larger symbols with
error bars correspond to the observed column densities (or limits)
over the integration range {$162\leq\vlsr\leq192$\,\kms}.}
\label{fig:pie}
\end{figure*}

\clearpage
\begin{figure*}[ht!]
\figurenum{11}
\epsscale{0.8}
\begin{center}
\rotatebox{0}{\plotone{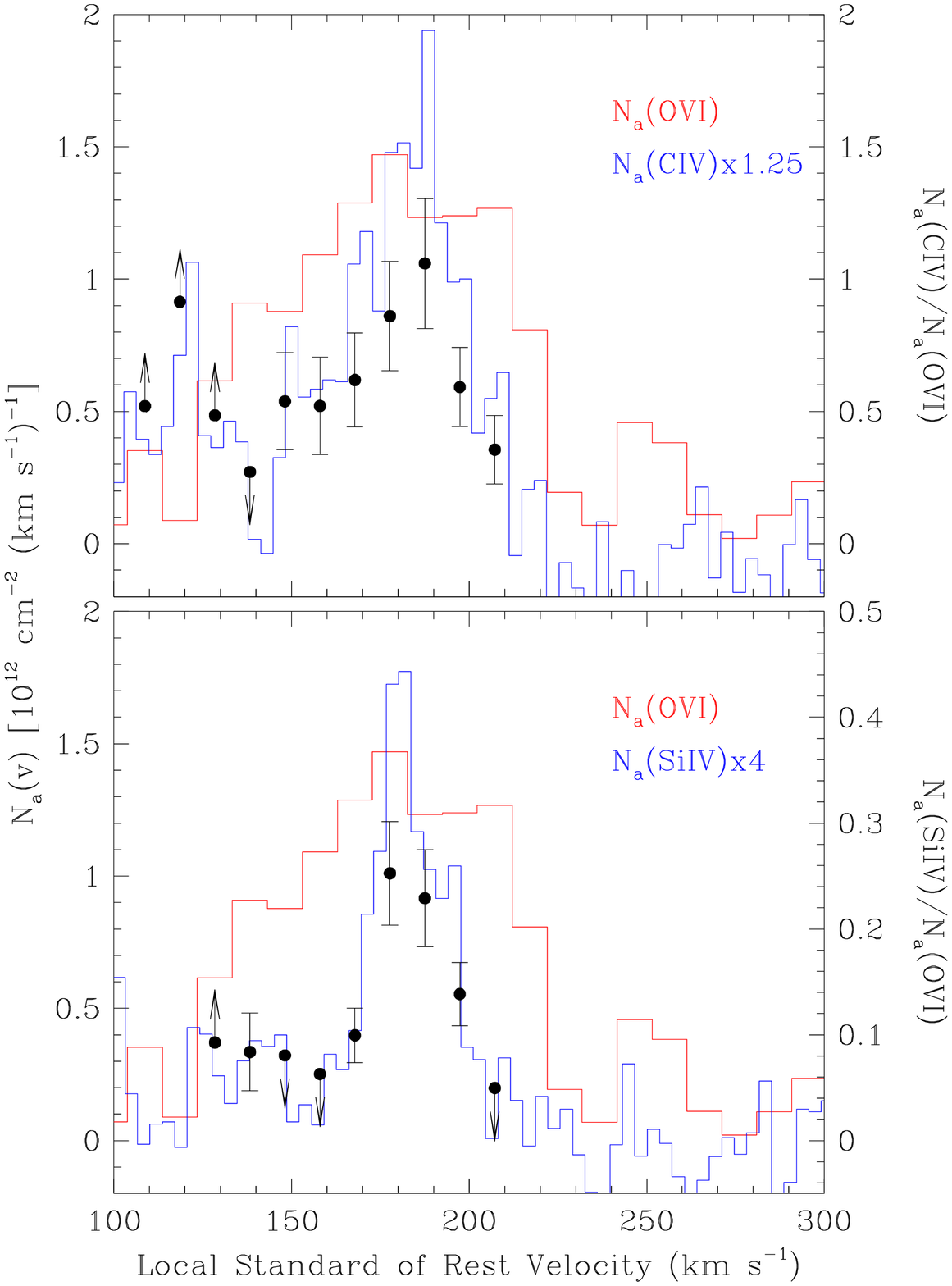}}
\end{center}
\vglue -0.5in
\protect\caption[ACD Comparison 2]{In the above panels, we overlay
the apparent column density profiles of {\ion{O}{6}} (shaded or red
histogram) over the apparent column density profiles of
{\ion{C}{4}$\lambda1548$} (unshaded or blue histogram, top panel) and
{\ion{Si}{4}$\lambda1393$} (unshaded or blue histogram, bottom
panel). In each panel, we also plot the apparent column density ratio
of the more moderate ionization species relative to {\ion{O}{6}}
(black points, with scales on the right axis).}
\label{fig:acdcomp2}
\end{figure*}

\clearpage
\begin{figure*}[ht!]
\figurenum{12}
\epsscale{0.8}
\begin{center}
\rotatebox{0}{\plotone{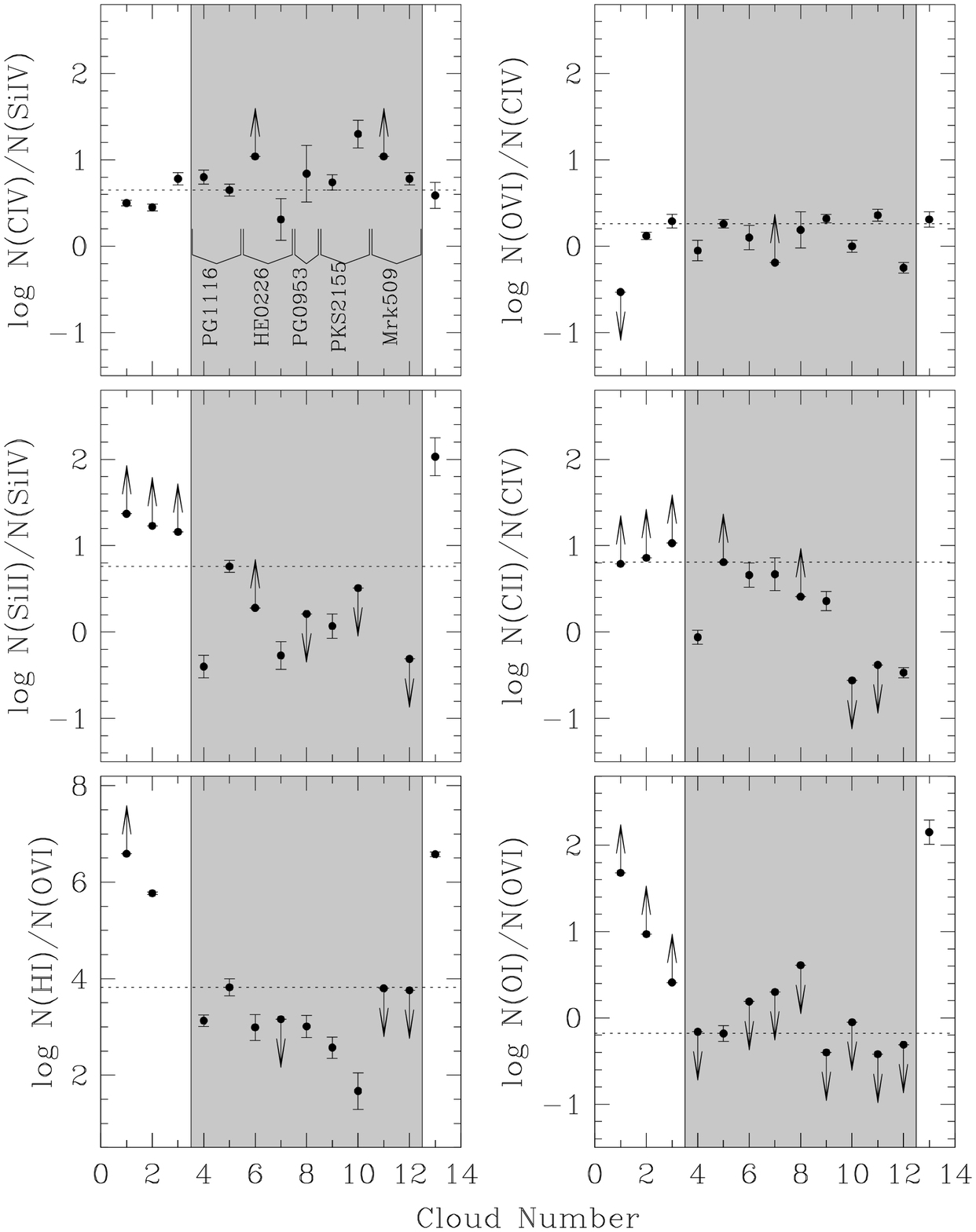}}
\end{center}
\vglue -0.5in \protect\caption[Column Density Ratios]{This figure
shows various column density ratios for the clouds in the
{PG\,$1116+215$} sight line (clouds 1--5), as well as the high
velocity clouds toward {HE\,$0226-4110$} {\citep[clouds
6--7,][]{fox05}}, {PG\,$0953+414$} {\citep[cloud 8,][]{fox05}},
{PKS\,$2155-304$} {\citep[clouds 9--10,][]{csg04}}, {Mrk\,509}
{\citep[clouds 11--12,][]{csg04}}, and Complex C {\citep[cloud
13,][]{csg03,fox04,sembach04b}}. See Table~\ref{tab:cloudkey} for
further information on the clouds. The shaded region highlights the
column density ratios for the highly-ionized HVCs studied in detail
thus far. The dashed horizontal line marks the ratio for the
{$\vlsr=+184$\,\kms} HVC in the {PG\,$1116+215$} sight line (cloud
5).} \label{fig:coldensrat}
\end{figure*}

\clearpage


\begin{deluxetable}{lrcccccc}
\tabletypesize{\scriptsize}
\tablewidth{0pc}
\tablecaption{$v_{\rm{LSR}}\approx+100$\,\kms\ HVC Measurements\tablenotemark{a}}
\tablehead
{
\colhead{Ion} & \colhead{Transition} & \colhead{$\log f \lambda$} & \colhead{$\langle{v_{\rm{LSR}}}\rangle$\tablenotemark{b}} & \colhead{$b$\tablenotemark{b}} & \colhead{$\log \acd$\tablenotemark{c}} & \colhead{$W_{\lambda}$\tablenotemark{d}} & \colhead{Integration Range} \\
              & \colhead{(\AA)}      &                            & \colhead{(\kms)}                                          & \colhead{(\kms)}               & \colhead{(cm$^{-2}$)}                  & \colhead{(m\AA)}                         & \colhead{(\kms)}
}
\startdata
\ion{ H}{1}\tablenotemark{e}
            & $1025.722$ & $ 1.909$ & $115.0\pm 1.6\pm 5.0$ & $21.5\pm 1.0$ & $14.94\pm0.04\pm0.08$ & $178\pm 5\pm32$ &  85 to 140 \\
            & $ 972.537$ & $ 1.450$ & $114.0\pm 2.3\pm 5.0$ & $23.1\pm 1.5$ & $15.31\pm0.06\pm0.09$ & $166\pm 6\pm31$ &  85 to 140 \\
            & $ 949.743$ & $ 1.122$ & $111.9\pm 1.5\pm 5.0$ & $22.1\pm 1.1$ & $15.69\pm0.05\pm0.08$ & $164\pm 4\pm30$ &  85 to 140 \\
            & $ 937.803$ & $ 0.864$ & $110.9\pm 1.7\pm 5.0$ & $24.1\pm 0.8$ & $16.00\pm0.04\pm0.09$ & $164\pm 3\pm30$ &  85 to 140 \\
            & $ 930.748$ & $ 0.652$ & $108.6\pm 1.6\pm 5.0$ & $23.0\pm 1.0$ & $16.15\pm0.05\pm0.09$ & $161\pm 3\pm30$ &  85 to 140 \\
            & $ 926.226$ & $ 0.470$ & $106.8\pm 1.5\pm 5.0$ & $21.9\pm 0.6$ & $16.21\pm0.04\pm0.08$ & $148\pm 3\pm26$ &  85 to 140 \\
            & $ 923.150$ & $ 0.311$ & $104.7\pm 1.3\pm 5.0$ & $21.3\pm 1.0$ & $16.38\pm0.05\pm0.09$ & $145\pm 4\pm26$ &  85 to 140 \\
            & $ 920.963$ & $ 0.170$ & $100.7\pm 1.0\pm 5.0$ & $20.8\pm 0.9$ & $16.53\pm0.05\pm0.11$ & $134\pm 4\pm25$ &  85 to 140 \\
            & $ 919.351$ & $ 0.043$ & $108.4\pm 1.4\pm 5.0$ & $22.5\pm 0.8$ & $16.51\pm0.04\pm0.08$ & $135\pm 4\pm25$ &  85 to 140 \\
            & $ 918.129$ & $-0.072$ & $104.0\pm 1.1\pm 5.0$ & $22.4\pm 0.9$ & $16.43\pm0.04\pm0.10$ & $105\pm 7\pm21$ &  85 to 140 \\ \hline
\ion{ C}{1}\tablenotemark{f}
            & $1656.928$ & $ 2.367$ & \nodata               & \nodata       & $<12.99$              & $< 36$          & \nodata  \\
            & $1277.245$ & $ 2.225$ & \nodata               & \nodata       & $<13.12$              & $< 17$          & \nodata  \\
            & $ 945.191$ & $ 2.157$ & \nodata               & \nodata       & $<13.38$              & $< 27$          & \nodata  \\
            & $1560.309$ & $ 2.082$ & \nodata               & \nodata       & $<13.17$              & $< 28$          & \nodata  \\
            & $1328.833$ & $ 2.077$ & \nodata               & \nodata       & $<13.14$              & $< 18$          & \nodata  \\ \hline
\ion{ C}{2} & $1334.532$ & $ 2.233$ & $ 98.9\pm 2.8\pm 1.5$ & $15.3\pm 5.0$ & $13.28\pm0.07\pm0.08$ & $ 31\pm 5\pm 6$ &  85 to 140 \\
\ion{C}{2}* & $1335.708$ & $ 2.186$ & \nodata               & \nodata       & $<12.97$              & $< 17$          & \nodata  \\ \hline
\ion{ C}{3} & $ 977.020$ & $ 2.869$ & $112.6\pm 1.3\pm 5.0$ & $23.1\pm 0.8$ & $13.71\pm0.04\pm0.09$ & $143\pm 5\pm28$ &  85 to 140\\ \hline
\ion{ C}{4} & $1548.204$ & $ 2.468$ & $111.3\pm 1.1\pm 1.5$ & $21.7\pm 0.7$ & $13.34\pm0.03\pm0.04$ & $ 73\pm 5\pm 7$ &  85 to 140 \\
            & $1550.781$ & $ 2.167$ & $107.3\pm 3.2\pm 1.5$ & $21.2\pm 2.3$ & $13.41\pm0.09\pm0.04$ & $ 46\pm 9\pm 5$ &  85 to 140 \\ \hline
\ion{ N}{1} & $1200.710$ & $ 1.715$ & \nodata               & \nodata       & $<13.66$              & $< 26$          & \nodata  \\ \hline
\ion{ N}{2} & $1083.994$ & $ 2.080$ & \nodata               & \nodata       & $<13.48$              & $< 34$          & \nodata  \\ \hline
\ion{ N}{5} & $1238.821$ & $ 2.286$ & \nodata               & \nodata       & $<12.98$              & $< 21$          & \nodata  \\
            & $1242.804$ & $ 1.985$ & \nodata               & \nodata       & $<13.21$              & $< 17$          & \nodata  \\ \hline
\ion{ O}{1} & $1302.168$ & $ 1.796$ & \nodata               & \nodata       & $<13.13$              & $< 10$          & \nodata  \\
            & $1304.858$ & $ 1.795$ & \nodata               & \nodata       & $<13.37$              & $< 17$          & \nodata  \\ \hline
\ion{ O}{6} & $1031.926$ & $ 2.136$ & \nodata               & \nodata       & $13.29\pm0.12\pm0.15$ & $ 20\pm 6\pm 7$ &  85 to 140 \\ \hline
\ion{Mg}{1} & $2026.477$ & $ 2.360$ & \nodata               & \nodata       & $<13.33$              & $< 74$          & \nodata  \\
            & $1707.061$ & $ 0.873$ & \nodata               & \nodata       & $<14.60$              & $< 45$          & \nodata  \\ \hline
\ion{Mg}{2} & $2796.354$ & $ 3.236$ & \nodata               & \nodata       & $<12.18$              & $< 66$          & \nodata  \\
            & $2803.532$ & $ 2.933$ & \nodata               & \nodata       & $<12.55$              & $< 78$          & \nodata  \\ \hline
\ion{Si}{2} & $1260.422$ & $ 3.172$ & \nodata               & \nodata       & $12.14\pm0.11\pm0.06$ & $ 19\pm 5\pm 3$ &  85 to 140 \\
            & $1193.290$ & $ 2.842$ & \nodata               & \nodata       & $<12.58$              & $< 28$          & \nodata  \\
            & $1190.416$ & $ 2.541$ & \nodata               & \nodata       & $<12.92$              & $< 30$          & \nodata  \\
            & $1526.707$ & $ 2.308$ & \nodata               & \nodata       & $<12.86$              & $< 22$          & \nodata  \\
            & $1304.370$ & $ 2.051$ & \nodata               & \nodata       & $<13.11$              & $< 17$          & \nodata  \\ \hline
\ion{Si}{3} & $1206.500$ & $ 3.294$ & $103.8\pm 0.7\pm 1.5$ & $18.6\pm 0.8$ & $12.78\pm0.02\pm0.07$ & $ 81\pm 4\pm15$ &  85 to 140 \\ \hline
\ion{Si}{4} & $1393.760$ & $ 2.854$ & $110.5\pm 2.5\pm 1.5$ & $23.3\pm 1.5$ & $12.54\pm0.07\pm0.08$ & $ 28\pm 4\pm 5$ &  85 to 140 \\
            & $1402.773$ & $ 2.552$ & \nodata               & \nodata       & $<12.55$              & $< 15$          & \nodata  \\ \hline
\ion{ P}{2} & $1152.818$ & $ 2.451$ & \nodata               & \nodata       & $<12.85$              & $< 20$          & \nodata  \\
            & $1532.533$ & $ 0.667$ & \nodata               & \nodata       & $<14.63$              & $< 27$          & \nodata  \\ \hline
\ion{ S}{1} & $1425.030$ & $ 2.251$ & \nodata               & \nodata       & $<12.59$              & $<  9$          & \nodata  \\
            & $1295.653$ & $ 2.052$ & \nodata               & \nodata       & $<13.08$              & $< 16$          & \nodata  \\ \hline
\ion{ S}{2} & $1250.578$ & $ 0.832$ & \nodata               & \nodata       & $<14.32$              & $< 17$          & \nodata  \\ \hline
\ion{Fe}{2} & $2382.765$ & $ 2.882$ & \nodata               & \nodata       & $<12.43$              & $< 45$          & \nodata  \\
            & $2600.173$ & $ 2.793$ & \nodata               & \nodata       & $<12.50$              & $< 43$          & \nodata  \\
            & $2344.214$ & $ 2.427$ & \nodata               & \nodata       & $<13.01$              & $< 54$          & \nodata  \\
            & $2586.650$ & $ 2.252$ & \nodata               & \nodata       & $<13.10$              & $< 55$          & \nodata  \\
            & $1608.451$ & $ 1.968$ & \nodata               & \nodata       & $<13.34$              & $< 31$          & \nodata  \\
            & $2374.461$ & $ 1.871$ & \nodata               & \nodata       & $<13.55$              & $< 54$          & \nodata  \\
            & $2260.780$ & $ 0.742$ & \nodata               & \nodata       & $<14.63$              & $< 50$          & \nodata  \\
            & $2249.877$ & $ 0.612$ & \nodata               & \nodata       & $<14.62$              & $< 34$          & \nodata
\enddata
\tablenotetext{a}{Uncertainties on measured values are reported at
{$1\sigma$} confidence. For integrated apparent column densities and
equivalent widths, two uncertainties are quoted. The first is the
estimated {$1\sigma$} uncertainty from the quadrature addition of statistical and continuum
placement errors. The second is the uncertainty
resulting from changing the width of the integration range by 10\,\kms.}
\tablenotetext{b}{Velocity centroids and $b$-values are derived using
the apparent optical depth method (see equation 3) and are reported
only if the equivalent width exceeds five times its error. The second
errors quote for the velocity centroid reflect the uncertainty in the
absolute wavelength calibration ($\sim5$\,\kms\ for FUSE,
$\sim1.5$\,\kms\ for E140M, and $\sim2.5$\,\kms\ for E230M). These do
not affect the measurements of $b$-values, which constitute
differential measurements about the centroid.}
\tablenotetext{c}{Apparent column density upper limits are {$3\sigma$}
confidence estimates derived from direct integration over the
integration range 85--140\,\kms.}
\tablenotetext{d}{Equivalent width limits are {$3\sigma$} confidence
estimates over the integration range 85--140\,\kms.}
\tablenotetext{e}{Measurements on the Lyman series lines are highly
uncertain due to blending with lower velocity absorption.}
\tablenotetext{f}{For the {1277.245\,\AA}, {1328.833\,\AA}, and
{1656.928\,\AA} lines, we adopt the self-consistent values of
{$f\lambda$} derived by \citet{jt01}.}
\label{tab:hvc100measure}
\end{deluxetable}

\begin{deluxetable}{lrcccccc}
\tabletypesize{\scriptsize}
\tablewidth{0pc}
\tablecaption{{${v_{\mathrm{LSR}}}\approx+184$\,\kms} HVC
Measurements\tablenotemark{a}}
\tablehead {
\colhead{Ion} & \colhead{Transition} & \colhead{$\log f \lambda$} & \colhead{$\langle{v_{\mathrm{LSR}}}\rangle$\tablenotemark{b}} & \colhead{$b$\tablenotemark{b}} & \colhead{$\log \acd$\tablenotemark{c}} & \colhead{$W_{\lambda}$\tablenotemark{d}} & \colhead{Integration Range} \\
& \colhead{(\AA)} & & \colhead{(\kms)} & \colhead{(\kms)} &
\colhead{(cm$^{-2}$)} & \colhead{(m\AA)} & \colhead{(\kms)}}
\startdata
\ion{ H}{1} & $1025.722$ & $ 1.909$ & $180.2\pm 2.0\pm 5.0$ & $36.3\pm 1.2$ & $15.16\pm0.04\pm0.05$ & $291\pm 6\pm31$ & 140 to 230 \\
            & $ 972.537$ & $ 1.450$ & $177.9\pm 2.3\pm 5.0$ & $33.9\pm 1.3$ & $15.50\pm0.05\pm0.04$ & $259\pm 6\pm22$ & 140 to 230 \\
            & $ 949.743$ & $ 1.122$ & $177.4\pm 2.2\pm 5.0$ & $34.0\pm 1.5$ & $15.84\pm0.05\pm0.04$ & $256\pm 7\pm23$ & 140 to 230 \\
            & $ 937.803$ & $ 0.864$ & $175.5\pm 1.7\pm 5.0$ & $32.5\pm 1.1$ & $16.16\pm0.04\pm0.03$ & $245\pm 6\pm19$ & 140 to 230 \\
            & $ 930.748$ & $ 0.652$ & $177.0\pm 2.0\pm 5.0$ & $32.2\pm 1.3$ & $16.22\pm0.04\pm0.04$ & $235\pm 6\pm22$ & 140 to 230 \\
            & $ 926.226$ & $ 0.470$ & $178.3\pm 1.7\pm 5.0$ & $29.4\pm 1.2$ & $16.37\pm0.05\pm0.02$ & $221\pm 7\pm15$ & 140 to 230 \\
            & $ 923.150$ & $ 0.311$ & $186.2\pm 1.6\pm 5.0$ & $32.1\pm 1.5$ & $16.49\pm0.05\pm0.03$ & $226\pm 6\pm20$ & 140 to 230 \\
            & $ 920.963$ & $ 0.170$ & $184.6\pm 1.5\pm 5.0$ & $29.4\pm 1.5$ & $16.61\pm0.05\pm0.02$ & $213\pm 7\pm17$ & 140 to 230 \\
            & $ 919.351$ & $ 0.043$ & $179.6\pm 1.2\pm 5.0$ & $26.3\pm 1.4$ & $16.75\pm0.05\pm0.02$ & $199\pm 8\pm12$ & 140 to 230 \\
            & $ 918.129$ & $-0.072$ & $182.5\pm 1.6\pm 5.0$ & $27.6\pm 1.7$ & $16.73\pm0.06\pm0.02$ & $183\pm11\pm12$ & 140 to 230 \\ \hline
\ion{ C}{1}\tablenotemark{e}
            & $1656.928$ & $ 2.367$ & \nodata               & \nodata       & $<13.09$              & $< 46$          & \nodata  \\
            & $1277.245$ & $ 2.225$ & \nodata               & \nodata       & $<13.24$              & $< 22$          & \nodata  \\
            & $ 945.191$ & $ 2.157$ & \nodata               & \nodata       & $<13.55$              & $< 38$          & \nodata  \\
            & $1560.309$ & $ 2.082$ & \nodata               & \nodata       & $<13.30$              & $< 35$          & \nodata  \\
            & $1328.833$ & $ 2.077$ & \nodata               & \nodata       & $<13.23$              & $< 23$          & \nodata  \\ \hline
\ion{ C}{2} & $1334.532$ & $ 2.233$ & $192.3\pm 0.7\pm 1.5$ & $22.8\pm 0.7$ & $14.55\pm0.03\pm0.01$ & $261\pm 5\pm 9$ & 140 to 230 \\
            & $1335.708$ & $ 2.186$ & \nodata               & \nodata       & $<13.08$              & $< 22$          & \nodata  \\ \hline
\ion{ C}{3} & $ 977.020$ & $ 2.869$ & $174.9\pm 1.5\pm 5.0$ & $30.8\pm 1.3$ & $13.94\pm0.03\pm0.03$ & $225\pm 8\pm19$ & 140 to 230 \\ \hline
\ion{ C}{4} & $1548.204$ & $ 2.468$ & $178.3\pm 1.9\pm 1.5$ & $21.6\pm 2.4$ & $13.74\pm0.04\pm0.01$ & $153\pm12\pm 3$ & 140 to 230 \\
            & $1550.781$ & $ 2.167$ & $183.7\pm 3.2\pm 1.5$ & $22.7\pm 3.8$ & $13.70\pm0.06\pm0.01$ & $ 82\pm11\pm 1$ & 140 to 230 \\ \hline
\ion{ N}{1} & $1200.223$ & $ 2.018$ & \nodata               & \nodata       & $<13.47$              & $< 32$          & \nodata  \\
            & $1200.710$ & $ 1.715$ & \nodata               & \nodata       & $<13.72$              & $< 34$          & \nodata  \\ \hline
\ion{ N}{2} & $1083.994$ & $ 2.080$ & $181.0\pm 2.6\pm 5.0$ & $24.7\pm 2.4$ & $14.25\pm0.06\pm0.01$ & $117\pm14\pm 3$ & 140 to 230 \\ \hline
\ion{ N}{5} & $1238.821$ & $ 2.286$ & \nodata               & \nodata       & $<13.10$              & $< 27$          & \nodata  \\
            & $1242.804$ & $ 1.985$ & \nodata               & \nodata       & $<13.32$              & $< 22$          & \nodata  \\ \hline
\ion{ O}{1} & $1302.168$\tablenotemark{f}
                         & $ 1.796$ & $192.8\pm 1.3\pm 1.5$ & $22.0\pm 1.9$ & $14.12\pm0.02\pm0.01$ & $ 75\pm 4\pm 2$ & 140 to 230 \\
            &            &          & $180.2\pm 1.4\pm 1.5$ & $11.9\pm 3.5$ & $13.82\pm0.03\pm0.07$ & $ 37\pm 3\pm 7$ & 140 to 192 \\
\ion{O}{1}* & $1304.858$ & $ 1.795$ & \nodata               & \nodata       & $<13.46$              & $< 21$          & \nodata  \\ \hline
\ion{ O}{6} & $1031.926$ & $ 2.136$ & $182.2\pm 2.1\pm 5.0$ & $32.7\pm 1.5$ & $14.00\pm0.03\pm0.02$ & $ 99\pm 7\pm 4$ & 140 to 230 \\ \hline
\ion{Mg}{1} & $2026.477$ & $ 2.360$ & \nodata               & \nodata       & $<13.43$              & $< 96$          & \nodata  \\
            & $1707.061$ & $ 0.873$ & \nodata               & \nodata       & $<14.71$              & $< 63$          & \nodata  \\ \hline
\ion{Mg}{2} & $2796.354$ & $ 3.236$ & $184.4\pm 1.4\pm 2.5$ & $16.5\pm 2.2$ & $13.19\pm0.06\pm0.01$ & $319\pm26\pm 8$ & 140 to 230 \\
            & $2803.532$ & $ 2.933$ & $186.2\pm 3.0\pm 2.5$ & $14.3\pm 6.0$ & $13.17\pm0.07\pm0.01$ & $208\pm33\pm 6$ & 140 to 230 \\ \hline
\ion{Si}{2} & $1260.422$ & $ 3.172$ & $184.6\pm 0.6\pm 1.5$ & $17.4\pm 0.8$ & $13.50\pm0.03\pm0.01$ & $185\pm 6\pm 6$ & 140 to 230 \\
            & $1193.290$ & $ 2.842$ & $185.0\pm 1.4\pm 1.5$ & $17.9\pm 2.0$ & $13.59\pm0.05\pm0.01$ & $145\pm10\pm 1$ & 140 to 230 \\
            & $1190.416$ & $ 2.541$ & $178.5\pm 2.7\pm 1.5$ & $ 4.2\pm21.0$ & $13.56\pm0.08\pm0.01$ & $ 71\pm12\pm 3$ & 140 to 230 \\
            & $1526.707$ & $ 2.308$ & $184.8\pm 1.3\pm 1.5$ & $11.9\pm 3.4$ & $13.82\pm0.04\pm0.01$ & $108\pm 8\pm 2$ & 140 to 230 \\
            & $1304.370$ & $ 2.051$ & $179.7\pm 2.0\pm 1.5$ & $ 8.7\pm 7.7$ & $13.85\pm0.04\pm0.01$ & $ 68\pm 7\pm 2$ & 140 to 230 \\ \hline
\ion{Si}{3} & $1206.500$ & $ 3.294$ & $181.4\pm 0.6\pm 1.5$ & $20.8\pm 0.6$ & $13.37\pm0.02\pm0.01$ & $193\pm 4\pm 6$ & 140 to 230 \\ \hline
\ion{Si}{4} & $1393.760$ & $ 2.854$ & $181.0\pm 1.5\pm 1.5$ & $23.6\pm 1.9$ & $13.07\pm0.03\pm0.01$ & $ 82\pm 6\pm 3$ & 140 to 230 \\
            & $1402.773$ & $ 2.552$ & $179.7\pm 2.8\pm 1.5$ & $28.3\pm 2.6$ & $13.15\pm0.05\pm0.03$ & $ 55\pm 6\pm 3$ & 140 to 230 \\ \hline
\ion{ P}{2} & $1152.818$ & $ 2.451$ & \nodata               & \nodata       & $<13.03$              & $< 29$          & \nodata  \\
            & $1532.533$ & $ 0.667$ & \nodata               & \nodata       & $<14.75$              & $< 36$          & \nodata  \\ \hline
\ion{ S}{1} & $1425.030$ & $ 2.251$ & \nodata               & \nodata       & $<12.69$              & $< 11$          & \nodata  \\
            & $1295.653$ & $ 2.052$ & \nodata               & \nodata       & $<13.19$              & $< 21$          & \nodata  \\ \hline
\ion{ S}{2} & $1250.578$ & $ 0.832$ & \nodata               & \nodata       & $<14.46$              & $< 21$          & \nodata  \\ \hline
\ion{Fe}{2} & $2382.765$ & $ 2.882$ & $185.5\pm 1.4\pm 2.5$ & $20.0\pm 2.0$ & $13.41\pm0.04\pm0.01$ & $240\pm18\pm 4$ & 140 to 230 \\
            & $2600.173$ & $ 2.793$ & $182.7\pm 1.4\pm 2.5$ & $14.8\pm 2.9$ & $13.41\pm0.05\pm0.01$ & $186\pm18\pm11$ & 140 to 230 \\
            & $2344.214$ & $ 2.427$ & \nodata               & \nodata       & $13.36\pm0.09\pm0.02$ & $ 92\pm22\pm 7$ & 140 to 230 \\
            & $2586.650$ & $ 2.252$ & \nodata               & \nodata       & $13.47\pm0.10\pm0.03$ & $ 88\pm23\pm 9$ & 140 to 230 \\
            & $1608.451$ & $ 1.968$ & \nodata               & \nodata       & $<13.49$              & $< 39$          & \nodata  \\
            & $2374.461$ & $ 1.871$ & \nodata               & \nodata       & $<13.70$              & $< 71$          & \nodata  \\
            & $2260.780$ & $ 0.742$ & \nodata               & \nodata       & $<14.73$              & $< 63$          & \nodata  \\
            & $2249.877$ & $ 0.612$ & \nodata               & \nodata       & $<14.73$              & $< 45$          & \nodata
\enddata
\tablenotetext{a}{Uncertainties on measured values are reported at
{$1\sigma$} confidence. For integrated apparent column densities and
equivalent widths, two uncertainties are quoted. The first is the
estimated {$1\sigma$} uncertainty from the quadrature addition of statistical and continuum
placement errors. The second is the uncertainty
resulting from changing the width of the integration range by 10\,\kms.}
\tablenotetext{b}{Velocity centroids and $b$-values are derived
using the apparent optical depth method (see equation 3) and are
reported only if the equivalent width exceeds five times its error.
The second errors quote for the velocity centroid reflect the
uncertainty in the absolute wavelength calibration
($\sim5$\,{\,km\,s$^{-1}$}\ for FUSE, $\sim1.5$\,{\,km\,s$^{-1}$}\
for E140M, and $\sim2.5$\,{\,km\,s$^{-1}$}\ for E230M). These do not
affect the measurements of $b$-values, since those constitute differential measurements
about the centroid.}
\tablenotetext{c}{Apparent column density upper limits are
{$3\sigma$} confidence estimates derived from direct integration
over the integration range 140--230\,\kms.}
\tablenotetext{d}{Equivalent width limits are {$3\sigma$} confidence
estimates over the integration range 140--230\,\kms.}
\tablenotetext{e}{For the {1277.245\,\AA}, {1328.833\,\AA}, and
{1656.928\,\AA} lines, we adopt the self-consistent values of
{$f\lambda$} derived by \citet{jt01}.}
\tablenotetext{f}{Two entries are listed for the
{\ion{O}{1}$\lambda1302.169$} transition corresponding to two chosen
integration ranges. The HVC profile appears to be blended with an
unrelated absorption feature redward of the HVC. This feature may be a
weak intergalactic {Lyman $\alpha$} absorption line at
{$z\approx0.0719$} {\citep[see ][]{sembach04b}}. The first row includes
this feature in the integration, while the second row limits the
integration to exclude the feature (cutting off the integration at
${v_{\mathrm{LSR}}}=+192$\,\kms). See \S\ref{sec:184oicol} for a detailed
examination of the {\ion{O}{1}} column density for this HVC.}
\label{tab:hvc184measure}
\end{deluxetable}

\begin{deluxetable}{lrcccccc}
\tabletypesize{\tiny}
\tablewidth{0pc}
\tablecaption{Galactic Halo and Intermediate-Velocity Gas Measurements\tablenotemark{a}}
\tablehead {
\colhead{Ion} & \colhead{Transition} & \colhead{$\log f \lambda$} & \colhead{$\langle{v_{\mathrm{LSR}}}\rangle$\tablenotemark{b}} & \colhead{$b$\tablenotemark{b}} & \colhead{$\log \acd$\tablenotemark{c}} & \colhead{$W_{\lambda}$\tablenotemark{d}} & \colhead{Integration Range} \\
& \colhead{(\AA)} & & \colhead{(\AA)} & \colhead{(\kms)} & \colhead{(cm$^{-2}$)} & \colhead{(m\AA)} & \colhead{(\kms)}}
\startdata
\ion{ H}{1}\tablenotemark{e}
            & $ 930.748$ & $ 0.652$ & $  5.2\pm 3.0\pm 5.0$ & $75.2\pm 1.8$ & $16.63\pm0.03\pm0.02$ & $524\pm 6\pm27$ & -100 to  85 \\
            & $ 926.226$ & $ 0.470$ & $  3.6\pm 3.1\pm 5.0$ & $70.5\pm 1.8$ & $16.83\pm0.03\pm0.02$ & $515\pm 7\pm22$ & -100 to  85 \\
            & $ 923.150$ & $ 0.311$ & $  6.6\pm 3.5\pm 5.0$ & $70.6\pm 2.1$ & $16.96\pm0.03\pm0.02$ & $509\pm 8\pm21$ & -100 to  85 \\
            & $ 920.963$ & $ 0.170$ & $  8.9\pm 3.4\pm 5.0$ & $70.2\pm 2.1$ & $17.11\pm0.03\pm0.03$ & $501\pm 9\pm21$ & -100 to  85 \\
            & $ 919.351$ & $ 0.043$ & $ -0.3\pm 3.9\pm 5.0$ & $69.7\pm 2.3$ & $17.21\pm0.03\pm0.01$ & $505\pm 9\pm18$ & -100 to  85 \\
            & $ 918.129$ & $-0.072$ & $-11.9\pm 3.7\pm 5.0$ & $72.2\pm 2.4$ & $17.38\pm0.03\pm0.02$ & $532\pm 8\pm28$ & -100 to  85 \\ \hline
\ion{ C}{1}\tablenotemark{f}
            & $1656.932$ & $ 2.367$ & \nodata               & \nodata       & $<13.29$              & $< 73$          & \nodata  \\
            & $1277.245$ & $ 2.225$ & \nodata               & \nodata       & $<13.38$              & $< 34$          & \nodata  \\
            & $ 945.191$ & $ 2.157$ & \nodata               & \nodata       & $<13.71$              & $< 53$          & \nodata  \\
            & $1560.309$ & $ 2.082$ & \nodata               & \nodata       & $<13.47$              & $< 55$          & \nodata  \\
            & $1328.833$ & $ 2.077$ & \nodata               & \nodata       & $<13.44$              & $< 42$          & \nodata  \\ \hline
\ion{ C}{2} & $1334.532$ & $ 2.233$ & $ -2.6\pm 1.5\pm 1.5$ & $59.8\pm 1.0$ & $15.04\pm0.02\pm0.01$ & $666\pm 6\pm 7$ & -100 to  85 \\ \hline
\ion{ C}{3} & $ 977.020$ & $ 2.869$ & $  3.9\pm 2.0\pm 5.0$ & $66.4\pm 1.4$ & $14.33\pm0.02\pm0.01$ & $508\pm 8\pm20$ & -100 to  85 \\ \hline
\ion{ C}{4} & $1548.204$ & $ 2.468$ & $-12.2\pm 1.9\pm 1.5$ & $58.6\pm 1.5$ & $14.17\pm0.02\pm0.01$ & $401\pm14\pm 5$ & -100 to  85 \\
            & $1550.781$ & $ 2.167$ & $-14.3\pm 3.9\pm 1.5$ & $61.3\pm 3.2$ & $14.17\pm0.03\pm0.01$ & $234\pm18\pm 3$ & -100 to  85 \\ \hline
\ion{ N}{1} & $1199.550$ & $ 2.200$ & $-29.4\pm 1.7\pm 1.5$ & $29.2\pm 3.0$ & $14.67\pm0.03\pm0.01$ & $282\pm14\pm 1$ & -100 to  85 \\
            & $1200.710$ & $ 1.715$ & $-26.1\pm 1.9\pm 1.5$ & $30.9\pm 3.2$ & $15.08\pm0.03\pm0.01$ & $247\pm17\pm 4$ & -100 to  85 \\ \hline
\ion{ N}{2} & $1083.994$ & $ 2.080$ & $ -9.0\pm 2.8\pm 5.0$ & $59.6\pm 2.3$ & $14.88\pm0.03\pm0.01$ & $411\pm19\pm 6$ & -100 to  85 \\ \hline
\ion{ N}{5} & $1238.821$ & $ 2.286$ & \nodata               & \nodata       & $<13.14$              & $< 31$          & \nodata  \\
            & $1242.804$ & $ 1.985$ & \nodata               & \nodata       & $<13.48$              & $< 33$          & \nodata  \\ \hline
\ion{ O}{1} & $1302.168$ & $ 1.796$ & $-21.2\pm 0.8\pm 1.5$ & $46.5\pm 0.7$ & $15.24\pm0.01\pm0.01$ & $467\pm 5\pm 2$ & -100 to  85 \\ \hline
\ion{ O}{6} & $1031.926$ & $ 2.136$ & $ 18.5\pm 4.0\pm 5.0$ & $42.2\pm 6.6$ & $14.16\pm0.03\pm0.01$\tablenotemark{g}
                                                                                                    & $136\pm14\pm 1$ & -100 to  85 \\ \hline
\ion{Mg}{1} & $2026.477$ & $ 2.360$ & \nodata               & \nodata       & $<13.59$              & $<180$          & \nodata  \\
            & $1707.061$ & $ 0.873$ & \nodata               & \nodata       & $<14.86$              & $< 89$          & \nodata  \\ \hline
\ion{Mg}{2} & $2796.354$ & $ 3.236$ & $ -7.6\pm 2.6\pm 2.5$ & $57.1\pm 1.8$ & $13.90\pm0.03\pm0.01$ & $1218\pm32\pm 2$ & -100 to  85 \\
            & $2803.532$ & $ 2.933$ & $ -9.5\pm 2.8\pm 2.5$ & $55.9\pm 2.0$ & $14.11\pm0.03\pm0.01$ & $1091\pm38\pm 3$ & -100 to  85 \\ \hline
\ion{Si}{2} & $1260.422$ & $ 3.172$ & $ -7.6\pm 1.7\pm 1.5$ & $57.9\pm 1.2$ & $14.03\pm0.02\pm0.01$ & $598\pm 7\pm 4$ & -100 to  85 \\
            & $1193.290$ & $ 2.842$ & $-10.2\pm 2.2\pm 1.5$ & $55.0\pm 1.6$ & $14.25\pm0.02\pm0.01$ & $501\pm12\pm 2$ & -100 to  85 \\
            & $1190.416$ & $ 2.541$ & $-17.9\pm 2.5\pm 1.5$ & $56.8\pm 2.1$ & $14.44\pm0.03\pm0.01$ & $465\pm15\pm 4$ & -100 to  85 \\
            & $1526.707$ & $ 2.308$ & $-18.7\pm 1.3\pm 1.5$ & $45.4\pm 1.2$ & $14.71\pm0.02\pm0.01$ & $525\pm10\pm 1$ & -100 to  85 \\
            & $1304.370$ & $ 2.051$ & $-21.8\pm 0.9\pm 1.5$ & $42.9\pm 1.2$ & $14.95\pm0.02\pm0.01$ & $427\pm 8\pm 4$ & -100 to  85 \\ \hline
\ion{Si}{3} & $1206.500$ & $ 3.294$ & $ -0.3\pm 1.3\pm 1.5$ & $61.6\pm 0.8$ & $13.92\pm0.01\pm0.01$ & $590\pm 5\pm 6$ & -100 to  85 \\ \hline
\ion{Si}{4} & $1393.760$ & $ 2.854$ & $-18.9\pm 1.3\pm 1.5$ & $48.6\pm 1.5$ & $13.68\pm0.01\pm0.01$ & $293\pm 8\pm 2$ & -100 to  85 \\
            & $1402.773$ & $ 2.552$ & $ -9.5\pm 1.9\pm 1.5$ & $43.9\pm 2.5$ & $13.71\pm0.02\pm0.01$ & $182\pm 7\pm 4$ & -100 to  85 \\ \hline
\ion{ P}{2} & $1152.818$ & $ 2.451$ & $-25.5\pm 8.5\pm 5.0$ & $43.2\pm13.3$ & $13.59\pm0.08\pm0.01$ & $ 76\pm15\pm 1$ & -100 to  85 \\
            & $1532.533$ & $ 0.667$ & \nodata               & \nodata       & $<14.74$              & $< 41$          & \nodata  \\ \hline
\ion{ S}{1} & $1425.030$ & $ 2.251$ & \nodata               & \nodata       & $<12.84$              & $< 18$          & \nodata  \\
            & $1295.653$ & $ 2.052$ & \nodata               & \nodata       & $<13.41$              & $< 34$          & \nodata  \\ \hline
\ion{ S}{2} & $1250.578$ & $ 0.832$ & $-38.9\pm 3.0\pm 1.5$ & $49.6\pm 4.5$ & $15.49\pm0.03\pm0.03$ & $167\pm10\pm11$ & -100 to  85 \\ \hline
\ion{Fe}{2} & $2382.765$ & $ 2.882$ & $-12.4\pm 2.2\pm 2.5$ & $55.7\pm 1.6$ & $14.13\pm0.02\pm0.01$ & $924\pm23\pm 3$ & -100 to  85 \\
            & $2600.173$ & $ 2.793$ & $-13.3\pm 1.7\pm 2.5$ & $51.7\pm 1.4$ & $14.22\pm0.02\pm0.01$ & $966\pm23\pm10$ & -100 to  85 \\
            & $2344.214$ & $ 2.427$ & $-21.2\pm 1.7\pm 2.5$ & $42.0\pm 1.9$ & $14.55\pm0.03\pm0.01$ & $760\pm28\pm 3$ & -100 to  85 \\
            & $2586.650$ & $ 2.252$ & $-23.9\pm 1.9\pm 2.5$ & $38.6\pm 2.3$ & $14.61\pm0.04\pm0.01$ & $689\pm29\pm 6$ & -100 to  85 \\
            & $1608.451$ & $ 1.968$ & $-27.5\pm 1.8\pm 1.5$ & $40.1\pm 2.6$ & $14.79\pm0.03\pm0.01$ & $373\pm16\pm 2$ & -100 to  85 \\
            & $2374.461$ & $ 1.871$ & $-28.4\pm 2.2\pm 2.5$ & $40.8\pm 3.3$ & $14.86\pm0.03\pm0.01$ & $513\pm30\pm 7$ & -100 to  85 \\
            & $2260.780$ & $ 0.742$ & $-21.1\pm 7.9\pm 2.5$ & $44.5\pm12.0$ & $15.16\pm0.06\pm0.01$ & $134\pm23\pm 3$ & -100 to  85 \\
            & $2249.877$ & $ 0.612$ & \nodata               & \nodata       & $15.04\pm0.11\pm0.03$ & $ 73\pm24\pm 5$ & -100 to  85
\enddata
\tablenotetext{a}{Uncertainties on measured values are reported at
{$1\sigma$} confidence. For integrated apparent column densities
and equivalent widths, two uncertainties are quoted. The first is
the estimated {$1\sigma$} uncertainty from the quadrature addition
of statistical and continuum placement errors. The second is the
uncertainty resulting from changing the width of the integration
range by 10\,\kms.}
\tablenotetext{b}{Velocity centroids and $b$-values are derived
using the apparent optical depth method (see equation 3) and are
reported only if the equivalent width exceeds five times its error.
The second errors quote for the velocity centroid reflect the
uncertainty in the absolute wavelength calibration
($\sim5$\,{\,km\,s$^{-1}$}\ for FUSE, $\sim1.5$\,{\,km\,s$^{-1}$}\
for E140M, and $\sim2.5$\,{\,km\,s$^{-1}$}\ for E230M). These do not
affect the measurements of $b$-values, since those constitute differential measurements
about the centroid.}
\tablenotetext{c}{Apparent column density upper limits are
{$3\sigma$} confidence estimates derived from direct integration
over the integration range -100 to +85\,\kms.}
\tablenotetext{d}{Equivalent width limits are reported to
{$3\sigma$} confidence over the integration range -100 to +85\,\kms.}
\tablenotetext{e}{The Lyman {$\beta-\eta$} absorption lines at low
velocities are contaminated by geocoronal emission. Consequently, we do
not present measurements of these lines. Moreover, the apparent
increases in the equivalent widths of the Lyman $\kappa$~(919.351\,\AA)
and Lyman $\lambda$~(918.129\,\AA) lines are due to blends in the
region of the high order Lyman series with higher order Lyman series
absorption, high velocity Lyman series absorption features, and
{\ion{O}{1}} absorption.}
\tablenotetext{f}{For the {1277.245\,\AA}, {1328.833\,\AA}, and
{1656.928\,\AA} lines, we adopt the self-consistent values of {$f\lambda$}
derived by \citet{jt01}.}
\tablenotetext{g}{We note that {\citet{savage03}} find {$\log
N$(\ion{O}{6})$=14.26\pm0.03$} over the velocity range {-55\,\kms}
to {+115\,\kms}, which agrees well with the listed value if the
column density of the {+100\,\kms} component listed in
Table~\ref{tab:hvc100measure} is included with our estimate.}
\label{tab:haloivcmeasure}
\end{deluxetable}

\begin{deluxetable}{lccccc}
\tabletypesize{\footnotesize}
\tablewidth{0pc}
\tablecaption{Adopted Column Densities}
\tablehead {
& \multicolumn{5}{c}{$\log [N~(\mathrm{cm}^{-2})]$\tablenotemark{a}} \\
& \multicolumn{5}{c}{\hrulefill} \\
\colhead{Ion} &
\colhead{$-44$\,\kms} &
\colhead{$-7$\,\kms} &
\colhead{$+56$\,\kms} &
\colhead{$100$\,\kms} &
\colhead{$184$\,\kms}
}
\startdata
\ion{ C}{1} & $13.23\pm0.08$   & $<12.87$         & $<12.80$         & $<12.83$         & $<12.94$         \\
\ion{ C}{2} & $>14.56$         & $>14.68$         & $>14.39$         & $ 13.28\pm 0.07$ & $>14.55$         \\
\ion{ C}{3} & $>13.78$         & $>13.88$         & $>13.77$         & $>13.71$         & $>13.94$         \\
\ion{ C}{4} & $ 13.77\pm 0.03$ & $ 13.82\pm 0.03$ & $ 13.36\pm 0.05$ & $ 13.34\pm 0.03$ & $ 13.74\pm 0.04$ \\
\ion{ N}{1} & $>14.75$         & $>14.79$         & $<13.65$         & $<13.66$         & $<13.47$         \\
\ion{ N}{2} & $>14.44$         & $>14.54$         & $ 14.11\pm 0.06$ & $<13.48$         & $>14.25$         \\
\ion{ N}{5} & $<12.78$         & $<12.84$         & $<12.91$         & $<12.91$         & $<13.10$         \\
\ion{ O}{1} & $>14.92$         & $>14.91$         & $>14.06$         & $<13.13$         & $ 13.82_{-0.03}^{+0.09}$\tablenotemark{b} \\
\ion{ O}{6} & $<13.24$         & $ 13.94\pm 0.03$ & $ 13.65\pm 0.06$ & $ 13.29\pm 0.12$ & $ 14.00\pm 0.03$ \\
\ion{Mg}{1} & $<13.43$         & $<13.47$         & $<13.06$         & $<13.33$         & $<13.43$         \\
\ion{Mg}{2} & $>13.70$         & $>13.76$         & $>13.20$         & $<12.18$         & $ 13.18\pm0.05$  \\
\ion{Si}{2} & $>14.64$         & $>14.60$         & $>13.74$         & $ 12.14\pm0.11$  & $ 13.85\pm0.04$  \\
\ion{Si}{3} & $>13.45$         & $>13.53$         & $>13.35$         & $>12.78$         & $>13.37$         \\
\ion{Si}{4} & $ 13.27\pm 0.01$ & $ 13.37\pm 0.02$ & $ 12.58\pm 0.05$ & $ 12.54\pm 0.07$ & $ 13.09_{-0.03}^{+0.06}$ \\
\ion{ P}{2} & $ 13.38\pm 0.10$ & $ 13.07\pm0.10$  & $<12.95$         & $<12.85$         & $<13.03$         \\
\ion{ S}{1} & $<12.62$         & $<12.59$         & $<12.55$         & $<12.59$         & $<12.69$         \\
\ion{ S}{2} & $>15.12$         & $ 14.95\pm 0.03$ & $<13.94$         & $<14.32$         & $<14.46$         \\
\ion{Fe}{2} & $ 14.92\pm0.06$  & $ 14.68\pm0.11$  & $ 13.30\pm 0.03$ & $<12.43$         & $ 13.46\pm 0.09$
\enddata
\tablenotetext{a}{Uncertainties on measured values are reported at
{1$\sigma$} confidence, and include statistical and continuum placement uncertainties only.
Column density upper limits are reported at {$3\sigma$} confidence. Lower limits are quoted
at {$1\sigma$} confidence (see text). The integration ranges for the components
were: $-100\leq{v_{\mathrm{LSR}}}\leq-25$\,\kms\ for ${v_{\mathrm{LSR}}}\approx-44$\,\kms;
$-25\leq{v_{\mathrm{LSR}}}\leq+37$\,\kms\ for ${v_{\mathrm{LSR}}}\approx-7$\,\kms;
$+37\leq{v_{\mathrm{LSR}}}\leq+85$\,\kms\ for ${v_{\mathrm{LSR}}}\approx+56$\,\kms;
$+85\leq{v_{\mathrm{LSR}}}\leq+140$\,\kms\ for ${v_{\mathrm{LSR}}}\approx+100$\,\kms;
$+140\leq{v_{\mathrm{LSR}}}\leq+230$\,\kms\ for ${v_{\mathrm{LSR}}}\approx+184$\,\kms.}
\tablenotetext{b}{See Table~\ref{tab:hvc184measure} and text (\S\ref{sec:184oicol}) for
details regarding our adopted {\ion{O}{1}} column density and
errors.}
\label{tab:coladopt}
\end{deluxetable}


\begin{deluxetable}{lccc}
\tabletypesize{\tiny} \tablewidth{0pc} \tablecaption{Column Density
Ratios and Relative Abundance Constraints for High Ionization
Species\tablenotemark{a}} \tablehead {
\colhead{Model\tablenotemark{b}/Component} &
\colhead{$\acd$(\ion{Si}{4})/$\acd$(\ion{O}{6})} &
\colhead{$\acd$(\ion{C}{4})/$\acd$(\ion{O}{6})} &
\colhead{$\acd$(\ion{N}{5})/$\acd$(\ion{O}{6})} } \startdata
IVC @ ${v_{\mathrm{LSR}}}\approx -44$\,\kms & $>1.07$        & $>3.39$        & \nodata \\
ISM @ ${v_{\mathrm{LSR}}}\approx  -7$\,\kms & $ 0.27\pm0.02$ & $ 0.76\pm0.07$ & $<0.08$ \\
IVC @ ${v_{\mathrm{LSR}}}\approx +56$\,\kms & $ 0.09\pm0.02$ & $ 0.51\pm0.09$ & $<0.18$ \\
HVC @ ${v_{\mathrm{LSR}}}\approx+100$\,\kms & $ 0.17\pm0.05$ & $ 1.12\pm0.25$ & $<0.42$ \\
HVC @ ${v_{\mathrm{LSR}}}\approx+184$\,\kms & $ 0.12\pm0.01$ & $ 0.55\pm0.06$ & $<0.11$ \\
\hline
Radiative Cooling\tablenotemark{c}       & 0.001--0.013 & 0.036--0.170 & 0.053--0.090 \\
Conductive Interfaces\tablenotemark{d}   & 0.003--0.058 & 0.042--0.930 & 0.065--0.510 \\
Turbulent Mixing Layers\tablenotemark{e} & 0.087--1.290 & 1.050--6.920 & 0.170--0.580 \\
Shock Ionization\tablenotemark{f}        & 0.002--0.180 & 0.016--1.040 & 0.031--0.051
\enddata
\tablenotetext{a}{Uncertainties on values are quoted at {$1\sigma$} confidence, while ranges are quote at {$3\sigma$} confidence.}
\tablenotetext{b}{The collision ionization models considered are
described in {\citet[][radiative cooling]{heck02}},
{\citet[][conductive interfaces]{bbf90}}, {\citet[][turbulent mixing
layers]{ssb93}}, {\citet[][shock ionization]{ds96}}. {\citet{fox04}}
summarize the expected ratios for solar and Complex C abundance
patterns and the model assumptions used compare these ratios. The ranges quoted
here assume solar relative abundances, and the model assumptions adopted by
{\citet{fox04}} are summarized below.}
\tablenotetext{c}{The quoted ratios assume gas cooling from a temperature
{$10^6$\,K} with a cooling flow velocity of {100\,\kms} for isobaric
and isochoric (and intermediate) conditions.}
\tablenotetext{d}{The quoted ratios assume magnetic field orientations in the range
{0--85$^\circ$}, and interface ages in the range {$10^5$--$10^7$\,yrs}.}
\tablenotetext{e}{The quoted ratios assume gas with entrainment velocities in the
range {25--100\,\kms} and mixing-layer temperatures in the range
{1--3$\times 10^5$\,K}.}
\tablenotetext{f}{The quoted ratios assume shock velocities in the range
{150-500\,\kms}, and magnetic parameters in the range {0--4\,$\mu$G cm$^{-3/2}$}.}
\label{tab:cine}
\end{deluxetable}

\begin{deluxetable}{rcccc}
\tabletypesize{\small} \tablewidth{0pc} \tablecaption{Model
Parameters for Lyman Series Fit\tablenotemark{a}}
\tablehead {
\colhead{Velocity}      & \colhead{$b$(\ion{H}{1})} & \colhead{$\log N$(\ion{H}{1})} & \colhead{$b$(\ion{O}{1})} & \colhead{$\log N$(\ion{O}{1})} \\
\colhead{(km~s$^{-1}$)} & \colhead{(km~s$^{-1}$)}   & \colhead{(cm$^{-2}$)}          & \colhead{(km~s$^{-1}$)}   & \colhead{(cm$^{-2}$)}
}
\startdata
$-43.6\pm0.5$   & $15.5$\tablenotemark{b} & $19.83$\tablenotemark{b} & $12.7\pm0.1$ & $16.49$\tablenotemark{b} \\
$ -8.2\pm0.8$   & $14.7$\tablenotemark{b} & $19.70$\tablenotemark{b} & $ 8.8\pm0.1$ & $16.36$\tablenotemark{b} \\
$ 26.7\pm1.0$   & $37.0\pm3.1$            & $17.29$\tablenotemark{b} & $13.0\pm3.8$ & $13.95\pm0.07$ \\
$ 55.93\pm0.07$ & $ 9.9\pm0.3$            & $17.58$\tablenotemark{b} & $ 3.8\pm0.2$ & $14.24\pm0.02$ \\
$100$\tablenotemark{b}
                & $21.9\pm1.7$            & $16.42\pm0.02$           & \nodata      & \nodata        \\ \hline
$z=0.0719$      & $ 9.6\pm1.53$           & $12.81\pm0.04$ &
\nodata      & \nodata        \\ \hline
$183.3\pm0.2$\tablenotemark{c}
                & $ \begin{array}{c} 25.9\pm0.6 \\ 14.2\pm0.1 \end{array}$ & $\begin{array}{c} 16.73\tablenotemark{b} \\ 18.3\tablenotemark{b} \end{array} $
                &  $ 3.9\pm0.3$ & $13.82$\tablenotemark{b}
\enddata
\tablenotetext{a}{Uncertainties are quoted at {$1\sigma$}
confidence and contain statistical and continuum placement errors
only. The fitting process assumed a constant Gaussian instrumental
line spread function with a FWHM resolution of {20\,\kms}.}
\tablenotetext{b}{These values were held fixed in the fitting
process. Thus no uncertainty is quoted. See text for details on
the derivation of these values.}
\tablenotetext{c}{We performed two least-squares fits to the data,
holding the {\ion{H}{1}} column density fixed in each case, and
allowing the component velocity and {\ion{H}{1}} Doppler width to
vary. In the first fit (see Figure~\ref{fig:lyman}, panel a), we
fixed {$N$(\ion{H}{1})} at the value derived from integrating the
Lyman~$\lambda$\ line over the velocity range 140--230\,\kms. In
the second fit (see Figure~\ref{fig:lyman}, panel c), we fixed
{$N$(\ion{H}{1})} at the column density limit quoted by
{\citet{wakker03}} from the non-detection of 21\,cm emission. See
text for further details of the fitting procedure.}
 \label{tab:hifit}
\end{deluxetable}

\begin{deluxetable}{lll}
\tablewidth{0pc}
\tablecaption{Adopted Solar Abundances}
\tablehead {
\colhead{Element} & \colhead{A$_\odot$\tablenotemark{a}} &
\colhead{Reference} }
\startdata
 C & 8.39  & \citet{ala02} \\
 N & 7.931 & \citet{hol01} \\
 O & 8.66  & \citet{asplund04} \\
Mg & 7.538 & \citet{hol01} \\
Al & 6.48  & \citet{ag89} \\
Si & 7.536 & \citet{hol01} \\
 P & 5.57  & \citet{ag89} \\
 S & 7.27  & \citet{ag89} \\
Fe & 7.448 & \citet{hol01}
\enddata
\tablenotetext{a}{We adopt the standard notation for abundances:
A$_\odot$(X)$=12+\log [N$(X)$/N$(H)$]_\odot$.}
\label{tab:solabund}
\end{deluxetable}

\begin{deluxetable}{crrrrr}
\tabletypesize{\scriptsize} \tablewidth{0pc} \tablecaption{Logarithmic
Column Density Ratios\tablenotemark{a}}
\tablehead {
\colhead{Sight line} & \multicolumn{2}{c}{PKS\,$2155-304$\tablenotemark{b}} & \multicolumn{2}{c}{Mrk\,509\tablenotemark{b}} & \\
\colhead{Ratio/Velocity} & \colhead{--140\,\kms} & \colhead{--270\,\kms} & \colhead{--240\,\kms} & \colhead{--300\,\kms} & \colhead{Complex C\tablenotemark{b,c}} }
\startdata
$\acd$(\ion{H}{1})/$\acd$(\ion{O}{6})   & $2.57_{-0.14}^{+0.22}$ & $1.67_{-0.23}^{+0.38}$ & $<3.80$                & $<3.76$                 & $6.38_{-0.05}^{+0.05}$\tablenotemark{d} \\
$\acd$(\ion{O}{1})/$\acd$(\ion{O}{6})   & $<-0.40$               & $<-0.05$               & $<-0.42$               & $<-0.31$                & $2.15\pm0.14$ \\
$\acd$(\ion{Si}{2})/$\acd$(\ion{Si}{4}) & $0.07_{-0.13}^{+0.14}$ & $<0.51$                & \nodata                & $<-0.31$                & $2.03_{-0.20}^{+0.22}$ \\
$\acd$(\ion{C}{2})/$\acd$(\ion{C}{4})   & $0.36_{-0.08}^{+0.11}$  & $<-0.56$               & $<-0.38$               & $-0.47_{-0.06}^{+0.06}$ & \nodata \\
$\acd$(\ion{C}{4})/$\acd$(\ion{Si}{4})  & $0.74_{-0.09}^{+0.08}$ & $1.30_{-0.16}^{+0.12}$ & $>1.04$                & $0.78_{-0.07}^{+0.06}$   & $0.59_{-0.15}^{+0.12}$ \\
$\acd$(\ion{C}{4})/$\acd$(\ion{N}{5})   & $>0.40$                & $>0.47$                & $>0.45$                & $>0.94$                 & $>0.07$ \\
$\acd$(\ion{O}{6})/$\acd$(\ion{C}{4})   & $0.32_{-0.05}^{+0.05}$ &
$0.00_{-0.07}^{+0.06}$ & $0.36_{-0.07}^{+0.06}$ &
$-0.25_{-0.06}^{+0.05}$ & $0.31_{-0.09}^{+0.08}$ \\ \hline\hline
Sight line                            & \multicolumn{2}{c}{HE\,$0226-4110$\tablenotemark{e}} & PG\,$0953+414$\tablenotemark{e} & \\
Ratio/Velocity                        & +145\,\kms    & +200\,\kms     & +120\,\kms    \\ \hline
$\acd$(\ion{H}{1})/$\acd$(\ion{O}{6})   & $2.99\pm0.27$ & $<3.16$        & $3.01\pm0.23$ \\
$\acd$(\ion{O}{1})/$\acd$(\ion{O}{6})   & $<0.19$       & $<0.30$        & $<0.61$       \\
$\acd$(\ion{Si}{2})/$\acd$(\ion{Si}{4}) & $>0.28$       & $-0.27\pm0.16$ & $<0.21$       \\
$\acd$(\ion{C}{2})/$\acd$(\ion{C}{4})   & $0.66\pm0.14$ & $0.67\pm0.19$  & $>0.41$       \\
$\acd$(\ion{C}{4})/$\acd$(\ion{Si}{4})  & $>1.04$       & $0.31\pm0.24$  & $0.84\pm0.33$ \\
$\acd$(\ion{C}{4})/$\acd$(\ion{N}{5})   & $>0.06$       & $>0.20$        & $>-0.19$      \\
$\acd$(\ion{O}{6})/$\acd$(\ion{C}{4})   & $0.10\pm0.14$ & $>-0.19$ &
$0.09\pm0.21$ \\ \hline\hline
Sight Line & \multicolumn{5}{c}{PG\,$1116+215$} \\
Ratio/Velocity                        & --44\,\kms     & --7\,\kms &
+56\,\kms      & +100\,\kms      & +184\,\kms \\ \hline
$\acd$(\ion{H}{1})/$\acd$(\ion{O}{6})   & $>6.59$        & $
5.77\pm0.03$ &  \nodata       & $ 3.13\pm0.012$\tablenotemark{f}
                                                                                                           & $ 3.82\pm0.18$ \\
$\acd$(\ion{O}{1})/$\acd$(\ion{O}{6})   & $>1.68$        & $>0.97$        & $>0.41$        & $<-0.16$        & $-0.18_{-0.04}^{+0.09}$ \\
$\acd$(\ion{Si}{2})/$\acd$(\ion{Si}{4}) & $>1.37$        & $>1.23$        & $>1.16$        & $-0.40\pm0.13$  & $ 0.76\pm0.07$ \\
$\acd$(\ion{C}{2})/$\acd$(\ion{C}{4})   & $>0.79$        & $>0.86$        & $>1.03$        & $ -0.06\pm0.08$ & $>0.81$ \\
$\acd$(\ion{C}{4})/$\acd$(\ion{Si}{4})  & $ 0.50\pm0.03$ & $ 0.45\pm0.04$ & $ 0.78\pm0.07$ & $  0.80\pm0.08$ & $ 0.65\pm0.07$ \\
$\acd$(\ion{C}{4})/$\acd$(\ion{N}{5})   & $>0.99$        & $>0.98$        & $>0.45$        & $ >0.43$        & $>0.64$ \\
$\acd$(\ion{O}{6})/$\acd$(\ion{C}{4})   & $<-0.53$       &
$0.12\pm0.04$ & $ 0.29\pm0.08$ & $ -0.05\pm0.12$ & $0.26\pm0.05$
\enddata
\tablenotetext{a}{Uncertainties are quoted at {$1\sigma$} confidence, while limits are quoted at {$3\sigma$} confidence.}
\tablenotetext{b}{The column density ratios for these HVCs were taken from {\cite{csg04}}. See their Table 5 for notes.}
\tablenotetext{c}{Column density ratios for Complex C are for the
sight line toward PG\,$1259+593$, from {\citet{csg03}}. We update
the {$N$(\ion{O}{1})$/N$(\ion{O}{6})} ratio using the {\ion{O}{1}}
column density reported by {\citet{sembach04a}},
$(7.2\pm2.1)\times 10^{15}~\mathrm{cm}^{-2}$, and the
{\ion{O}{6}} column density reported by {\citet{fox04}},
$(5.13\pm0.47)\times 10^{13}~\mathrm{cm}^{-2}$.}
\tablenotetext{d}{\citet{sembach03} report $N$(\ion{H}{1})$/N$(\ion{O}{6}) for a
number of sight lines through Complex C ranging from
{$4\times10^4$} (toward Mrk\,506) to {$1.7\times10^6$} (toward
PG\,$1259+593$).}
\tablenotetext{e}{The column density ratios for these HVC were taken from Tables 4, 5, and 9 of {\citet{fox05}}.}
\tablenotetext{f}{For the {\ion{H}{1}} column density of the {${v_{\mathrm{LSR}}}\approx+100$\,\kms} component,
we assume the value from the Lyman series fit ($\log N$(\ion{H}{1})$=16.42\pm0.02$; see Table~\ref{tab:hifit}),
which is consistent with the upper limit on the {\ion{O}{1}} column density and a solar metallicity.}
\label{tab:cdr}
\end{deluxetable}

\begin{deluxetable}{rrrrr}
\tabletypesize{\small}
\tablewidth{0pc}
\tablecaption{Cloud Key}
\tablehead {
\colhead{Number} & \colhead{Sight Line} & \colhead{Velocity} & \colhead{Gal. Long.} & \colhead{Gal. Lat.} \\
                 &                      & \colhead{(\kms)}   & \colhead{($^\circ$)} & \colhead{($^\circ$)}
}
\startdata
 1 & PG\,$1116+215$  & $ -44$ & $223\fdg36$ & $+68\fdg21$ \\
 2 & PG\,$1116+215$  & $  -7$ & $223\fdg36$ & $+68\fdg21$ \\
 3 & PG\,$1116+215$  & $ +56$ & $223\fdg36$ & $+68\fdg21$ \\
 4 & PG\,$1116+215$  & $+100$ & $223\fdg36$ & $+68\fdg21$ \\
 5 & PG\,$1116+215$  & $+184$ & $223\fdg36$ & $+68\fdg21$ \\
 6 & HE\,$0226-4110$ & $+145$ & $253\fdg94$ & $-65\fdg77$ \\
 7 & HE\,$0226-4110$ & $+200$ & $253\fdg94$ & $-65\fdg77$ \\
 8 & PG\,$0953+414$  & $+120$ & $179\fdg79$ & $+51\fdg71$ \\
 9 & PKS\,$2155-304$ & $-140$ & $ 17\fdg73$ & $-52\fdg25$ \\
10 & PKS\,$2155-304$ & $-270$ & $ 17\fdg73$ & $-52\fdg25$ \\
11 & Mrk\,509        & $-240$ & $ 35\fdg97$ & $-29\fdg86$ \\
12 & Mrk\,509        & $-300$ & $ 35\fdg97$ & $-29\fdg86$ \\
13 & PG\,$1259+593$  & $-110$ & $120\fdg56$ & $+58\fdg05$
\enddata
\label{tab:cloudkey}
\end{deluxetable}

\end{document}